\documentclass[a4paper,12pt,fleqn]{article}
\usepackage[DIV13]{typearea}
\usepackage{amsmath}
\usepackage{amssymb}
\usepackage{amsfonts}
\usepackage{bbm} 
\usepackage[footnotesize]{caption2}
\usepackage{graphicx} 
\usepackage[bbgreekl]{MyBbol}
\usepackage{accents}
\usepackage{mathrsfs}
\usepackage[center,footnotesize,hang]{subfigure}
\usepackage{bigstrut}
\usepackage{url}

\newcommand{\CenterObject}[1]{\ensuremath{\vcenter{\hbox{#1}}}}
\newcommand{\CenterEps}[2][1]{\ensuremath{\vcenter{\hbox{\includegraphics[scale=#1]{#2.eps}}}}}

\newcommand{\CiteSeeSaw}{\cite{Minkowski:1977sc,Yanagida:1980,Glashow:1979vf,Gell-Mann:1980vs,Mohapatra:1980ia}}

\newcommand{\eV}{\ensuremath{\,\mathrm{eV}}}

\newcommand{\GeV}{\ensuremath{\,\mathrm{GeV}}}
\newcommand{\TeV}{\ensuremath{\,\mathrm{TeV}}}

\def\D{\mathrm{d}} 
\def\I{\mathrm{i}}
\def\SU{\mathrm{SU}}
\def\U{\mathrm{U}}
\def\ChargeC{\mathrm{C}}                
\newcommand{\SuperField}[1]{\bbsymbol{#1}}  
\newcommand{\NuR}{N_\mathrm{R}}
\DeclareMathOperator{\re}{Re}
\DeclareMathOperator{\im}{Im}

\DeclareMathOperator{\Tr}{Tr}
\DeclareMathOperator{\diag}{diag}
\newcommand{\RaiseBrace}[1]{\raise1.5pt\hbox{$\displaystyle#1$}}

\newcommand{\package}[1]{\href{http://www.ph.tum.de/~rge/}{\tt #1}}
\newcommand{\function}[1]{{\tt #1}}
\newcommand{\param}[1]{{\tt #1}}

\newcommand{\optparam}[1]{{\tt\em #1}}

\allowdisplaybreaks[1]

\begin{document}

\begin{titlepage}

\ \vspace*{-15mm}
\begin{flushright}
TUM-HEP-576/05\\
DESY 05-013\\
SHEP/0504
\end{flushright}
\vspace*{5mm}

\begin{center}
{\Huge\sffamily\bfseries 
Running Neutrino Mass Parameters\\[2mm] 
in See-Saw Scenarios
}
\\[10mm]
{\large
Stefan Antusch\footnote{E-mail: \texttt{santusch@hep.phys.soton.ac.uk}}$^{(a)}$,
J\"{o}rn Kersten\footnote{E-mail: \texttt{joern.kersten@desy.de}}$^{(b)}$,
Manfred Lindner\footnote{E-mail: \texttt{lindner@ph.tum.de}}$^{(c)}$,\\[1mm]
Michael Ratz\footnote{E-mail: \texttt{mratz@th.physik.uni-bonn.de}}$^{(d)}$, and
Michael Andreas Schmidt\footnote{E-mail: \texttt{mschmidt@ph.tum.de}}$^{(c)}$}
\\[5mm]
{\small\textit{$^{(a)}$
Department of Physics and Astronomy,
University of Southampton, \\ 
Southampton, SO17 1BJ, United Kingdom
}}
\\[3mm]
{\small\textit{$^{(b)}$Deutsches Elektronen-Synchrotron DESY, 
Notkestra{\ss}e 85, 22603 Hamburg, Germany}}\\[3mm]
{\small\textit{$^{(c)}$
Physik-Department T30, 
Technische Universit\"{a}t M\"{u}nchen\\ 
James-Franck-Stra{\ss}e,
85748 Garching, Germany
}}
\\[3mm]
{\small\textit{$^{(d)}$
Physikalisches Institut der Universit\"at Bonn,\\
Nussallee 12, 53115 Bonn, Germany.}}
\end{center}
\vspace*{1.0cm}

\begin{abstract}
\noindent
We systematically analyze quantum corrections in see-saw scenarios, including
effects from above as well as below the see-saw scales. We derive approximate
renormalization group equations for  neutrino masses, lepton mixings and CP
phases,  yielding an analytic understanding and a simple estimate of the size of
the effects.   Even for hierarchical masses, they often exceed the precision of
future experiments. Furthermore, we provide a software package allowing for a
convenient numerical renormalization group analysis, with heavy singlets being 
integrated out successively at their mass thresholds.  We also discuss
applications to model building and related topics. 
\end{abstract}

\end{titlepage}

\newpage
\setcounter{footnote}{0}

\section{Introduction}
  
The observed smallness of neutrino masses finds an attractive explanation in the see-saw
mechanism \CiteSeeSaw. The light neutrino masses are, at tree-level, given by
the famous see-saw relation
\begin{equation}\label{eq:SeeSaw}
 m_\nu\,=\,-
 (m_\nu^\mathrm{Dirac})^T\,M^{-1}\,m_\nu^\mathrm{Dirac}\;.
\end{equation}
This relation emerges from integrating out heavy, singlet neutrinos with mass
matrix $M$. The Dirac neutrino mass $m_\nu^\mathrm{Dirac}$ is proportional to
the neutrino Yukawa coupling $Y_\nu$. Clearly, the see-saw operates at high
energy scales while its implications are measured by experiments at low scales.
Therefore, the neutrino masses given by Eq.~\eqref{eq:SeeSaw} are  subject to
quantum corrections, i.e.\ they are modified by renormalization group (RG)
running.

The running of neutrino masses and lepton mixing angles has 
been investigated intensively in the literature. 
For non-hierarchical neutrino mass spectra, RG effects can be very 
large and they can have interesting implications for model building. 
For example, lepton mixing angles can be magnified 
\cite{Balaji:2000au,Miura:2000bj,Antusch:2002fr,%
Mohapatra:2003tw,Hagedorn:2004ba}, bimaximal 
mixing at high energy can be compatible with low-energy experiments  
\cite{Antusch:2002hy,Miura:2003if,Shindou:2004tv} or the small mass 
splittings can be generated from exactly 
degenerate light neutrinos 
\cite{Chankowski:2000fp,Chun:2001kh,Chen:2001gk,Joshipura:2002xa,%
Joshipura:2002gr,Singh:2004zu}. 
On the other hand, facing the high precision of future neutrino  experiments,
rather small RG corrections are important as well. For  instance, deviations
from $\theta_{13}=0$  or maximal mixing $\theta_{23}=\pi/4$  are induced by RG
effects \cite{Antusch:2003kp,Mei:2004rn,Antusch:2004yx} also for a hierarchical
spectrum. 
However, in most studies only the running of the dimension 5 operator  has been
considered,  which is only  appropriate for the energy range below the mass
scale of the heavy singlets. 

The importance of including the effects from energy ranges above and between these
mass thresholds when analyzing RG effects in GUT models has been 
pointed out in 
\cite{Tanimoto:1995bf,Casas:1999tp,Casas:1999ac,King:2000hk,Antusch:2002rr,%
Antusch:2002hy,Antusch:2002fr,Miura:2003if,Shindou:2004tv,Mei:2004rn}.  
They are typically at least as important as the effects from below the
thresholds since  the relevant couplings, i.e.\ the entries of $Y_\nu$, can be
of order one, regardless of $\tan\beta$.\footnote{Large entries of $Y_\nu$ could
be important in models of gauge-Yukawa unification (see, e.g., 
\cite{Gogoladze:2003pp}), and may even be important for precision gauge
unification in the MSSM \cite{Casas:2000pa}.} 
Previous studies have investigated the RG effects above the 
see-saw scales mainly numerically.  

In this paper we derive formulae which allow to understand  the running of the
neutrino parameters above the see-saw scales analytically. We further provide
a software package for analyzing the RG evolution (with correct
treatment of non-degenerate
see-saw scales) numerically. We apply our  results to investigate consequences
of the running above the see-saw scales  for model building and leptogenesis and
compare the size of RG corrections to the precision of future experiments. 

The paper is organized as follows: In Sec.~\ref{sec:RunningInSeeSaw}, we review
how the predictions for neutrino masses can be evolved from the GUT scale to the
electroweak scale.  Sec.~\ref{sec:AnalyticalFormulae} is dedicated to the
analytic understanding of RG effects in see-saw scenarios with special emphasis
on the range between $M_\mathrm{GUT}$ and the highest see-saw scale. In
Sec.~\ref{sec:RunningBetween}, we analyze the running between the see-saw scales
in more detail. Sec.~\ref{sec:MathematicaPackages} contains a brief description
of the accompanying Mathematica packages for numerical RG analyses (a detailed
documentation is available at \url{http://www.ph.tum.de/~rge/}). In
Sec.~\ref{sec:Applications}, we discuss applications to model building and
related topics. Alternatives to the simplest see-saw scenario are briefly
discussed in  Sec.~\ref{sec:AlternativeScenarios}.  Finally,
Sec.~\ref{sec:Conclusions} contains our conclusions.

\section{Running Neutrino Masses in See-Saw Scenarios}
\label{sec:RunningInSeeSaw}

In this section,  we discuss how to obtain the RG evolution of neutrino masses,
starting from initial conditions at a 
very high energy scale.\footnote{In the following we will refer to this high 
energy scale as $M_\mathrm{GUT}$, although it can be any other scale where 
additional new physics, apart from the heavy singlet neutrinos, has to be 
taken into account.}
An important
technical issue is that the heavy singlet neutrinos involved in the see-saw
mechanism have to be integrated out one by one. Thus, one has to consider a
series of effective theories \cite{King:2000hk,Antusch:2002rr}.
We will focus on the SM and the MSSM amended by three singlet neutrinos
$N_\mathrm{R}^i$ or three singlet superfields $\SuperField{\nu}_i$,
respectively. 
The discussion can be applied to other scenarios, such as multi-Higgs models, 
and a different number of singlets in a straightforward way.

We consider the Lagrangian of the SM extended by 
\begin{equation}\label{eq:ExtSM}
        \mathscr{L}^\nu = 
        -\overline{\NuR} Y_\nu \ell_\mathrm{L} \widetilde\phi^\dagger
        -\frac{1}{2} \overline{\NuR} M \NuR^\ChargeC
        +\text{h.c.} \;,
\end{equation}
where
$\ell_\mathrm{L}:=(\ell_\mathrm{L}^1,\ell_\mathrm{L}^2,\ell_\mathrm{L}^3)^T$
denotes the left-handed lepton doublets, $\phi$ is the Higgs doublet and
$\widetilde\phi = i\tau^2 \phi^*$ its charge conjugate. The superscript
$\ChargeC$ denotes charge conjugation of fermion fields, and
$N_\mathrm{R}^\ChargeC:=(N_\mathrm{R})^\ChargeC$.  In the supersymmetric case,
$\phi$ is replaced by the Higgs doublet $H_u$ coupling to the up-type quarks.

In order to define mass and mixing parameters as functions of the
renormalization scale $\mu$ above the highest see-saw scale, we consider the
effective light neutrino mass matrix
\begin{equation} \label{eq:mnuFullTheory}
  m_\nu(\mu)\, =\, -\frac{v^2}{2} \, Y_\nu^T(\mu)\, M^{-1}(\mu) \,Y_\nu(\mu) \;,
\end{equation}
where $Y_\nu$ and $M$ are $\mu$-dependent.  The relevant Higgs vev is
$v=246\GeV$ in the SM and $v=246\GeV\cdot\sin\beta$ in the MSSM.%
\footnote{
As indicated in Eq.~(\ref{eq:mnuFullTheory}), we do not take into account the 
running of the Higgs vev. 
In principle, $v$ runs as well, so that $m_\nu$ actually does not yield
 the physical neutrino masses.  However, the evolution of $v$ depends on
 the renormalization scheme and on the definition of the Higgs mass, see
 e.g.\ \cite{Kielanowski:2003jg}, so that there is no straightforward
 definition of a neutrino mass with a running vev.  In any case, the
 mixing angles and phases are independent of the value of $v$.
This definition has shown appropriate for the
applications discussed in this paper, such as leptogenesis.
}
$m_\nu$ is the mass matrix of the three light neutrinos as obtained from
block-diagonalizing the complete $6\times6$ neutrino mass matrix,
following the standard see-saw calculation.  
The scale-dependent mixing parameters are obtained from $m_\nu(\mu)$
and the running charged lepton Yukawa matrix $Y_e(\mu)$.
In Sec.~\ref{sec:AnalyticalFormulae} we are going to analyze the 
energy dependence of the parameters in the lepton sector such as 
neutrino masses, lepton mixing angles and CP phases above the 
highest see-saw scale analytically. Therefore, we will make use of the 
RGE for the composite quantity $m_\nu$, calculated from those for
$Y_\nu$ and $M$
\cite{Grzadkowski:1987tf,Grzadkowski:1987wr,Casas:1999tp,Casas:1999ac}.
It is given by
\begin{equation} \label{eq:Betamnu}
        16\pi^2 \, \frac{\D m_\nu}{\D t} =
        (C_e \, Y_e^\dagger Y_e + C_\nu \, Y_\nu^\dagger Y_\nu)^T \, m_\nu +
        m_\nu \, (C_e \, Y_e^\dagger Y_e + C_\nu \, Y_\nu^\dagger Y_\nu) + 
        \bar\alpha \, m_\nu
\end{equation}
with $t:=\ln(\mu/\mu_0)$,
\begin{subequations} \label{eq:CeCnu}
\begin{alignat}{2}
        C_e &= -\frac{3}{2} \:,\; C_\nu=\frac{1}{2} && \;\; \text{in the SM,}
\\
        C_e &= C_\nu = 1 && \;\; \text{in the MSSM,}
\end{alignat}
\end{subequations}
and (with $Y_e$, $Y_d$ and $Y_u$ being the Yukawa matrices of charged leptons,
down- and up-type quarks, respectively)%
\footnote{We use GUT charge normalization for the gauge coupling $g_1$.}
\begin{subequations}
\begin{eqnarray}
        \bar\alpha_\mathrm{SM} &=&
        -\frac{9}{10} g_1^2 -\frac{9}{2} g_2^2 + 
                2\Tr\big(Y_\nu^\dagger Y_\nu + Y_e^\dagger Y_e
                + 3 Y_d^\dagger Y_d + 3 Y_u^\dagger Y_u \big) ,
\\
        \bar\alpha_\mathrm{MSSM} &=&
        -\frac{6}{5} g_1^2 - 6 g_2^2 + 
                2\Tr\left(Y_\nu^\dagger Y_\nu 
                + 3 Y_u^\dagger Y_u \right) .
        \label{eq:alphabarMSSM}
\end{eqnarray}
\end{subequations}

The RGE \eqref{eq:Betamnu} governs only the evolution of the light neutrino
mass matrix above the highest see-saw scale, 
which is given by the mass eigenvalue $M_3$ of the heaviest singlet $N_\mathrm{R}^3$.
For $\mu<M_3$, we obtain the correct RG evolution by
integrating out $N_\mathrm{R}^3$. 
This leads to the appearance of
an effective neutrino mass operator
\begin{equation} \label{eq:Kappa3}
    \mathscr{L}_\kappa\, =\,
    \frac{1}{4} \, \accentset{(3)}{\kappa}_{fg} \; 
    (\overline{\ell_\mathrm{L}^\ChargeC}{}^f\cdot \phi) \;
    (\ell_{\mathrm{L}}^g\cdot\phi)
    + \text{h.c.} \;,
\end{equation}
where  $f,g \in \{1,2,3\}$ are family indices and where the dot indicates the 
$\mathrm{SU}(2)_\mathrm{L}$-invariant contractions.    The coefficient of this
operator is obtained by the (tree-level) matching condition\footnote{
We do not discuss finite threshold corrections, which arise due to the fact
that the singlet neutrinos do not decouple abruptly \cite{Pich:1998xt}.  The
resulting uncertainty in the low-energy results is typically not larger than
that due to two-loop effects.  In the \package{REAP} software package described in
Sec.~\ref{sec:MathematicaPackages}, the corrections can be implemented
approximately by integrating out $N_\mathrm{R}^3$ slightly below $M_3$.
}
\begin{equation}
 \accentset{(3)}{\kappa}_{gf}\,=\,
 2(Y_\nu^T)_{g3}\,M^{-1}_3\,(Y_\nu)_{3f}\;,
\end{equation}
which is imposed at $\mu=M_3$. 
This expression is
specified in the mass basis for the singlets,
i.e.\ in the basis where $M$ is diagonal. Let
us mention that finding the matching scale properly requires some care as the
mass matrix $M$ (and consequently the eigenvalue $M_3$) itself is subject to the
RG evolution.
As a consequence, for scales below $M_3$ the effective neutrino mass matrix 
can be described as a sum of two contributions,
\begin{equation}\label{eq:EffNuMass3}
    m_\nu \,=\,
    -\frac{v^2}{4} \RaiseBrace{\bigl(}
    \, \accentset{(3)}{\kappa} + 
    2 \, \accentset{(3)}{Y}_\nu^T \accentset{(3)}{M}^{-1}
     \accentset{(3)}{Y}_\nu \RaiseBrace{\bigr)} \;.
\end{equation}
The $2\times3$ Yukawa matrix $\accentset{(3)}{Y}_\nu$ is obtained by simply removing 
the last row of $Y_\nu$ in the basis where $M$ is diagonal. 
The $2\times2$ mass matrix $\accentset{(3)}{M}$ is found from $M$ by removing
the last row and column.
By construction, $m_\nu$ is a continuous function of the renormalization scale.
The RG evolution of the second term on the right-hand side of
Eq.~\eqref{eq:EffNuMass3} is governed by Eq.~\eqref{eq:Betamnu} with $Y_\nu$
replaced by $\accentset{(3)}{Y}_\nu$. The running of the first term, on the
other hand, is determined by the RGE \cite{Antusch:2002rr}
\begin{equation} \label{eq:RGEForKappaBetweenThresholds}
        16\pi^2 \, \frac{\D \,\accentset{(3)}{\kappa}}{\D t} =
        \RaiseBrace{\bigl(}
         C_e \, Y_e^\dagger Y_e +
         C_\nu \, \accentset{(3)}{Y}_\nu^\dagger \accentset{(3)}{Y}_\nu
         \RaiseBrace{\bigr)}^T \, \accentset{(3)}{\kappa} +
        \accentset{(3)}{\kappa} \, \RaiseBrace{\bigl(}
         C_e \, Y_e^\dagger Y_e + C_\nu \,
         \accentset{(3)}{Y}_\nu^\dagger \accentset{(3)}{Y}_\nu
         \RaiseBrace{\bigr)} + 
        \accentset{(3)}{\bar\alpha} \; \accentset{(3)}{\kappa}
\end{equation}
with $C_e$ and $C_\nu$ as in Eqs.~\eqref{eq:CeCnu} 
\cite{Chankowski:1993tx,Babu:1993qv,Antusch:2001ck,Antusch:2001vn}, and
\begin{subequations}
\begin{eqnarray}
    \accentset{(3)}{\bar\alpha}_\mathrm{SM} &=&
    -3 g_2^2 +
    2\Tr\!\RaiseBrace{\bigl(}
     \accentset{(3)}{Y}^{\dagger}_\nu \accentset{(3)}{Y}_\nu +
     Y_e^\dagger Y_e + 3 Y_d^\dagger Y_d + 3 Y_u^\dagger Y_u
    \RaiseBrace{\bigr)} + 
    \lambda \;,\label{eq:alphanSM}
\\
    \accentset{(3)}{\bar\alpha}_\mathrm{MSSM} &=&
    -\frac{6}{5} g_1^2 - 6 g_2^2 +
    2\Tr\!\RaiseBrace{\bigl(}
     \accentset{(3)}{Y}^{\dagger}_\nu \accentset{(3)}{Y}_\nu +
     3 Y_u^\dagger Y_u
    \RaiseBrace{\bigr)} \;.\label{eq:alphanMSSM}
\end{eqnarray}
\end{subequations}

\begin{figure}
 \centerline{\includegraphics[scale=1.2]{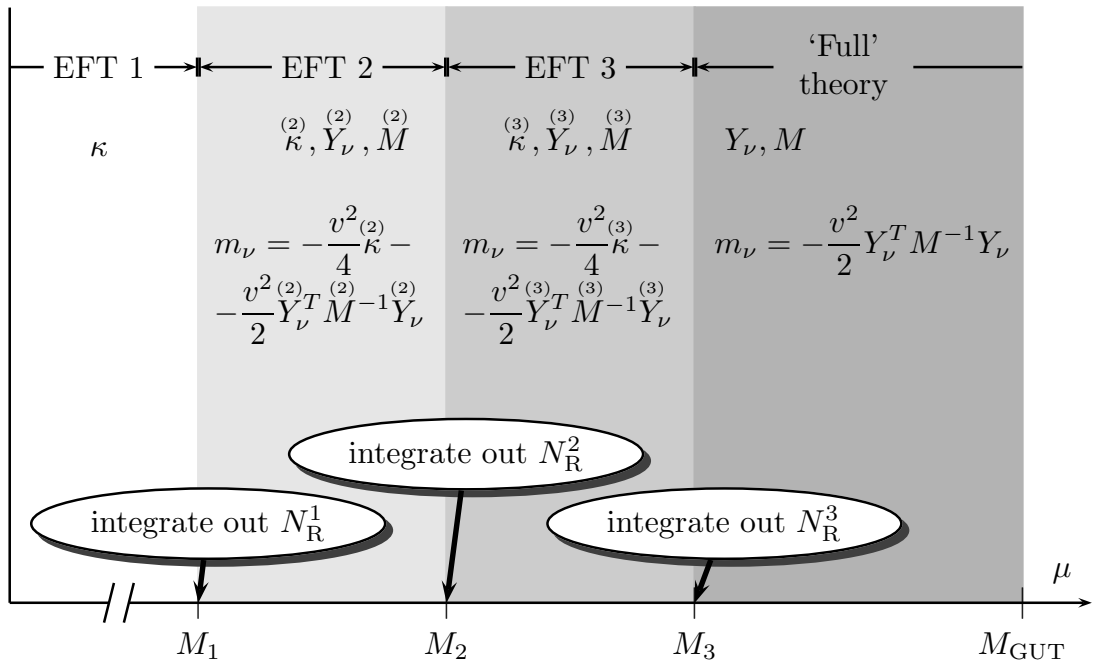}}
 \caption{Validity ranges of the effective theories (EFTs) in the renormalization
 scale $\mu$. At a scale close to the mass thresholds $M_i$, the  EFTs are
 related by matching conditions. Although we show this illustration for 3 heavy
 singlets, it is straightforward to generalize it to an arbitrary number
 (cf.\ \cite{Antusch:2002rr}).}
 \label{fig:EFTRanges3x3}
\end{figure}

One can now evolve the effective neutrino mass matrix down to
the scale $M_2$ and repeat the matching procedure there. From integrating out 
$N_\mathrm{R}^2$ at $\mu=M_2$, the Yukawa
matrix gets further reduced and the effective neutrino mass operator receives 
an additional contribution. After a subsequent RG evolution to $\mu = M_1$, 
the procedure is repeated for $N_\mathrm{R}^1$. 
The emerging  effective theories, as well as the quantities
relevant to neutrino masses in each of them, are illustrated in
Fig.~\ref{fig:EFTRanges3x3}.

In summary, the running of the effective neutrino mass matrix $m_\nu$  above and
between the see-saw scales  is given by the running of two parts,
\begin{equation}\label{eq:EffNuMass}
    m_\nu \,=\,
    -\frac{v^2}{4} \RaiseBrace{\bigl(}
    \, \accentset{(n)}{\kappa} + 
    2 \, \accentset{(n)}{Y}_\nu^T \accentset{(n)}{M}^{-1}
     \accentset{(n)}{Y}_\nu \RaiseBrace{\bigr)} \;.
\end{equation}
where $n$ labels the effective theory (cf.~Fig.~\ref{fig:EFTRanges3x3}). In the
SM and the MSSM, the 1-loop $\beta$-functions for $m_\nu$  in the various
effective theories can be summarized as
\begin{eqnarray}\label{eq:BetaEffNuMass}
16 \pi^2 \, 
\frac{\D\accentset{(n)}{X}}{\D t} & = &\vphantom{\frac{1}{2}} 
\RaiseBrace{\bigl(}C_e Y_e^\dagger Y_e + C_\nu \accentset{(n)}{Y}^\dagger_\nu   
   \accentset{(n)}{Y}_\nu\RaiseBrace{\bigr)}^T \:\accentset{(n)}{X}
 + \accentset{(n)}{X} \, \RaiseBrace{\bigl(}C_e Y_e^\dagger Y_e + 
 C_\nu \accentset{(n)}{Y}^\dagger_\nu\accentset{(n)}{Y}_\nu\RaiseBrace{\bigr)} 
 +\accentset{(n)}{\bar\alpha}_X \accentset{(n)}{X}
 \; ,
\end{eqnarray}
where $\accentset{(n)}{X}$ stands for $\accentset{(n)}{\kappa}$ or for 
$2\, \accentset{(n)}{Y}_\nu^T \accentset{(n)}{M}^{-1} \accentset{(n)}{Y}_\nu$, 
respectively. The coefficients $C_i$ and $\bar\alpha_i$ are listed in  
Tab.~\ref{tab:BetaEffNuMass}. 

\clearpage
 
\begin{table}
\centering
\begin{tabular}{|l|l||c|c|l|}
\hline 
Model & $\accentset{(n)}{X}$ $\vphantom{\frac{1}{2}}$ &
$\!C_e\!$ & $\!C_\nu\!$ & flavour-trivial term $\accentset{(n)}{\bar\alpha}_X$\\
\hline 
SM & $\accentset{(n)}{\kappa}$ \(\vphantom{\sqrt{\big|}^C}\)&   
$\!-\tfrac{3}{2}\!$ & $\tfrac{1}{2}$ & 
$2\Tr \RaiseBrace{\bigl(} 
\accentset{(n)}{Y}^{\dagger}_\nu \accentset{(n)}{Y}_\nu
  +Y_e^\dagger Y_e +3Y_d^\dagger Y_d +3Y_u^\dagger Y_u 
  \RaiseBrace{\bigr)}
  -3 g_2^2+\lambda$
\\
SM & $2\, \accentset{(n)}{Y}_\nu^T \accentset{(n)}{M}^{-1}
\accentset{(n)}{Y}_\nu \!\!$ 
$\vphantom{\frac{1}{2}}$\(\vphantom{\sqrt{\big|}^C}\)&
$\!-\tfrac{3}{2}\!$ & $\tfrac{1}{2}$ & 
$2\Tr \RaiseBrace{\bigl(} 
\accentset{(n)}{Y}^{\dagger}_\nu \accentset{(n)}{Y}_\nu
  +Y_e^\dagger Y_e +3Y_d^\dagger Y_d +3Y_u^\dagger Y_u 
  \RaiseBrace{\bigr)}
  -\tfrac{9}{10} g_1^2 - \tfrac{9}{2} g_2^2$
\\
\hline
MSSM & $\accentset{(n)}{\kappa}$ $\vphantom{\frac{1}{2}}$\(\vphantom{\sqrt{\big|}^C}\)&
$1$ & $1$ &
$2\Tr \RaiseBrace{\bigl(} 
\accentset{(n)}{Y}^{\dagger}_\nu \accentset{(n)}{Y}_\nu
  +3Y_u^\dagger Y_u   \RaiseBrace{\bigr)}
  -\tfrac{6}{5} g_1^2 - 6 g_2^2$
\\
MSSM & $2\, \accentset{(n)}{Y}_\nu^T \accentset{(n)}{M}^{-1}
\accentset{(n)}{Y}_\nu \!\!$ 
 $\vphantom{\frac{1}{2}}$&
 $1$ & $1$ & 
$2\Tr \RaiseBrace{\bigl(} 
\accentset{(n)}{Y}^{\dagger}_\nu \accentset{(n)}{Y}_\nu
  +3Y_u^\dagger Y_u   \RaiseBrace{\bigr)}
  -\tfrac{6}{5} g_1^2 - 6 g_2^2$
\\[1mm]
\hline 
\end{tabular}   
\caption{\label{tab:BetaEffNuMass}
Coefficients of the $\beta$-functions of Eq.~\eqref{eq:BetaEffNuMass}, which govern the
running of the effective neutrino mass matrix in minimal see-saw models.  
}
\end{table}

\section{Analytic Understanding of the RG Evolution}
\label{sec:AnalyticalFormulae}
The methods of 
\cite{Babu:1987im,Grzadkowski:1987tf,Casas:1999tg,Antusch:2003kp} can be used to derive
differential equations for the running of the neutrino masses, mixing
angles and CP phases in the see-saw scenario.  In this section, we
concentrate on the full theory above the highest see-saw scale.
The corresponding differential equations for the running below the
see-saw scales have been discussed in
\cite{Chankowski:1999xc,Casas:1999tg,Antusch:2003kp}.
We abbreviate the
flavour-dependent terms in the RGE \eqref{eq:Betamnu} by
\begin{equation} \label{eq:DefP}
        P := C_e \, Y_e^\dagger Y_e + C_\nu \, Y_\nu^\dagger Y_\nu \;.
\end{equation}
Due to the appearance of the neutrino Yukawa couplings, the running depends on
more parameters than below the see-saw scale. In particular, since the see-saw
formula does not allow to determine $Y_\nu$ uniquely from the light neutrino
mass matrix, the running is no longer determined by  (the RG extrapolation of)
low-energy parameters only. Moreover, $Y_e^\dagger Y_e$ and $Y_\nu^\dagger
Y_\nu$ are not simultaneously diagonalizable in general.  As a consequence, the
RG evolution generates off-diagonal entries in the charged lepton Yukawa
couplings, even if one starts in a basis where they are diagonal (cf.\ the RGEs
in App.~\ref{app:RGEs}).  This is also different from the situation below the
see-saw scale and makes the results more complicated.

In a given basis, $Y_e^\dagger Y_e$ and $m_\nu$ can be diagonalized by unitary
matrices, $U_e$ and $U_\nu$, respectively.  The lepton mixing matrix is given by
$U_\mathrm{MNS} = U_e^\dagger U_\nu$.  Keeping the basis fixed, both matrices
change with the renormalization scale, so that the RGEs of the mixing parameters
consist of two parts, one coming from the RG change of $U_e$, and the other from
the change of $U_\nu$. We will refer to these as $U_e$ and $U_\nu$ contribution
in the following.\footnote{One might wonder whether it is possible to simplify
the situation by working in the basis where $P$ is diagonal. This is not the
case, since the $U_e$ contribution depends on a different linear combination of
$Y_e^\dagger Y_e$ and $Y_\nu^\dagger Y_\nu$.} Further details and the
derivation of the formulae are given in App.~\ref{app:AnalyticFormulae}.

We will first discuss the $U_\nu$ contribution, which is often dominant. An
important result is that in the RGEs above the see-saw scale, the same mass
squared differences appear in the denominators as below the see-saw scale, so
that
\begin{subequations} \label{eq:FirstMessage}
\begin{eqnarray}
    \Delta\theta_{12}, \Delta\varphi_1, \Delta\varphi_2, \Delta\delta
    &\propto& \frac{1}{\Delta m^2_\mathrm{sol}} \;,
\\*
    \Delta\theta_{13}, \Delta\theta_{23}
    &\propto& \frac{1}{\Delta m^2_\mathrm{atm}} \;,
\end{eqnarray}
\end{subequations}
where, as usual, $\Delta m^2_\mathrm{atm} := m_3^2-m_2^2$ and $\Delta
m^2_\mathrm{sol} := m_2^2-m_1^2$.%
\footnote{For specific textures, this
observation has been made in \cite{Antusch:2002hy,Antusch:2002fr}. The result
can also be obtained by using the formulae of \cite{Casas:1999tg}.}  Thus,
$\theta_{12}$ and the phases generically still run faster than $\theta_{13}$ and
$\theta_{23}$.  Besides, the running is suppressed by a strong normal mass
hierarchy, as it is the case below $M_1$.   For the unphysical phases%
\footnote{The term ``unphysical phases'' is somewhat misleading here,
 since the distinction between physical and unphysical parameters is not
 completely trivial in the full theory, cf.\ App.~\ref{app:UnphysPhases}.}, 
we find a
generically larger change $\Delta\delta_e \propto 1/\Delta m^2_\mathrm{sol}$,
while $\Delta\delta_\mu, \Delta\delta_\tau \propto 1/\Delta m^2_\mathrm{atm}$.

Often, the evolution will be dominated by a single element of $P$. Then, the
derivatives of the masses and mixing parameters are given by this element times
the corresponding entry in the tables of Sec.~\ref{sec:RunningMasses} and
App.~\ref{app:Tables}. We will discuss an example in
Sec.~\ref{sec:HierarchicalYnu}. Of course, if several entries of $P_{fg}$ are
relevant, one obtains the  analytic description by simply adding up their
contributions. The tables are given in the basis where $Y_e$ is diagonal and
where the unphysical phases in the MNS matrix are zero (cf.\
Apps.~\ref{app:GeneralStrategy} and \ref{app:UnphysPhases}).  In order to keep
the expressions reasonably short, we only present the first order of the
expansion in the small CHOOZ angle $\theta_{13}$.  We furthermore use the
abbreviation 
\begin{equation} \label{eq:DefZeta}
    \zeta := \frac{\Delta m^2_\mathrm{sol}}{\Delta m^2_\mathrm{atm}} \;.
\end{equation}
Its current best-fit value is $\zeta\approx0.038$ \cite{Maltoni:2004ei}.
Note that this value is
measured at low energy.  It can change significantly, if the running of the mass
eigenvalues is not a simple rescaling.

The tables in the appendix show that the  numerators of the RGEs are of the
order of $m_i^2$ in the generic case, i.e.\ if there are no significant
cancellations.  Then, the generic enhancement and suppression factors  given in
Tab.~\ref{tab:EnhancementFactorsAngles} yield a first estimate of the RG change
of the mixing angles. In particular, they allow to understand analytically when
the evolution is enhanced or suppressed compared to the naive estimate
\begin{equation}
 \Delta\theta_{ij}^\mathrm{naive}
 \,=\,
 \frac{1}{16\pi^2} \, P_{fg} \times 
 \ln\frac{M_\mathrm{GUT}}{M_*} \;,
\end{equation}
where $P_{fg}$ is assumed to dominate the running and $M_*$ is the corresponding
see-saw scale.  The analogous factors for the CP phases are given in
Tab.~\ref{tab:EnhancementFactorsPhases}. The size of quantum corrections can
thus be estimated by multiplying $\Delta\theta_{ij}^\mathrm{naive}$ with the
corresponding enhancement or suppression factor. As the mass hierarchy is weaker
in the neutrino sector than in the quark sector, the change of the mixing
parameters in the MNS matrix is larger than that of the ones in the CKM matrix. 
\begin{table}[htbp]\setlength{\bigstrutjot}{4pt}
  \centering
  \begin{tabular}{|l||c|c|c||c|c|c||c|c|c|}\hline
&\multicolumn{3}{|c||}{$\Dot\theta_{12}$}&\multicolumn{3}{|c||}{$\Dot\theta_{13}$}&\multicolumn{3}{|c|}{$\Dot\theta_{23}$}\bigstrut\\\hline
& d. & n.h. &i.h. &d. & n.h. &i.h. &d. & n.h. &i.h.\bigstrut\\\hline
$P_{11}$ & $\frac{m^2}{\Delta m^2_\text{sol}}$ & $1$ & $\zeta^{-1}$
&$\mathcal{O}(\theta_{13})$ & $\mathcal{O}(\theta_{13})$ & $\mathcal{O}(\theta_{13})$
&$\mathcal{O}(\theta_{13})$ & $\mathcal{O}(\theta_{13})$ & $\mathcal{O}(\theta_{13})$
\bigstrut\\
$P_{22}$ &  $\frac{m^2}{\Delta m^2_\text{sol}}$ & $1$ & $\zeta^{-1}$
&$\frac{m^2}{\Delta m^2_\text{atm}}$ & $\sqrt{\zeta}$ & $\mathcal{O}(\theta_{13})$
&$\frac{m^2}{\Delta m^2_\text{atm}}$ & $1$ & $1$
\bigstrut\\
$P_{33}$ &  $\frac{m^2}{\Delta m^2_\text{sol}}$ & $1$ & $\zeta^{-1}$
&$\frac{m^2}{\Delta m^2_\text{atm}}$ & $\sqrt{\zeta}$ & $\mathcal{O}(\theta_{13})$
&$\frac{m^2}{\Delta m^2_\text{atm}}$ & $1$ & $1$
\bigstrut\\
$\re P_{21}$ &  $\frac{m^2}{\Delta m^2_\text{sol}}$ & $1$ & $\zeta^{-1}$
&$\frac{m^2}{\Delta m^2_\text{atm}}$ & $1$ & $1$
&$\frac{m^2}{\Delta m^2_\text{atm}}$ & $\sqrt{\zeta}$ & $\mathcal{O}(\theta_{13})$
\bigstrut\\
$\re P_{31}$ &  $\frac{m^2}{\Delta m^2_\text{sol}}$ & $1$ & $\zeta^{-1}$
&$\frac{m^2}{\Delta m^2_\text{atm}}$ & $1$ & $1$
&$\frac{m^2}{\Delta m^2_\text{atm}}$ & $\sqrt{\zeta}$ & $\mathcal{O}(\theta_{13})$
\bigstrut\\
$\re P_{32}$ &  $\frac{m^2}{\Delta m^2_\text{sol}}$ & $1$ & $\zeta^{-1}$
&$\frac{m^2}{\Delta m^2_\text{atm}}$ & $\sqrt{\zeta}$ & $\mathcal{O}(\theta_{13})$
&$\frac{m^2}{\Delta m^2_\text{atm}}$ & $1$ & $1$
\bigstrut\\
$\im P_{21}$ &  $\frac{m^2}{\Delta m^2_\text{sol}}$ & $\mathcal{O}(\theta_{13})$ & $\zeta^{-1}$
&$\frac{m^2}{\Delta m^2_\text{atm}}$ & $1$ & $1$
&$\frac{m^2}{\Delta m^2_\text{atm}}$ & $\sqrt{\zeta}$ & $\mathcal{O}(\theta_{13})$
\bigstrut\\
$\im P_{31}$ &   $\frac{m^2}{\Delta m^2_\text{sol}}$ & $\mathcal{O}(\theta_{13})$ & $\zeta^{-1}$
&$\frac{m^2}{\Delta m^2_\text{atm}}$ & $1$ & $1$
&$\frac{m^2}{\Delta m^2_\text{atm}}$ & $\sqrt{\zeta}$ & $\mathcal{O}(\theta_{13})$
\bigstrut\\
$\im P_{32}$ &  $\mathcal{O}(\theta_{13})$ & $\mathcal{O}(\theta_{13})$ & $\mathcal{O}(\theta_{13})$
&$\frac{m^2}{\Delta m^2_\text{atm}}$ & $\sqrt{\zeta}$ & $\mathcal{O}(\theta_{13})$
&$\frac{m^2}{\Delta m^2_\text{atm}}$ & $\sqrt{\zeta}$ & $\mathcal{O}(\theta_{13})$
\bigstrut\\\hline
\end{tabular}
\caption{Generic enhancement and suppression factors for the evolution
 of the angles, yielding an estimate of the size of the RG effect.  The 
 table entries correspond to the terms in the mixing parameter RGEs with
 the coefficient given by the first column.
 A `$1$' indicates that there is no generic enhancement or
        suppression. `d.' stands for a degenerate neutrino mass spectrum, i.e.\ $\Delta
        m_\mathrm{atm}^2\ll m_1^2\sim m_2^2\sim m_3^2\sim m^2$. `n.h.' denotes a
        normally hierarchical spectrum, i.e.\ $m_1\ll m_2\ll m_3$, and
        `i.h.' means an inverted hierarchy, i.e.\ $m_3\ll m_1\lesssim m_2$.}
\label{tab:EnhancementFactorsAngles}
\end{table}

The RG evolution can deviate significantly from the generic estimate, if
cancellations occur. For example, for non-zero $\varphi_1-\varphi_2$, the
running of $\theta_{12}$ usually gets damped (as it is the case below
the see-saw scales \cite{Haba:2000tx}).  Such effects can be understood from the
complete formulae in App.~\ref{app:Tables}.  However, care should be taken when
estimating the RG effects for special phase configurations with extreme
cancellations, such as $\varphi_1-\varphi_2=\pi$, as terms proportional to
$\theta_{13}$ (which are neglected in our formulae) can become important then.

\subsection{Running of the Mixing Angles} \label{sec:RunningAngles}

From the generic enhancement and suppression factors for the evolution of the
solar angle in Tab.~\ref{tab:EnhancementFactorsAngles}, we see that all terms in
$\Dot\theta_{12}$ are enlarged by  $m^2/\Delta m^2_\mathrm{sol}$ for
quasi-degenerate masses.  Thus, there will be large RG effects, if the different
terms do not cancel each other.  The term involving $\im P_{32}$ is an
exception, because its leading order is proportional to $\theta_{13}$, so that
it only plays a role in special cases.  In the case of a strong normal
hierarchy, there is no enhancement.  However, for a moderate hierarchy where the
square of the lightest neutrino mass is small compared to  $\Delta
m^2_\mathrm{atm}$ but larger than $\Delta m^2_\mathrm{sol}$ the running is still
enhanced by $m_1^2/\Delta m^2_\mathrm{sol}$.  This is similar for an inverted
hierarchy, where the evolution is generically enhanced by $\zeta^{-1}$, because
the masses $m_1$ and $m_2$ are almost degenerate.  Thus, the RG change of
$\theta_{12}$ is generically large for an inverted hierarchy and for a
degenerate spectrum, and small for a normal hierarchy.  This conclusion is
unchanged compared to the region below the see-saw scale.

The enhancement and suppression factors of $\theta_{13}$ are similar to those of
$\theta_{23}$.  The evolution of both angles does not depend on $P_{11}$ for
$\theta_{13}=0$.  The terms proportional to the other $P_{fg}$ are enhanced by
$m^2/\Delta m^2_\mathrm{atm}$ in the degenerate case, so that we expect
significant effects here as well.  However, as already mentioned, they are
usually smaller than those for $\theta_{12}$.  For both hierarchical spectra,
the running is slow.  For a diagonal $P$ and an inverted hierarchy with $m_3=0$,
$\theta_{13}$ does not run at all, if it vanishes at some energy, as it is the
case below the see-saw scale \cite{Grimus:2004cj}.  However, this is no longer
true if $P_{21}$ or $P_{31}$ is non-zero.

As far as the dependence of the RGEs on the mixing parameters is concerned, we
find from Tab.~\ref{tab:RGcorrections4angles} that the terms in the RGEs which
are proportional to the diagonal elements of $P$ exhibit basically the same
behavior as the RGEs below the see-saw scale \cite{Antusch:2003kp}. The running
of $\theta_{12}$ and $\theta_{23}$ is damped by non-zero Majorana phases, while
the situation is more complicated for $\theta_{13}$.  In particular, the value
of the Dirac phase in the case $\theta_{13}=0$ is determined by the condition
that $\Dot\delta$ remain finite.  Additionally, the running is suppressed if the
mixing angles are small, as it is the case in the quark sector.  (This is
another reason why the leptonic mixings run faster than the quark mixings
\cite{LRS}.)

If the diagonal elements are equal, their contributions to the RGEs cancel
exactly.  This follows from the fact that the mixing angles do not change under
the RG, if $P$ is the identity matrix and thus does not distinguish between the
flavours.  Of course, this statement holds also for the RGEs of the CP phases. 
It provides a consistency check for the results.

Interesting new effects occur for non-zero off-diagonal elements in $P$. Some of
their coefficients in the RGEs do not vanish for vanishing mixings, so that
non-zero mixing angles are generated radiatively. Because of this, it is
possible to reach low-energy parameter regions that are compatible with
experiment even if the neutrino mass matrix is diagonal at the GUT scale
\cite{Hagedorn:2004ba}.  This is in striking contrast to the region below the
see-saw scale and to the quark sector. The terms proportional to the real parts
of the off-diagonal $P_{fg}$ exhibit the same dependence on the Majorana phases
as the diagonal elements.  Some of them are suppressed for large angles
$\theta_{12}$ and $\theta_{23}$.  For example, the $\re P_{23}$ contribution to
$\Dot\theta_{23}$ vanishes for maximal atmospheric mixing.  The influence of the
imaginary parts has quite a different dependence on the mixing parameters, in
particular on the Majorana phases. The corresponding terms become maximal for
non-vanishing phases, for instance for $\varphi_1-\varphi_2=\pi/2$ in the case
of $\theta_{12}$. Thus, the usual damping of the running by non-zero Majorana
phases does not always take place above the see-saw scales.  However, the
maximal damping for $\varphi_1-\varphi_2=\pi$ (or $\varphi_i=\pi$ in the case of
$\theta_{23}$) still occurs, since the coefficients of $\im P_{fg}$ are zero
then.  Some examples for the running with large imaginary entries in $P$ will be
discussed in Sec.~\ref{sec:RunningDespitePhases}.

\subsection{Running of the Phases}
\label{sec:RunningOfPhases}

The CP phases show a fast running in general. The corresponding generic
enhancement and suppression factors are given in
Tab.~\ref{tab:EnhancementFactorsPhases}.
As for the RGE of the Dirac phase $\delta$, there is always a term
proportional to $\theta_{13}^{-1}$, which is further enhanced for a
degenerate spectrum. This implies that the running of $\delta$ is in general
significant for small $\theta_{13}$, irrespectively of the
hierarchy.\footnote{Note, however, that in measurable quantities
$\delta$ appears always in combination with $\sin\theta_{13}$, so that the RG
change of predictions for experiments may not be significant.}
For $\theta_{13}=0$, $\delta$ and $\Dot\delta$ are undefined.  However,
it is possible to define an analytic continuation yielding a smooth
evolution \cite{Antusch:2003kp}.
In addition, for the degenerate or inversely hierarchical spectrum, the running
of $\delta$ gets enhanced by terms proportional to $m^2/\Delta m^2_\mathrm{sol}$
or $\zeta^{-1}$, respectively.
The coefficients of $P_{fg}$ in $\Dot\delta$ are given in Tab.~\ref{tab:RGcorrections4Delta},
from where one obtains the RGE as 
$\Dot\delta = \theta_{13}^{-1} \Dot\delta^{(-1)} + \Dot\delta^{(0)} +
 \mathscr{O}(\theta_{13})$.

\begin{table}\setlength{\bigstrutjot}{4pt}
  \centering
  \begin{tabular}{|l||c|c|c||c|c|c|}\hline
&\multicolumn{3}{|c||}{$\Dot\varphi_i$}&\multicolumn{3}{|c|}{$\Dot\delta$}\bigstrut\\\hline
& d. & n.h. &i.h. &d. & n.h. &i.h.\bigstrut\\\hline
$P_{11}$ & $\frac{m^2}{\Delta m^2_\text{sol}}$& $\mathcal{O}(\theta_{13})$ &
$\zeta^{-1}$
& $\frac{m^2}{\Delta m^2_\text{sol}}$ & $\sqrt{\zeta}$ & $\zeta^{-1}$
\bigstrut\\
$P_{22}$ & $\frac{m^2}{\Delta m^2_\text{sol}}$ & $\sqrt{\zeta}$ & $\zeta^{-1}$
&$\frac{m^2}{\Delta m^2_\text{atm}}\theta_{13}^{-1}+ \frac{m^2}{\Delta
  m^2_\text{sol}}$ &$\sqrt{\zeta}\theta_{13}^{-1}$&$\zeta^{-1}$
\bigstrut\\
$P_{33}$ & $\frac{m^2}{\Delta m^2_\text{sol}}$ & $\sqrt{\zeta}$ &$\zeta^{-1}$
&$\frac{m^2}{\Delta m^2_\text{atm}}\theta_{13}^{-1}+ \frac{m^2}{\Delta
  m^2_\text{sol}}$ & $\sqrt{\zeta}\theta_{13}^{-1}$&$\zeta^{-1}$
\bigstrut\\
$\re P_{21}$ & $\frac{m^2}{\Delta m^2_\text{sol}}$ & $\sqrt{\zeta}$ & $\zeta^{-1}$
&$\frac{m^2}{\Delta m^2_\text{atm}}\theta_{13}^{-1}+ \frac{m^2}{\Delta
  m^2_\text{sol}}$ & $\theta_{13}^{-1}$&$\theta_{13}^{-1}+\zeta^{-1}$
\bigstrut\\
$\re P_{31}$ & $\frac{m^2}{\Delta m^2_\text{sol}}$ & $\sqrt{\zeta}$ & $\zeta^{-1}$
&$\frac{m^2}{\Delta m^2_\text{atm}}\theta_{13}^{-1}+ \frac{m^2}{\Delta
  m^2_\text{sol}}$ &   $\theta_{13}^{-1}$&$\theta_{13}^{-1}+\zeta^{-1}$
\bigstrut\\
$\re P_{32}$ & $\frac{m^2}{\Delta m^2_\text{sol}}$ & $\sqrt{\zeta}$& $\zeta^{-1}$ &
$\frac{m^2}{\Delta m^2_\text{atm}}\theta_{13}^{-1}+ \frac{m^2}{\Delta
  m^2_\text{sol}}$ &$\sqrt{\zeta}\theta_{13}^{-1}$&$\zeta^{-1}$
\bigstrut\\
$\im P_{21}$ & $\frac{m^2}{\Delta m^2_\text{sol}}$ & $1$ & $\zeta^{-1}$
&$\frac{m^2}{\Delta m^2_\text{atm}}\theta_{13}^{-1}+ \frac{m^2}{\Delta
  m^2_\text{sol}}$ &  $\theta_{13}^{-1}$&$\theta_{13}^{-1}+\zeta^{-1}$
\bigstrut\\
$\im P_{31}$ & $\frac{m^2}{\Delta m^2_\text{sol}}$ & $1$ & $\zeta^{-1}$
& $\frac{m^2}{\Delta m^2_\text{atm}}\theta_{13}^{-1}+ \frac{m^2}{\Delta
  m^2_\text{sol}}$&   $\theta_{13}^{-1}$&$\theta_{13}^{-1}+\zeta^{-1}$
\bigstrut\\
$\im P_{32}$ & $\frac{m^2}{\Delta m^2_\text{atm}}$ & $1$ & $1$
&$\frac{m^2}{\Delta m^2_\text{atm}}\theta_{13}^{-1}+ \frac{m^2}{\Delta
  m^2_\text{atm}}$ &  $\sqrt{\zeta}\theta_{13}^{-1}$& $\zeta^{-1}$
\bigstrut\\
\hline
  \end{tabular}
  \caption{Generic enhancement and suppression factors for the evolution
   of the CP phases, yielding an estimate of the size of the RG effect.
   The table entries correspond to the terms in the mixing parameter
   RGEs with the coefficient given by the first column.
   A `$1$' indicates that there is no generic enhancement or suppression. `d.' denotes a degenerate neutrino mass spectrum, i.e.\ $\Delta
    m_\mathrm{atm}^2\ll m_1^2\sim m_2^2\sim m_3^2\sim m^2$. `n.h.' denotes a
    normally hierarchical mass spectrum, i.e.\ $m_1\ll m_2\ll m_3$, and
    `i.h.' means an inverted hierarchy, i.e.\ $m_3\ll m_1\lesssim m_2$.}
  \label{tab:EnhancementFactorsPhases}
\end{table}

The situation is similar for the Majorana phases.  By the same reasoning as for
the running of the solar angle, the generic RG effects are large for degenerate
masses and for an inverted hierarchy, while they are suppressed for a strong
normal hierarchy. The coefficients of $P_{fg}$ in $\Dot \varphi_i$ are given in
Tab.~\ref{tab:RGcorrections4MajPhases}. These formulae are also important to
understand the evolution of the mixing angles in some cases.  An example will be
discussed in Sec.~\ref{sec:RunningDespitePhases}.

\begin{table}\centering
  \begin{tabular}{|l|c|}\hline
    &$16\pi^2 \left(\Dot\varphi_1-\Dot\varphi_2\right)$\\\hline 

$ P_{11}$ &
$-4\mathcal{S}_{12}\cos 2\theta_{12}$ \\

 $P_{22}$ &
$4\mathcal{S}_{12}c_{23}^2 \cos 2\theta_{12}$ \\

 $P_{33}$ &
 $4 \mathcal{S}_{12}s_{23}^2\cos 2\theta_{12}$ \\
 
 $\re P_{21}$ &
 $-8\mathcal{S}_{12}c_{23}\cos 2\theta_{12}\cot 2\theta_{12}$ \\
 
 $\re P_{31}$ &
 $8\mathcal{S}_{12}s_{23}\cos 2\theta_{12} \cot 2\theta_{12}$ \\
 
 $\re P_{32}$ &
 $-4\mathcal{S}_{12} \cos 2\theta_{12}\sin 2\theta_{23}$ \\
 
 $\im P_{21}$ &
 $-4\mathcal{Q}^-_{12}c_{23} \cot 2\theta_{12}$ \\
 
 $\im P_{31}$ &
 $4\mathcal{Q}^-_{12} s_{23}\cot 2\theta_{12}$ \\
 
 $\im P_{32}$ &
 $0 $\\\hline
  \end{tabular}
\caption{Coefficients of $P_{fg}$ in the slope of the Majorana phase difference
 for $\theta_{13}=0$. The abbreviations $\mathcal{S}_{ij}$ and $\mathcal{Q}_{ij}^\pm$ depend on the mass eigenvalues and phases only, and enhance the
running for a degenerate mass spectrum since they are of the form
$f_{ij}(m_i,m_j,\varphi_1,\varphi_2)/(m_j^2-m_i^2)$.  They are listed in
Tab.~\ref{tab:Coefficients}. We use the abbreviations $c_{ij}=\cos\theta_{ij}$
and $s_{ij}=\sin\theta_{ij}$ (cf.\ App.~\ref{subsec:Conventions4MixingParameters}).}
\label{tab:RGcorrections4MajPhaseDiff}
\end{table}

The evolution of the Majorana phase difference is governed by a simple equation,
which can be read off from Tab.~\ref{tab:RGcorrections4MajPhaseDiff}. It
indicates strong running, since the slope is still inversely proportional to
$\Delta m^2_\mathrm{sol}$. However, in the case of equal Majorana phases, only
the imaginary entries in $P$ and terms proportional to $\theta_{13}$ contribute
to the running.  Besides, the contribution proportional to the real parts is
suppressed for large solar mixing.

If $Y_\nu^\dagger Y_\nu$ is close to the identity matrix, its contribution to
the running is very small, since the terms proportional to the diagonal entries
cancel approximately.  Then, only the contribution from $Y_e^\dagger Y_e$
remains, so that the evolution above the see-saw scales is essentially the same
as below. However, many GUT models suggest a hierarchical structure for $Y_\nu$
like for the other Yukawa matrices. Then the main contribution will be due to
$P_{33}$ and the next-to-leading contribution will be from $\re P_{32}$,
if $Y_\nu^\dagger Y_\nu$ is almost diagonal in the basis with
diagonal $Y_e^\dagger Y_e$.
Thus, the phase difference tends to decrease while running
down,\footnote{More accurately, it runs away from $\pi$ and towards either 0
 or $2\pi$, i.e.\ $|\varphi_1-\varphi_2|$ decreases for
 $|\varphi_1-\varphi_2|<\pi$ and increases for
 $|\varphi_1-\varphi_2|>\pi$.
}
as it is the case below the see-saw scales.

\subsection{Running of the Masses} \label{sec:RunningMasses}

Below the see-saw scales, the evolution of the mass eigenvalues is, to a good
approximation, described by a universal scaling caused by the
flavour-independent part of the RGE
\cite{Chankowski:1999xc,Casas:1999tg,Antusch:2003kp}. This flavour-independent
term, however, becomes smaller at high energies. In the MSSM, it can even cross
zero at some intermediate scale. Therefore, the flavour-dependent terms play a
more important role above the see-saw scales, the more so they can be larger if
the entries of $Y_\nu$ are order one.

We list the coefficients in the slope of the mass eigenvalues and of the $\Delta
m^2$s in Tab.~\ref{tab:RGcorrections4masses} and
Tab.~\ref{tab:RGcorrections4m2Ds}, respectively. Clearly, the RGE for each mass
eigenvalue is proportional to the mass eigenvalue itself. As a consequence, the
mass eigenvalues can never run from a finite value to zero or vice versa. In
other words, the rank of the effective neutrino mass matrix is conserved under
the renormalization group. In contrast, the mass squared differences can, in
principle, run through zero. This, however, requires a very high value of $m_1$.

\begin{table}
\centerline{
$
\begin{array}{|l|c|c|c|}
\hline
 & 16\pi^2 \, \Dot m_1/m_1 
 & 16\pi^2 \, \Dot m_2/m_2 
 & 16\pi^2 \, \Dot m_3/m_3 
\\ \hline
\bar\alpha & 1 & 1 & 1 \\
P_{11} & 2 c_{12}^2 & 2 s_{12}^2 & 0 \\
P_{22} & 2 s_{12}^2 c_{23}^2 & 2 c_{12}^2 c_{23}^2 & 2 s_{23}^2 \\
P_{33} & 2 s_{12}^2 s_{23}^2 & 2 c_{12}^2 s_{23}^2 & 2 c_{23}^2 \\
\re P_{21}
 & -2 \sin 2\theta_{12} c_{23} & 2 \sin 2\theta_{12} c_{23} & 0 \\
\re P_{31}
 & 2 \sin 2\theta_{12} s_{23} & -2 \sin 2\theta_{12} s_{23} & 0 \\
\re P_{32}
 & -2 \sin 2\theta_{23} s_{12}^2 & -2 \sin 2\theta_{23} c_{12}^2
 & 2\sin 2\theta_{23} \\
\im P_{21} & 0 & 0 & 0 \\
\im P_{31} & 0 & 0 & 0 \\
\im P_{32} & 0 & 0 & 0 \\
\hline
\end{array}
$}
\caption{Coefficients of $P_{fg}$ in the slope of the mass eigenvalues
 for $\theta_{13}=0$.}
\label{tab:RGcorrections4masses}
\end{table}

\begin{table}
\centerline{
$
\begin{array}{|l|c|c|}
\hline
 & 8\pi^2 \, \frac{\D}{\D t} \Delta m^2_\mathrm{sol}
 & 8\pi^2 \, \frac{\D}{\D t} \Delta m^2_\mathrm{atm}
 \rule[-6pt]{0pt}{18pt}
\\ \hline
\bar\alpha & \Delta m^2_\mathrm{sol} & \Delta m^2_\mathrm{atm} \\
P_{11}
 & 2 s_{12}^2 m_2^2 - 2 c_{12}^2 m_1^2
 & -2 s_{12}^2 m_2^2 \\
P_{22}
 & 2 c_{23}^2 \left[ c_{12}^2 m_2^2 - s_{12}^2 m_1^2 \right]
 & 2 s_{23}^2 m_3^2 - 2 c_{12}^2 c_{23}^2 m_2^2 \\
P_{33}
 & 2 s_{23}^2 \left[ c_{12}^2 m_2^2 - s_{12}^2 m_1^2 \right]
 & 2 c_{23}^2 m_3^2 - 2 c_{12}^2 s_{23}^2 m_2^2 \\
\re P_{21}
 & 2 \sin 2\theta_{12} c_{23} \left[ m_2^2 + m_1^2 \right]
 & -2 \sin 2\theta_{12} c_{23} m_2^2 \\
\re P_{31}
 & -2 \sin 2\theta_{12} s_{23} \left[ m_2^2 + m_1^2 \right]
 & 2 \sin 2\theta_{12} s_{23} m_2^2 \\
\re P_{32}
 & -2 \sin 2\theta_{23} \left[ c_{12}^2 m_2^2 - s_{12}^2 m_1^2 \right]
 & 2 \sin 2\theta_{23} \left[ m_3^2 + c_{12}^2 m_2^2 \right] \\
\im P_{21} & 0 & 0 \\
\im P_{31} & 0 & 0 \\
\im P_{32} & 0 & 0 \\
\hline
\end{array}
$}
\caption{Coefficients of $P_{fg}$ in the slope of the mass squared
 differences for $\theta_{13}=0$.}
\label{tab:RGcorrections4m2Ds}
\end{table}

The flavour-independent term in the MSSM is subject to large cancellations (cf.\
Eq.\ \eqref{eq:alphabarMSSM}). Note that the running of the mass eigenvalues
strongly depends on the top Yukawa coupling $y_t$, since the term $\bar\alpha$
contains $6y_t^2$, and on the gauge couplings, which run differently for
different SUSY breaking scales. This could, at least partially, explain why
there exist mutually inconsistent numerical results for the scaling
of the mass eigenvalues below the see-saw scales
\cite{Antusch:2003kp,Giudice:2003jh,Mei:2003gn}.

\begin{figure}[h]
\centerline{\includegraphics{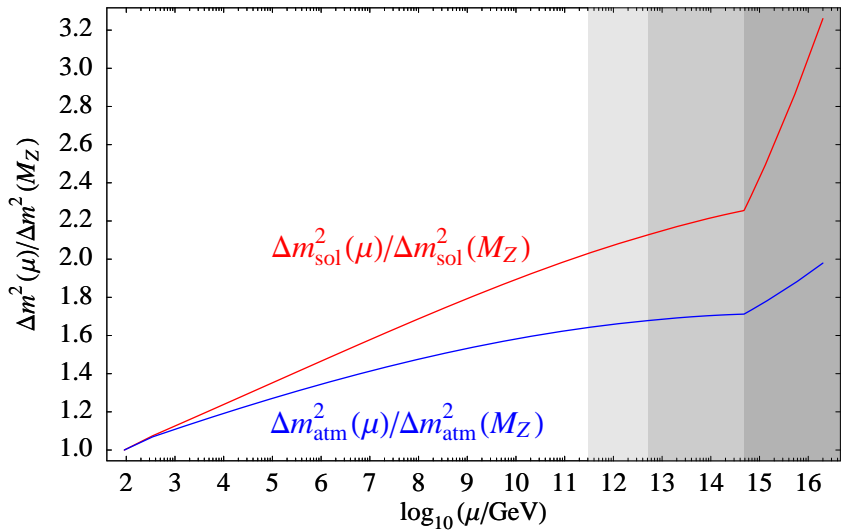}}
\caption{Example where the flavour-dependent terms dominate the running of the
mass eigenvalues for $M_3\le\mu\le M_\mathrm{GUT}$ in the MSSM. We use $Y_\nu=\diag(0.02,
0.1, 1)$ and $m_1= 0.04\,\mathrm{eV}$ at the GUT scale as well as a SUSY
breaking scale of $200\,\mathrm{GeV}$ and $\tan\beta=50$. $M$ is chosen such
that the low-energy parameters are compatible with experiment. The different
gray-shaded areas indicate the ranges of the effective theories (cf.\
Fig.~\ref{fig:EFTRanges3x3}).}
\label{fig:EvolutionDeltaM2}
\end{figure}

Between and above the see-saw scales, the running is strongly influenced by the
neutrino Yukawa couplings. In particular, depending on the size of the $Y_\nu$
entries, $\bar\alpha_\mathrm{MSSM}$ can turn negative or not. For order one
$Y_\nu$ entries, it typically stays positive. However, in such a situation,
$\bar\alpha_\mathrm{MSSM}$ becomes small so that $P$ can dominate the running.
Consider, for instance, the case of a dominant $P_{33}$ entry. Here, the
coefficient of $\Dot{m}_2$ is enhanced compared to the $\Dot{m}_1$ coefficient
by $(m_2/m_1)\,\cot^2 \theta_{12}$ (cf.\ Tab.~\ref{tab:RGcorrections4masses}).
In many cases  $\theta_{12}$ is at high scales much smaller than its low-energy
value, so that $m_2$ runs much faster than $m_1$. As a consequence, $\Delta
m^2_\mathrm{sol}$ can be significantly enhanced even for not too degenerate
spectra. A relatively drastic example is shown in
Fig.~\ref{fig:EvolutionDeltaM2}. Clearly, the discrepancy in the scaling of
$\Delta m^2_\mathrm{sol}$ and $\Delta m^2_\mathrm{atm}$ stems from the
flavour-dependent terms $P$. As $\tan\beta$ is large in this example, the
$P_{33}$ induced terms cause important effects already below the see-saw scale.
The dominant effect, however, is the running in the range $M_3\le\mu\le
M_\mathrm{GUT}$, i.e.\ over less than two orders of magnitude. By inspecting the
tables, we find that analogous features are present if other elements of $P$ are
large. In particular, one can enhance the evolution of  $\Delta
m^2_\mathrm{atm}$ as well. Therefore we expect many models which predict
realistic values for the masses at tree level to be ruled out by several
standard deviations due to RG effects.

If, on the other hand, the eigenvalues of $Y_\nu^\dagger Y_\nu$ are much smaller
than 1, $\bar\alpha_\mathrm{MSSM}$ typically flips its sign. The entries of $P$
are now small if $\tan\beta$ is small, and for large $\tan\beta$ they are
dominated by $Y_e^\dagger Y_e$. Hence, for small $\tan\beta$,
$\bar\alpha_\mathrm{MSSM}$ still dominates the running of the masses (away from
its zero point). In contrast, for large $\tan\beta$, the contribution of $P$
(being now dominated by $Y_e^\dagger Y_e$) is of similar importance, as it is
the case for the running of the effective neutrino mass operator $\kappa$ at
high energies. Since $\bar\alpha$ can be negative at scales close to the GUT
scale now, the contributions from the diagonal entries in $P$ can decrease the
RG effects.  The off-diagonal entries again can both increase and decrease
them. 

Finally, let us mention that since the terms in $\Dot m_i$ involving the
imaginary part of $P$ are proportional to $\sin\theta_{13}$, they do not
contribute in the approximation of vanishing $\theta_{13}$. Clearly, in the SM,
$\bar\alpha$ dominates the running if $Y_\nu$ is small.

\subsection{$\boldsymbol{U_e}$ Contribution to the Running}

As mentioned in the beginning of this section, the RGE for $Y_e$ contains
non-diagonal terms above and between the thresholds, so that there is an
additional contribution to the running of the leptonic mixing angles and CP
phases.  In the see-saw scenario, the RGE for $Y_e$ above $M_3$ is given by
\begin{equation}\label{eq:dYedt}
    16\pi^2 \, \frac{\D Y_e}{\D t} =
    Y_e \, (D_e \, Y_e^\dagger Y_e + D_\nu \, Y_\nu^\dagger Y_\nu) +
    \alpha_e \, Y_e =:
    Y_e \, F + \alpha_e \, Y_e
\end{equation}
with
\begin{subequations} \label{eq:DeDnu}
\begin{alignat}{2}
    D_e &= \frac{3}{2} \:,\; D_\nu=-\frac{3}{2} && \;\; \text{in the SM,}
\\
    D_e &= 3 \:,\; D_\nu = 1 && \;\; \text{in the MSSM.}
\end{alignat}
\end{subequations}
As usual, $\alpha_e$ is flavour diagonal (cf. App.~\ref{app:RGEs}).
The resulting contributions to the evolution of the angles for vanishing
$\theta_{13}$ and $y_e,y_\mu \ll y_\tau$ are listed in
Tab.~\ref{tab:YeContributionAngles}.  They
can simply be added to the expressions discussed above 
(cf.\ App.~\ref{app:Combination}).
\begin{table}
\centering
$
\begin{array}{|l|c|c|c|}
\hline
 & 16\pi^2 \, \Dot\theta_{12}^{U_e}
 & 16\pi^2 \, \Dot\theta_{13}^{U_e}
 & 16\pi^2 \, \Dot\theta_{23}^{U_e}
\\ \hline
F_{11} & 0 & 0 & 0 \\ 
F_{22} & 0 & 0 & 0 \\ 
F_{33} & 0 & 0 & 0 \\ 
\re F_{21} & -c_{23} & -s_{23} \cos\delta & 0 \\ 
\re F_{31} & s_{23} & -c_{23} \cos\delta & 0 \\ 
\re F_{32} & 0 & 0 & -1 \\ 
\im F_{21} & 0 & -s_{23} \sin\delta & 0 \\ 
\im F_{31} & 0 & -c_{23} \sin\delta & 0 \\ 
\im F_{32} & 0 & 0 & 0 \\
\hline
\end{array}
$
\caption{Coefficients of $F_{fg}$ in the $U_e$ contribution to the slope
 of the mixing angles for $\theta_{13}=0$ and $y_e,y_\mu \ll y_\tau$.
}
\label{tab:YeContributionAngles}
\end{table}

In contrast to the latter, all non-zero terms in the $U_e$ contribution
have a generic enhancement factor of 1.  The reason for this is the
strong hierarchy among the charged lepton masses.  As a consequence,
the $U_e$ contribution is negligible compared to the $U_\nu$
contribution, if the relevant factor in 
Tab.~\ref{tab:EnhancementFactorsAngles} is much larger than 1.  If it is
close to 1, both contributions are generically of the same order of
magnitude.  The $U_e$ contribution can even be dominant if the factor is
small.  This is also possible, if cancellations occur between the
leading-order terms in the RGEs.

To get a feeling for the size of the effects discussed in this section,
let us consider a rough estimate.  We assume that the running is linear
on a logarithmic
scale, that it is dominated by a single entry $y$ in $Y_\nu$, which
is related to the light neutrino mass $m_3$ and the see-saw scale $M_3$
by $m_3 = \frac{v^2}{2} \frac{y^2}{M_3}$, and that the relevant term in
Tab.~\ref{tab:YeContributionAngles} is of the order of 1.  Then we find
\begin{equation}
    |\Delta\theta^{U_e}| \sim
    |\Dot\theta^{U_e}| \, \ln\frac{M_\mathrm{GUT}}{M_3} \sim
    D_\nu y^2 \,
     \Bigl( 0.027 + 0.006\, \ln\frac{m_3/0.1\eV}{y^2} \Bigr) \;.
\end{equation}
Thus, the change is small, but it can still be
relevant in the context of precision studies (e.g.\ the change of
$\theta_{13}$), if $y$ is large.

\section{Running between the See-Saw Scales} \label{sec:RunningBetween}
Between the see-saw scales, the singlets are partly integrated out, 
which implies that only a  
$(n\!-\!1)\times 3$ submatrix of the neutrino Yukawa matrix remains.
Therefore, we expect that the running between the thresholds caused by the 
neutrino Yukawa matrix can differ significantly from the running 
above or below them. 

\begin{figure}
\centering
\includegraphics{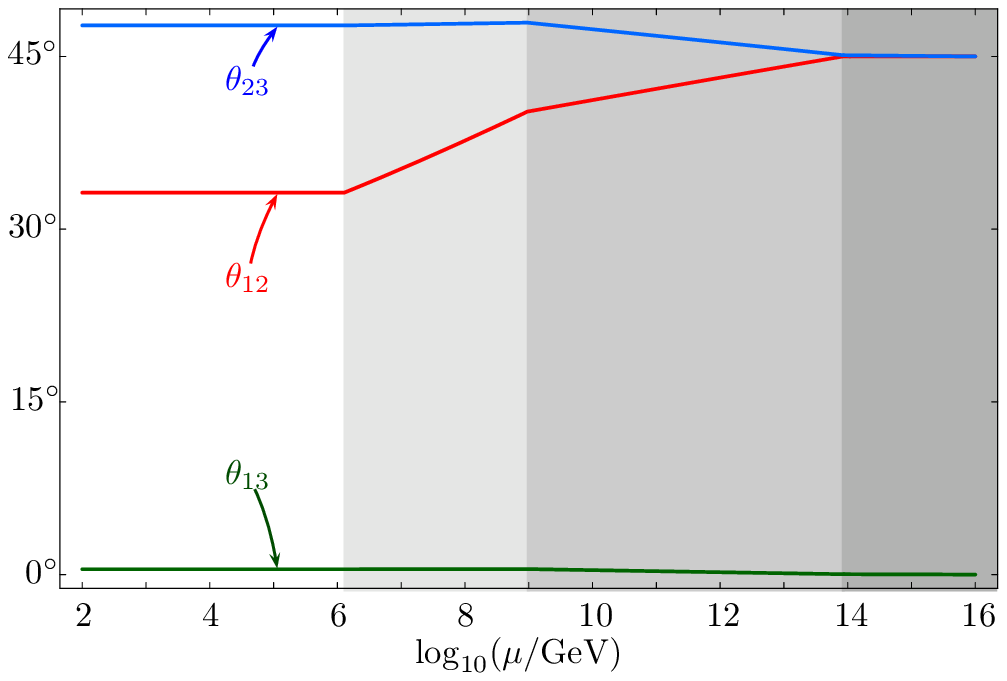}
 \caption{\label{fig:RunningFromFlavouBlindInt}
Running from maximal solar mixing at $M_\mathrm{GUT}$ to the experimentally preferred 
angle of
the LMA solution. The figure shows an example in the SM with a negative CP 
parity
for $m_2$ and a Yukawa matrix $Y_\nu = 0.5 \cdot 
\diag(\varepsilon^2, \varepsilon, 1)$ 
at $M_\mathrm{GUT}$ with
$\varepsilon=3.5 \cdot 10^{-3}$ and 
a normal mass hierarchy (from \cite{Antusch:2002hy}). 
The lightest neutrino has a mass  
of $0.004$~eV (at low energy).
The gray-shaded areas illustrate the validity ranges of the effective
theories emerging from integrating out the heavy singlet neutrinos.
 } 
\end{figure}

We now discuss the running due to the terms in the $\beta$-functions with a
 flavour structure proportional to the unit matrix.
Below and above the see-saw scales, they only cause a 
common scaling of the elements of the neutrino mass matrix  
and thus leave the mixing angles and phases unchanged. Between the thresholds,
however, the effective neutrino mass matrix consists of the two parts 
$\accentset{(n)}{\kappa}$ and
$2 \, \accentset{(n)}{Y}_\nu^T \accentset{(n)}{M}^{-1} \accentset{(n)}{Y}_\nu $, as shown in Eq.\
(\ref{eq:EffNuMass}). 
Here, the mixing angles and phases change in general, unless both parts are
scaled equally.   
From table \ref{tab:BetaEffNuMass}, we see that    
in the SM, the $\beta$-functions
 $\accentset{(n)}{\beta}_{{\kappa}}$ and 
$\accentset{(n)}{\beta}_{2 {Y}_\nu^T{M}^{-1} {Y}_\nu}$, have different
coefficients in the terms proportional to the gauge couplings and to
the Higgs self-coupling \cite{Antusch:2002rr}. 
This difference 
can be understood by 
looking at the corresponding diagrams of the ``full'' and the effective theory. 
For instance,  
the diagram for the correction to the effective vertex proportional to  
$\lambda$ and its 
counterpart with the
heavy singlet running in the loop are shown in figure 
\ref{fig:lambdaContrFullAndEffTheory}.     
Diagram (a) is UV divergent, whereas diagram (b) is UV finite. 
 We thus get no contribution proportional to $\lambda$ for the     
$\beta$-function of the composite operator. 
The situation is similar for some of the 
diagrams corresponding to the
vertex corrections proportional to the gauge couplings. 
Thus, in the SM,    
the RG scaling of the two parts $\accentset{(n)}{\kappa}$ and 
$2\, \accentset{(n)}{Y}_\nu^T \accentset{(n)}{M}^{-1} \accentset{(n)}{Y}_\nu$ 
of the effective mass matrix between the thresholds, caused by the 
interactions with trivial flavour structure, is different.    
This implies a running of the mixing angles and CP phases
in addition to the running of the mass eigenvalues.%
\footnote{To see this, let us assume that
 $U^T \RaiseBrace{\bigl(}\, \accentset{(n)}{\kappa} +
  2 \,\accentset{(n)}{Y}_\nu^T \accentset{(n)}{M}^{-1}
  \accentset{(n)}{Y}_\nu \RaiseBrace{\bigr)}\, U$
 is diagonal.  Then
 $U^T \RaiseBrace{\bigl(} a\,\accentset{(n)}{\kappa} +
  b \, 2 \,\accentset{(n)}{Y}_\nu^T \accentset{(n)}{M}^{-1}
  \accentset{(n)}{Y}_\nu \RaiseBrace{\bigr)}\, U$
 is in general only diagonal if $a=b$ (common scaling).
}
This effect can even give the dominant contribution to the
running of the mixing angles, as for instance in the example shown 
 in figure \ref{fig:RunningFromFlavouBlindInt} (from \cite{Antusch:2002hy}). 

\begin{figure}
\centering
    \subfigure[]{
                 \CenterEps[1]{efflambda}} \qquad
    \subfigure[]{
                 \CenterEps[1]{fulllambda}}     
                 \vspace{-0.2cm}        
 \caption{\label{fig:lambdaContrFullAndEffTheory}
 Figure (a) shows the diagram  which gives the 
contribution proportional to the Higgs self-coupling in the $\beta$-function of
the neutrino mass operator in the SM. Figure (b) shows its finite counterpart with the
heavy singlet running in the loop. The gray box labeled by $\kappa^i$
corresponds to the contribution to the effective neutrino mass operator from
integrating out the heavy singlet $N_\mathrm{R}^i$. 
 } 
\end{figure}

Due to the non-renormalization theorem in supersymmetric theories,
$\accentset{(n)}{\beta}_\kappa$ and  $\accentset{(n)}{\beta}_{2 Y_\nu^T M^{-1}
Y_\nu}$ are identical in the MSSM (see Tab.~\ref{tab:BetaEffNuMass} on
p.~\pageref{tab:BetaEffNuMass}), so that we can use the RGEs of
Sec.~\ref{sec:AnalyticalFormulae} between the see-saw scales as well.  In
particular, the enhanced running between the thresholds due to terms with a
trivial flavour structure does not occur.  Of course, the heavy degrees of
freedom have to be integrated out first, i.e.\ all parameters have to be
replaced by the effective ones between the thresholds.

\section{Mathematica Packages for Numerical RG Analyses} 
\label{sec:MathematicaPackages}
\subsection{Numerical Solution of the RGEs}

The Mathematica package \package{REAP} (Renormalization Group Evolution
of Angles and Phases) numerically solves the RGEs of the quantities relevant
for neutrino masses, for example the dimension 5 neutrino mass operator, the
Yukawa matrices and the gauge couplings.  The $\beta$-functions for the SM, the
MSSM and two Higgs doublet models with $\mathbb{Z}_2$ symmetry for FCNC
suppression (2HDM) with and without right-handed neutrinos are implemented.  In
addition, the same models are available for Dirac neutrinos.  New models can be
added by the user.  The heavy singlet neutrinos can be integrated out
automatically at the correct mass thresholds, as described in
Sec.~\ref{sec:RunningInSeeSaw}.\footnote{We do not consider SUSY threshold
corrections \cite{Chun:1999vb}, as they are usually numerically less important
\cite{Chankowski:2001hx}.} The software can also be applied to type II see-saw
models as long as one only considers the energy region below the additional
see-saw scale $M_\Delta$, where the new physics such as Higgs triplets only
leads to another contribution to the effective neutrino mass operator. The
package can be downloaded from \url{http://www.ph.tum.de/~rge/REAP/}.
Mathematica 5 is required.

\subsection{Extraction of Mixing Parameters from Mass Matrices}

The package \package{MixingParameterTools} (\package{MPT}) allows to extract the
physical lepton masses, mixing angles and CP phases from the mass matrices of
the neutrinos and the charged leptons. Thus, the running of the neutrino mass
matrix calculated by \package{REAP} can be translated into the running of the
mixing parameters and the mass eigenvalues. For the definition of the mixing
parameters, see App.~\ref{subsec:Conventions4MixingParameters} and the
documentation of the package. \package{MixingParameterTools} can also be useful
as a stand-alone application in order to study textures without running, and it
is not bound to the analysis of neutrino masses only but may be used for quark
and superpartner mass matrices as well.  Therefore, it can be obtained
separately from \package{REAP} at \url{http://www.ph.tum.de/~rge/MPT/}.

\subsection{Example Calculation}
The following simple example demonstrates how to use the Mathematica
packages to calculate the RG evolution of the neutrino mass matrix in
the MSSM extended by three heavy singlet neutrinos.  Of course, further
documentation is provided together with the packages.

\begin{enumerate}

\item The package corresponding to the model at the highest energy has
to be loaded.  All other packages needed in the course of the
calculation are loaded automatically.
(Note that ` is the backquote, which is used in opening quotation marks,
for example.)
\begin{verbatim}
  Needs["REAP`RGEMSSM`"]
\end{verbatim}

\item Next, we specify that we would like to use the MSSM with singlet
neutrinos and $\tan\beta=50$.  Furthermore, we set the SUSY breaking
scale to $200\GeV$ and use the SM as an effective theory below this
scale.
\begin{verbatim}
  RGEAdd["MSSM",RGEtan\[Beta]->50]
  RGEAdd["SM",RGECutoff->200]
\end{verbatim}

\item Now we have to provide the initial values.  For instance, let us
set the GUT-scale value of $\theta_{12}$ to $45^\circ$ and that of the
first Majorana phase to $50^\circ$.  Besides, we use a simple diagonal
pattern for the neutrino Yukawa matrix and the default values of the
package for the remaining parameters.
\begin{verbatim}
  RGESetInitial[2*10^16,
    RGE\[Theta]12->45 Degree,RGE\[Phi]1->50 Degree,
    RGEY\[Nu]->{{1,0,0},{0,0.5,0},{0,0,0.1}}]
\end{verbatim}

\item \function{RGESolve[\param{low},\param{high}]} solves the RGEs
between the energy scales low and high.  The heavy singlets are
integrated out automatically at their mass thresholds.
\begin{verbatim}
  RGESolve[100,2*10^16]
\end{verbatim}

\item Using \function{RGEGetSolution[\param{scale},\optparam{quantity}]}
we can query the value of the quantity given in the second argument at
the energy given in the first one.  For example, this returns the mass
matrix of the light neutrinos at $100\GeV$:
\begin{verbatim}
  MatrixForm[RGEGetSolution[100,RGEM\[Nu]]]
\end{verbatim}

\item To find the leptonic mass parameters, we use the function
\function{MNSParameters[\param{$m_\nu$},\param{$Y_e$}]} (which also
needs the Yukawa matrix of the charged leptons).  The results are given
in the order
$\{\{\theta_{12},\theta_{13},\theta_{23},\delta,\delta_e,\delta_\mu,
\delta_\tau,\varphi_1,\varphi_2\},
\{m_1,m_2,m_3\},\{y_e,y_\mu,y_\tau\}\}$.
\begin{verbatim}
  MNSParameters[
    RGEGetSolution[100,RGEM\[Nu]],RGEGetSolution[100,RGEYe]]
\end{verbatim}

\item Finally, we can plot the running of the mixing angles:
\begin{verbatim}
  Needs["Graphics`Graphics`"]
  mNu[x_]:=RGEGetSolution[x,RGEM\[Nu]]
  Ye[x_]:=RGEGetSolution[x,RGEYe]
  \[Theta]12[x_]:=MNSParameters[mNu[x],Ye[x]][[1,1]]
  \[Theta]13[x_]:=MNSParameters[mNu[x],Ye[x]][[1,2]]
  \[Theta]23[x_]:=MNSParameters[mNu[x],Ye[x]][[1,3]]
  LogLinearPlot[{\[Theta]12[x],\[Theta]13[x],\[Theta]23[x]},
   {x,100,2*10^16}]
\end{verbatim}

\end{enumerate}

\section{Applications}
\label{sec:Applications}
We now apply the analytical and numerical tools described in the previous
sections to some specific cases with interesting RG effects above, between
and below the see-saw scales within the conventional see-saw scenario.

\subsection{RG Effects for a Dominant $\boldsymbol{(Y_\nu)_{33}}$}
\label{sec:HierarchicalYnu}
Many unified models relate the Yukawa couplings of the different charged
fermions and the neutrinos, e.g.\ $Y_\nu\sim Y_u$ or $Y_\nu\sim Y_e$. For the
charged fermions, the quantities accessible through observation are $Y^\dagger
Y$, where $Y$ denotes the corresponding Yukawa matrix. It is convenient to work 
in the basis where $Y_u^\dagger Y_u$ and $Y_e^\dagger Y_e$ are diagonal and
positive, and the diagonal entries are ordered ascendingly. In this basis, all
three combinations $Y^\dagger Y$ have a dominant 33 entry. In this subsection,
we shall assume a similar pattern for $Y_\nu^\dagger Y_\nu$, i.e.\
$(Y_\nu^\dagger Y_\nu)_{33}\approx y_3^2\gg (Y_\nu^\dagger Y_\nu)_{ij\ne33}$.
Given such a hierarchy for $Y_\nu^\dagger Y_\nu$, the RG corrections
$\Delta\theta_{13}:=\theta_{13}(M_\mathrm{SUSY})-\theta_{13}(M_\mathrm{GUT})$
and $\Delta\theta_{23}$ can be approximated by 
\begin{eqnarray}
\Delta\theta_{13}
& \approx & 
        \frac{-1}{32\pi^2}
                        \left[C_e
                        y_\tau^2\,\ln\Bigl(\frac{M_\mathrm{GUT}}{M_\mathrm{SUSY}}\Bigr)
                        +C_\nu y_3^2\,
                        \ln\Bigl(\frac{M_\mathrm{GUT}}{M_*}\Bigr)
                \right] \,
        \sin 2\theta_{12} \, \sin 2\theta_{23}
        \times
\nonumber\\
&& {} \quad \times
        \frac{m_3}{\Delta m^2_\mathrm{atm} \left( 1+\zeta \right)}
        \left[
         m_1 \cos(\varphi_1-\delta) -
         \left( 1+\zeta \right) m_2 \, \cos(\varphi_2-\delta) -
         \zeta m_3 \, \cos\delta
        \right] \quad\;\;\;
\label{eq:AnalyticApproxT13} 
\\
\Delta\theta_{23}
& \approx &
        \frac{1}{32\pi^2}
                        \left[C_e y_\tau^2\,\ln\Bigl(\frac{M_\mathrm{GUT}}{M_\mathrm{SUSY}}\Bigr)
                        +C_\nu y_3^2 \,
                        \ln\Bigl(\frac{M_\mathrm{GUT}}{M_*}\Bigr)\right] 
        \, \sin 2\theta_{23}
   \times
                \nonumber\\
&& {} \quad \times                   
                        \frac{1}{\Delta m^2_\mathrm{atm}} 
        \left[
         c_{12}^2 \, |m_2\, e^{\I \varphi_2} + m_3|^2 +
         s_{12}^2 \, \frac{|m_1\, e^{\I \varphi_1} + m_3|^2}{1+\zeta}
        \right] ,
\label{eq:AnalyticApproxT23} 
\end{eqnarray}
where $M_*$ denotes the mass scale of the heavy neutrino(s) 
with the large Yukawa couplings.\footnote{
For the analytic estimates, we ignore complications due to the generically
non-degenerate see-saw scales \cite{Antusch:2002rr}. 
} 
To obtain these results, we read off the RGEs from 
Tab.~\ref{tab:RGcorrections4angles}, and integrated them with the approximation of
constant coefficients.
This is reasonably accurate, since the running of
$\theta_{13}$ and $\theta_{23}$ is almost linear on logarithmic scales
\cite{Antusch:2003kp}.%
\footnote{A comparison with numerical calculations shows that this is
 unchanged in the presence of neutrino Yukawa couplings.
}

In the SM, the term proportional to $y_\tau^2$ is negligible, since the 
Yukawa coupling is not enhanced by $\tan\beta$.  However, the $y_3^2$
contribution can be large, and it is not suppressed for small
$\tan\beta$.
Furthermore, except for $y_3$ and $M_*$, only (the RG extrapolation of)
low-energy parameters enter the expressions \eqref{eq:AnalyticApproxT13} and
\eqref{eq:AnalyticApproxT23}. 

In the case of the solar angle, the running is strongly non-linear when the  RG
change is large. Then, the approximation used in the above equations does  not
yield reliable results.  Even by integrating the RGE (assuming $\theta_{12}$ to
vary  but the other parameters to be constant), one arrives at an expression
which does not represent an accurate approximation in many cases because of the
running of $\Delta m^2_\mathrm{sol}$. Nevertheless, an inspection of the RGE
reveals several qualitative features of the running such as the damping
influence of the phases, as discussed in Sec.~\ref{sec:RunningAngles}.

The running of the Majorana phases may be regarded as encouraging for the
prospects of neutrinoless double $\beta$ decay experiments: it is known that
even if the mass eigenvalues are large enough to make a discovery in future
experiments possible, cancellations may strongly suppress the amplitude
\cite{Rodejohann:2000ne}. This can directly be seen from the fact that the
amplitude is governed by the effective neutrino mass
\begin{equation}
 \langle m_\nu\rangle
 \,=\,\left|  
        m_1 \, c_{12}^2 c_{13}^2 \, e^{\I \varphi_1} +
    m_2 \, s_{12}^2 c_{13}^2 \, e^{\I \varphi_2} +
    m_3 \, s_{13}^2 \, e^{2\I \delta}
  \right| \;,
\end{equation}
which is obviously suppressed if $\varphi_1-\varphi_2$ is close to $\pi$. 
However, for dominant $P_{33}$, the difference of Majorana phases is driven away
from $\pi$ at low energies due to RG effects 
(cf.\ the discussion in Sec.~\ref{sec:RunningOfPhases}). This implies that
cancellations tend to be avoided. Note that the contribution from $Y_e^\dagger
Y_e$, which persists below the see-saw scales, increases the effect
\cite{Antusch:2003kp}.

\subsection{Neutrino Yukawa Couplings with Two Large Entries}
\label{sec:ApproxFormulaeSHRND}
As another example, let us assume that the neutrino Yukawa matrix
contains two dominant entries, $(Y_\nu)_{33} \approx e^{-\I\gamma}
(Y_\nu)_{32} \approx y_3$ with an arbitrary phase $\gamma$, as it is the
case in many models where the large atmospheric mixing angle emerges
from $Y_\nu$ in the basis where $Y_e$ is diagonal.  Then 
$(Y_\nu^\dagger Y_\nu)_{33} \approx (Y_\nu^\dagger Y_\nu)_{22}$, which
causes a cancellation between the contributions proportional to these
terms in the RGEs of $\theta_{13}$ and $\theta_{23}$.  Thus, using the
same linear approximation as in Sec.~\ref{sec:HierarchicalYnu}, we
obtain the changes
\begin{eqnarray}
        \Delta\theta_{13}
& \approx & 
        \frac{-1}{32\pi^2}
        \left[
         C_e y_\tau^2\,\ln\Bigl(\frac{M_\mathrm{GUT}}{M_\mathrm{SUSY}}\Bigr)
          \sin 2\theta_{23} -
         2 C_\nu y_3^2 \cos\gamma \,
          \ln\Bigl(\frac{M_\mathrm{GUT}}{M_*}\Bigr) \cos 2\theta_{23}
        \right] \, \sin 2\theta_{12} \times
\nonumber\\*
&& {} \quad \times
        \frac{m_3}{\Delta m^2_\mathrm{atm} \left( 1+\zeta \right)}
        \left[
         m_1 \cos(\varphi_1-\delta) -
         \left( 1+\zeta \right) m_2 \, \cos(\varphi_2-\delta) -
         \zeta m_3 \, \cos\delta
        \right]
\nonumber\\*
&& {} +
        \frac{1}{16\pi^2}
        C_\nu y_3^2 \sin\gamma \,
        \ln\Bigl(\frac{M_\mathrm{GUT}}{M_*}\Bigr)
        \sin 2\theta_{12} \times
\nonumber\\*
&& {} \quad \times
        \frac{m_3}{\Delta m^2_\mathrm{atm} \left( 1+\zeta \right)}
        \left[
         m_1 \sin(\varphi_1-\delta) -
         \left( 1+\zeta \right) m_2 \, \sin(\varphi_2-\delta) +
         \zeta m_3 \, \sin\delta
        \right] \quad
\label{eq:ApproxDeltaT13SHRND} 
\\
        \Delta\theta_{23}
& \approx &
        \frac{1}{32\pi^2}
        \left[
         C_e y_\tau^2\,\ln\Bigl(\frac{M_\mathrm{GUT}}{M_\mathrm{SUSY}}\Bigr)
          \sin 2\theta_{23} -
         2 C_\nu y_3^2 \cos\gamma \,
          \ln\Bigl(\frac{M_\mathrm{GUT}}{M_*}\Bigr) \cos 2\theta_{23}
        \right] \times
\nonumber\\*
&& {} \quad \times                   
        \frac{1}{\Delta m^2_\mathrm{atm}} 
        \left[
         c_{12}^2 \, |m_2\, e^{\I \varphi_2} + m_3|^2 +
         s_{12}^2 \, \frac{|m_1\, e^{\I \varphi_1} + m_3|^2}{1+\zeta}
        \right]
\nonumber\\*
&& {} -
        \frac{1}{8\pi^2} C_\nu y_3^2 \sin\gamma \,
        \ln\Bigl(\frac{M_\mathrm{GUT}}{M_*}\Bigr) \,
        \frac{m_3}{\Delta m^2_\mathrm{atm}} 
        \left[
         c_{12}^2 \, m_2 \, \sin\varphi_2 +
         s_{12}^2 \frac{m_1 \sin\varphi_1}{1+\zeta}
        \right]
\nonumber\\*
&& {} +
        \frac{1}{16\pi^2} D_\nu y_3^2 \cos\gamma \,
        \ln\Bigl(\frac{M_\mathrm{GUT}}{M_*}\Bigr) \;.
\label{eq:ApproxDeltaT23SHRND} 
\end{eqnarray}
The change proportional to the real part of $P_{32}$ vanishes for
maximal atmospheric mixing.  Hence, the neutrino Yukawa couplings only
contribute significantly to the running of $\theta_{13}$ in this case,
if $(Y_\nu)_{32}$ has a large imaginary part and if the CP phases are
not close to 0 or $\pi$.  In $\Delta\theta_{23}$, they always play a
role by inducing off-diagonal elements in $Y_e^\dagger Y_e$, which leads
to the last term in Eq.~\eqref{eq:ApproxDeltaT23SHRND}.  This term is
actually dominant in the case of CP conservation and small $\tan\beta$.

\subsection{RG Corrections and Precision Measurements}

\begin{figure}
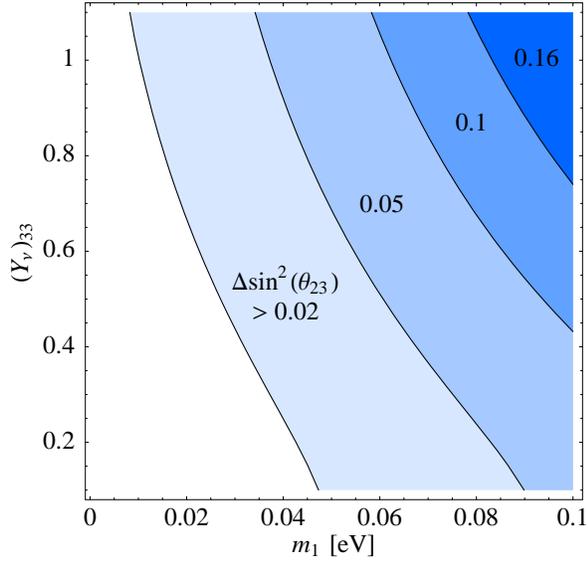
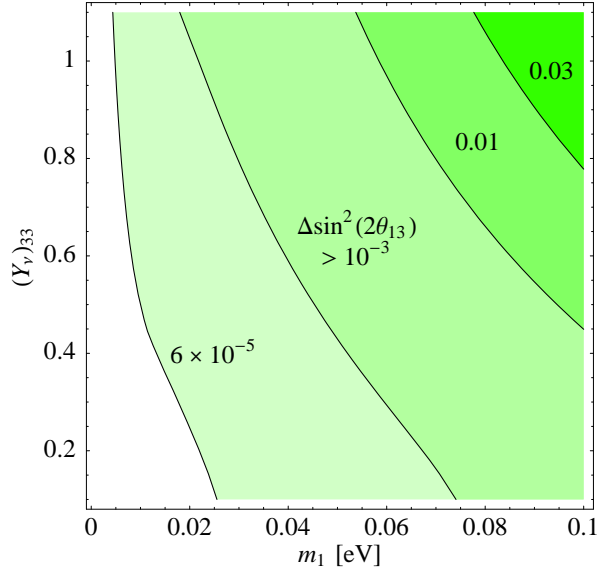
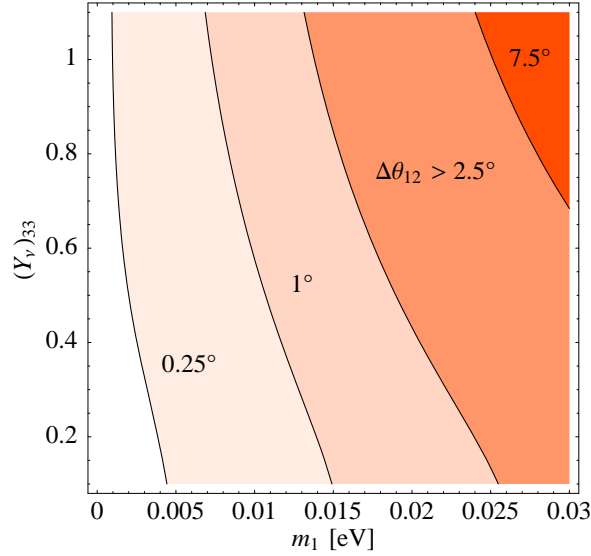

 \centering
 \subfigure[\label{fig:DeltaT23} RG induced change of $\sin^2 \theta_{23}$]{\CenterEps[0.8]{DeltaT23_C}}
 \hfil
 \subfigure[\label{fig:DeltaT13} RG induced change of $\sin^2 2\theta_{13}$]{\CenterEps[0.8]{DeltaT13_C}}
 \\
 \subfigure[\label{fig:DeltaT12} RG induced change of $\theta_{12}$]{\CenterEps[0.8]{DeltaT12_C}}
 \caption{\label{fig:DeltaRG} Estimated RG corrections to $\theta_{13}=0$,
 $\theta_{23}=\pi/4$ and  $\theta_{12}+\theta_\mathrm{C}=\pi/4$ with a large 
 $P_{33}$  in the MSSM with $\tan\beta=20$, $M_\mathrm{SUSY}=1\TeV$ and a normal
 neutrino mass ordering. For instance, $P_{33}$ can be generated  from the entry
 $(Y_\nu)_{33}$ in the neutrino Yukawa matrix, which was assumed here.  The
 running between the electroweak and the GUT scale has been calculated using the
 approximate formulae \eqref{eq:AnalyticApproxT13} and
 \eqref{eq:AnalyticApproxT23}. For producing the plots we have used    $\Delta
 m^2_\mathrm{atm}$ and $\Delta m^2_\mathrm{sol}$ at  $\mu=10^{14}$ GeV, which,
 for the considered parameter ranges, are about a factor $1.75$ larger than the 
 low energy values.    In Fig.~(a) and (c) the CP phases have been set to zero,
 and in Fig.~(b) $\varphi_1=0$ and $\varphi_2=\pi$ was assumed, leading to
 un-suppressed running.  Besides, the initial condition $\theta_{13}=0$ as well
 as the best-fit values for the remaining parameters have been used. }
\end{figure}

In this section, we will estimate the order of magnitude of RG effects
in see-saw models and compare it to the precision of future measurements
of neutrino mixing (see also
\cite{Mei:2004rn,Singh:2004zu,Ferrandis:2004vp,Kang:2005as} for related
works).  We shall first consider the effects of a large
$P_{33}$ as an example.  For instance, $P_{33}$ can be generated from
the entry $(Y_\nu)_{33}$.  Note that this is only an example.  RG
effects from different structures of $Y_\nu$ can be understood and
estimated using the analytic formulae of
Sec.~\ref{sec:AnalyticalFormulae}.  Graphically, the RG corrections
caused by $P_{33}$ in the MSSM with $\tan\beta=20$ are illustrated in
Fig.~\ref{fig:DeltaRG}.  We have assumed the initial values
$\theta_{13}=0$, $\theta_{23}=\pi/4$ and
$\theta_{12}+\theta_\mathrm{C}=\pi/4$ (where $\theta_\mathrm{C}$ is the
Cabibbo angle) at high energy, which may be especially interesting from
a theoretical point of view \cite{Raidal:2004iw,Minakata:2004xt,Frampton:2004vw}.
The changes of $\theta_{13}$ and $\theta_{23}$ have been calculated from
the approximations \eqref{eq:AnalyticApproxT13} and
\eqref{eq:AnalyticApproxT23}. 
We would like to stress that the mass squared differences are running
quantities as well and taking them as constant, as it was done in
Eqs.~\eqref{eq:AnalyticApproxT13} and \eqref{eq:AnalyticApproxT23},
restricts the accuracy of the estimates.
For producing the plots in Fig.~\ref{fig:DeltaRG}, we have used   
the values of $\Delta m^2_\mathrm{atm}$ and $\Delta m^2_\mathrm{sol}$ at
$\mu=10^{14}\GeV$.
For the considered parameter ranges and for $m_t(m_t)=175\GeV$ and 
$M_\mathrm{SUSY}=1\TeV$, the mass squared differences at $\mu=10^{14}\GeV$
are about a factor $1.75$ larger than the low-energy values. Note that their
running depends sensitively on the value of the top mass and on the SUSY 
breaking scale. 
The change of $\theta_{12}$ has also been
determined assuming a linear running, which is possible here because
only rather small neutrino masses and a moderate value of $\tan\beta$
are considered in the plot.  We have used those values for the Majorana
phases that do not damp the RG evolution, as well as best-fit values for
the oscillation parameters.
For the see-saw scale associated with the large Yukawa coupling, we have
used the approximation  
\begin{equation} \label{eq:ApproxMStar}
M_* \approx M_{33} \approx \frac{v^2}{2} (Y_\nu)_{33}^2 \, (m_\nu^{-1})_{33} \;.
\end{equation}
To justify this, let us reconstruct $M$ from $Y_\nu$ and $m_\nu$ using
the inverse of the see-saw formula \eqref{eq:mnuFullTheory},
$M = -\frac{v^2}{2} Y_\nu \, m_\nu^{-1} Y_\nu^T$,
for a dominant entry $(Y_\nu)_{33}$ in $Y_\nu$ and not too large
neutrino masses, $m_1 \lesssim 0.1\eV$.  In this case, one can see from
$m_\nu^{-1} = U_\nu \, \diag(m_1^{-1},m_2^{-1},m_3^{-1}) \, U_\nu^T$
that all entries of the inverse light neutrino mass matrix are usually
of the same order of magnitude.%
\footnote{Only for a narrow range in $m_1$ and a large difference of
 the Majorana phases, a suppression of the element $(m_\nu^{-1})_{33}$
 is possible.  Then, Eq.~\eqref{eq:ApproxMStar} may not be a good
 approximation.
}
Consequently, $M_{33}$ is dominated by the term proportional to
$(Y_\nu)_{33}^2$, i.e.\ the one given in Eq.~\eqref{eq:ApproxMStar}.
Furthermore, $M_{33}$ is the dominant entry in $M$, so that it is
approximately equal to the largest eigenvalue $M_3=M_*$.

We find that the RG changes are comparable to the sensitivities of planned
precision experiments (cf.~Tabs.~\ref{tab:SensitivitiesTheta13} and
\ref{tab:SensitivitiesTheta23}) in the  shaded parts of the parameter space,
providing a reason to be optimistic about the potential of these experiments to
find interesting results and to constrain model parameters.  Compared to the
change due to the charged lepton Yukawa couplings alone \cite{Antusch:2003kp},
the gray-shaded regions are expanded, since the contribution from the neutrino
Yukawa couplings has  the same sign in the case we considered. For a very strong
mass hierarchy, we find very small RG effects in our example.  One reason for
this is the decrease of the enhancement factors in the RGEs, as discussed in
Sec.~\ref{sec:RunningAngles}, but this is not the main effect.  What is more
important is the increase of $M_*$. From Eq.~\eqref{eq:ApproxMStar} we find that
it is roughly proportional to $m_1^{-1}$ for a strong hierarchy, so that it
becomes close to or even larger than $M_\mathrm{GUT}$.  Consequently, the RG
effects from $(Y_\nu)_{33}$ become negligible, and we are left with the change
proportional to $y_\tau^2$.  This change is small here, since we are using a
moderate value of $\tan\beta=20$.
\begin{table}
 \centerline{%
  \begin{tabular}{|c|c|c|c|c|c|c|}
  \hline
   Current & Beams & D-CHOOZ & T2K$+$NuMI & Reactor-II & JPARC-HK & NuFact-II
   \\
  \hline
   0.14 & 0.061 & 0.032 & 0.023 & 0.014 & $10^{-3}$ & $6\times 10^{-5}$\\
  \hline
  \end{tabular}
 }
 \caption{Current and expected sensitivities for $\sin^2 2\theta_{13}$
  at the 90\% CL \cite{Huber:2002mx,Huber:2003ak,Huber:2004ug}.  The
  entry ``Beams'' includes the conventional beam experiments MINOS,
  ICARUS and OPERA.  The last entry refers to an advanced stage neutrino
  factory with experiments at two different baselines.  The sensitivity
  of a first stage neutrino factory (``NuFact-I'') is similar to that of
  JPARC-HK.  For a description of the experiments and the assumptions
  used in the analysis, see
  \cite{Huber:2002mx,Huber:2003ak,Huber:2004ug} and references therein.
  The numbers should be treated with some care, since they depend on the
  true values of the other oscillation parameters, in particular
  $\Delta m^2_\mathrm{atm}$.
 }
 \label{tab:SensitivitiesTheta13}
\end{table}
\begin{table}
 \centerline{%
  \begin{tabular}{|c|c|c|c|c|}
  \hline
   Current & Beams & T2K$+$NuMI & JPARC-HK & NuFact-II
   \\
  \hline
   0.16 & 0.1 & 0.050 & 0.020 & 0.055 \\
  \hline
  \end{tabular}
 }
 \caption{Current and expected sensitivities for 
  $|0.5-\sin^2\theta_{23}|$ \cite{Antusch:2004yx}.
  The numbers are the minimal values required to exclude maximal mixing
  at the 90\% CL.  ``Current'' is the current limit from SuperKamiokande
  \cite{Ashie:2004mr}, ``Beams'' means conventional neutrino beams.  See
  \cite{Antusch:2004yx} and references therein
  for a description of the experiments and the analysis methods.  As
  in Tab.~\ref{tab:SensitivitiesTheta13}, the results depend on the
  true values of the other oscillation parameters.
 }
 \label{tab:SensitivitiesTheta23}
\end{table}

\begin{figure}
 \centering
 \subfigure[\label{fig:T23a}Evolution of the atmospheric mixing angle
  $\theta_{23}$]{
 \CenterEps[1]{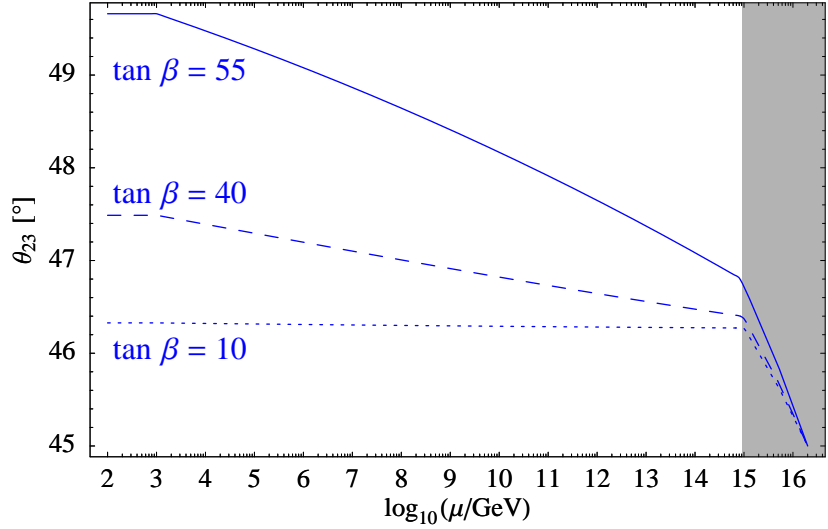}
 }\\
 \subfigure[\label{fig:T23b}Evolution of the atmospheric mass squared
  difference $\Delta m^2_\mathrm{atm}$]{
 \CenterEps[1]{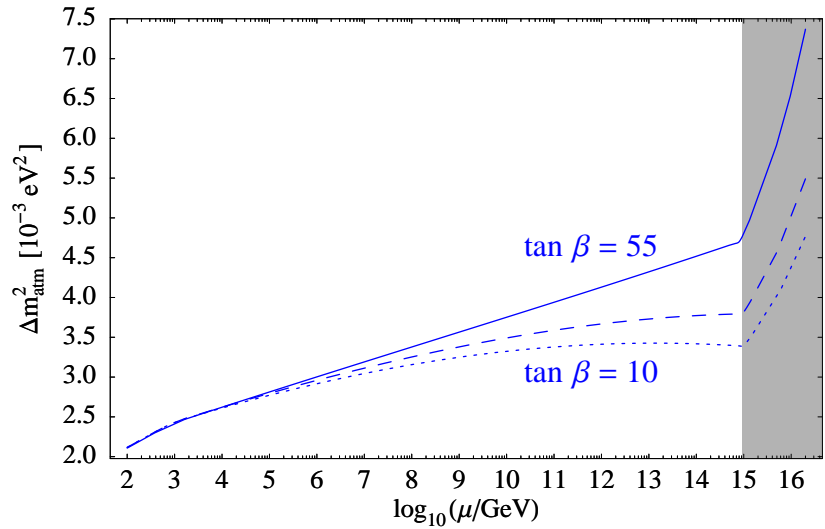}
 }
 \caption{\label{fig:Hierarchical_T23}
 Example for the running of $\theta_{23}$, Fig.~(a), and 
 $\Delta m^2_{\mathrm{atm}}$, Fig.~(b), for a hierarchical neutrino spectrum. 
 The plots show the RG evolution in the MSSM for  
 $\tan\beta = 55$ (solid lines), $40$ (dashed lines) and $10$ (dotted lines) 
 with $\theta_{23}=45^\circ$ at high energy and present best-fit values 
 for the other parameters as constraints at low energy. We have used  
 $(Y_\nu)_{33}=(Y_\nu)_{32}=1$ at the see-saw scale $M_3$ (in the basis where $M$
 and $Y_e$ are diagonal) as an example 
 (note that we use RL-convention for $Y_\nu$). 
 We have furthermore assumed that the right-handed neutrino with mass 
 $M_3$ dominates in the see-saw formulae, as in heavy sequential dominance 
 \cite{King:1998jw,King:1999mb}, which allows to approximately 
 calculate $M_3$ from $m_3$ in the hierarchical scheme.  
 To a good approximation, only one see-saw scale is relevant for the running in
 this case. The gray regions correspond to energies above this scale.  
 The evolution of $\Delta m^2_\mathrm{atm}$ depends quite sensitively on
 the value of the top mass and on the SUSY breaking scale.  We have used
 $m_t(m_t)=175\GeV$ and $M_\mathrm{SUSY}=1\TeV$.
 }
\end{figure}

In order to demonstrate that RG corrections from  $Y_\nu$ are not necessarily
negligible for a strongly hierarchical spectrum, let us consider another
example, where two elements of $Y_\nu$ are large. The evolution of the
atmospheric mixing angle and mass squared difference is shown in
Fig.~\ref{fig:Hierarchical_T23} for $\theta_{23}=\pi/4$ at high energy in the
MSSM with different values of $\tan\beta$ and a strong normal mass hierarchy. 
In this example, we have taken $(Y_\nu)_{33}=(Y_\nu)_{32}=1$ at $M_3$ and
assumed the other entries in $Y_\nu$ to be small in the basis where $M$ and
$Y_e$ are diagonal.  We have furthermore assumed that the right-handed neutrino
with mass $M_3$ dominates in the see-saw formula, as it is the case for  heavy
sequential dominance (HSD) \cite{King:1998jw,King:1999mb}.\footnote{
RG effects in this case have been discussed numerically in \cite{King:2000hk}, 
in agreement with our analytic results.  
}  
This allows to approximately calculate $M_3 \approx v^2
(Y_\nu)_{33}^2 \, m_3^{-1}$ with
$m_3 \approx \sqrt{\Delta m^2_\mathrm{atm}}$ in this case,
and to consider only one see-saw scale $M_*=M_3$ when discussing the running.
Eq.~\eqref{eq:ApproxDeltaT23SHRND} then simplifies to
\begin{equation} \label{eq:DeltaT23SHRNDHier}
  \Delta\theta_{23} \approx
  \frac{1}{32\pi^2} \,
   y_\tau^2\,\ln\Bigl(\frac{M_\mathrm{GUT}}{M_\mathrm{SUSY}}\Bigr)
   \left( 1 + 2 \sqrt{\zeta} c_{12}^2 \cos\varphi_2 \right) +
  \frac{1}{16\pi^2} \,
   \ln\Bigl(\frac{M_\mathrm{GUT}\sqrt{\Delta m^2_\mathrm{atm}}}{v^2}\Bigr)
  \;.
\end{equation}
The resulting change of $\theta_{23}$ is in the range of about
$[1^\circ,5^\circ]$.  Thus, even with a strong normal mass hierarchy,
the change of the mixing angles can be within the sensitivity of future long
baseline experiments.  The phase $\varphi_1$ is irrelevant due to
$m_1=0$, and $\varphi_2$ cannot cause a significant damping as it
appears together with the rather small quantity $\sqrt{\zeta}$.  In
Fig.~\ref{fig:Hierarchical_T23}, it has been set to 0.

As argued in Sec.~\ref{sec:ApproxFormulaeSHRND}, the running of $U_e$
(the second term in Eq.~\eqref{eq:DeltaT23SHRNDHier}) cannot be
neglected in this example, because the $U_\nu$ contribution is strongly
suppressed due to the cancellation between the terms proportional to
$P_{22}$ and $P_{33}$ and the vanishing of the term proportional to
$P_{23}$ for maximal atmospheric mixing and real $Y_\nu$.  Even without
cancellations, both contributions are generically of the same order of
magnitude for hierarchical neutrino masses.  Another lesson that can be
learned from this example is that a complete cancellation of the running
is very unlikely.  Hence, we always expect RG effects to be comparable
to the sensitivity of planned precision experiments if there are large
Yukawa couplings and if $Y_\nu$ and $Y_e$ are not simultaneously
diagonal.

\subsection{Large RG Effects Despite Phases}\label{sec:RunningDespitePhases}
\begin{figure}
 \centering
 \includegraphics{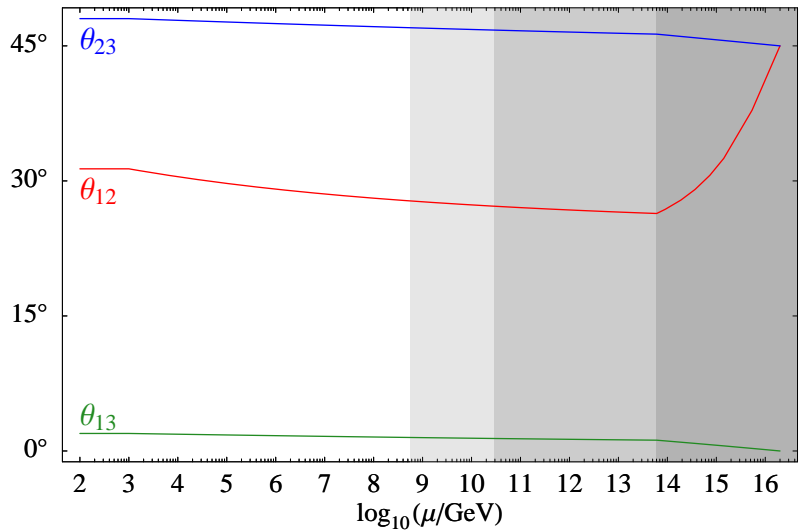}
 \caption{Fast running of the solar angle despite large Majorana phases
  $\varphi_1=\pi/2$, $\varphi_2=0$ in the MSSM with $\tan\beta=30$,
  $M_\mathrm{SUSY}=1\TeV$ and a
  normal mass hierarchy.  The evolution is dominated by the large
  imaginary part of $P_{31}$, see Eq.~\eqref{eq:YnuRunningDespitePhases}.
  Further initial conditions at the GUT scale 
  $M_\mathrm{GUT}=2\cdot10^{16}\GeV$ were bimaximal mixing,
  $m_1=0.08\eV$, $\Delta m^2_\mathrm{sol}=1.2\cdot10^{-4}\eV^2$, and
  $\Delta m^2_\mathrm{atm}=4\cdot10^{-3}\eV^2$.
 }
 \label{fig:RunningDespitePhases}
\end{figure}

The main new effect above the see-saw thresholds is the appearance of
off-diagonal terms in the Yukawa couplings.  As large off-diagonal
entries in the Yukawa matrices are postulated in a lot of fermion mass
models in order to explain the large lepton mixing angles, we expect an
important impact on the running in many cases.  As mentioned in
Sec.~\ref{sec:RunningAngles}, the effect of large imaginary entries in
$P$ is especially unusual, since their coefficients in the RGEs of the
mixing angles $\theta_{12}$ and $\theta_{23}$ vanish for zero Majorana
phases and become maximal if the
phases or their difference equal $\pi/2$.  Thus, a fast running is now
also possible for large Majorana phases.  A numerical example with
\begin{equation} \label{eq:YnuRunningDespitePhases}
    Y_\nu(M_\mathrm{GUT}) =
    \begin{pmatrix} 0.001 & 0 & 0 \\ 0 & 0.01 & 0 \\ -0.4\I & 0 & 0.5
    \end{pmatrix} \Rightarrow
    Y_\nu^\dagger Y_\nu(M_\mathrm{GUT}) =
    \begin{pmatrix} 0.16 & 0 & 0.2\I \\ 0 & 0.0001 & 0 \\ -0.2\I & 0 & 0.25
    \end{pmatrix} ,
\end{equation}
i.e.\ a large and purely imaginary $P_{31}$ (as usual given in the basis
where $Y_e$ is diagonal and all unphysical phases are zero) is shown
in Fig.~\ref{fig:RunningDespitePhases}.  We used the MSSM with
$\tan\beta=30$, $M_\mathrm{SUSY}=1\TeV$, a normal hierarchy,
$m_1=0.08\eV$, $\Delta m^2_\mathrm{sol}=1.2\cdot10^{-4}\eV^2$,
$\Delta m^2_\mathrm{atm}=4\cdot10^{-3}\eV^2$, $\varphi_1=\pi/2$, 
$\varphi_2=0$ and bimaximal mixing at the GUT scale 
$M_\mathrm{GUT}=2\cdot10^{16}\GeV$.
Reasonable values for the low-energy oscillation parameters are reached,
and $\Delta m^2_\mathrm{sol}$ stays positive.  The running of the solar
angle from maximal mixing to smaller values is caused by the term
proportional to $\im P_{31}$ in the RGE.  A negative value of $\im
P_{31}$ is required for $\Dot\theta_{12}>0$ (cf.\
Tab.~\ref{tab:RGcorrections4angles}), which is necessary to avoid
running to the ``dark side'' of the solar oscillation parameters
(corresponding to $\Delta m^2_\mathrm{sol} < 0$ with our conventions).
Alternatively, one could choose $\im P_{31} > 0$ and exchange the
initial phases, i.e.\ $\varphi_1=0$, $\varphi_2=\pi/2$.  The terms
proportional to the diagonal elements $P_{11}$ and $P_{33}$ do not play
a significant role here, since they have opposite signs and therefore
cancel approximately.  The example demonstrates that for sufficiently
large off-diagonal entries in $Y_\nu^\dagger Y_\nu$, it is possible to
avoid the requirement of an inverse hierarchy of the neutrino Yukawa
couplings which was found for diagonal $Y_\nu^\dagger Y_\nu$
\cite{Antusch:2002hy,Miura:2003if,Shindou:2004tv}.

\begin{figure}
 \centering
 \includegraphics{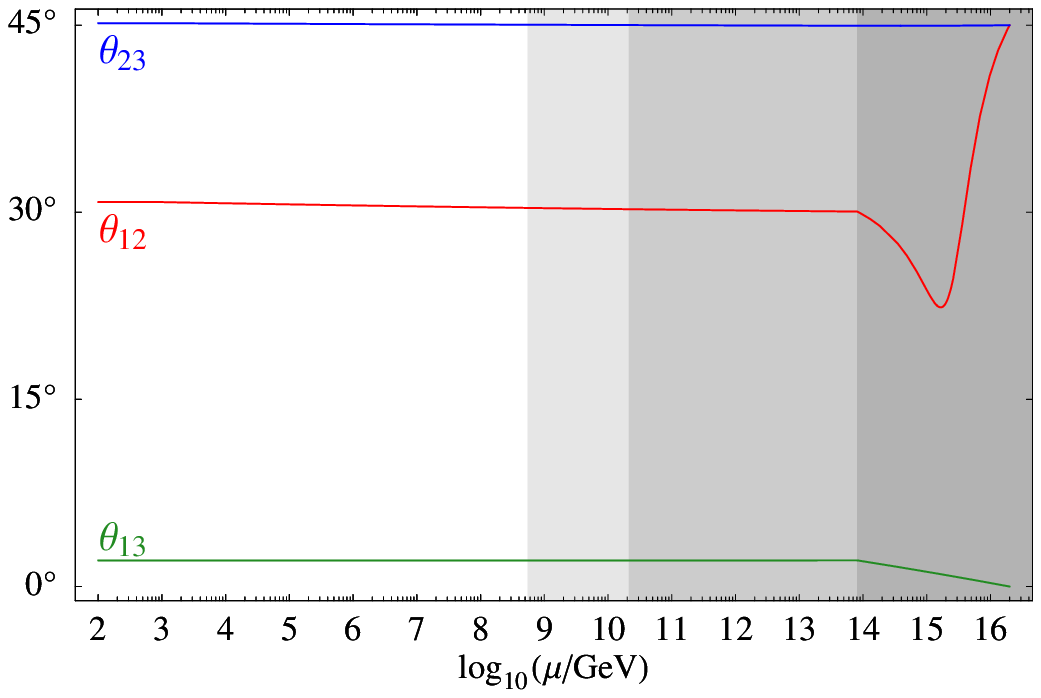}\\[2mm]
 \includegraphics{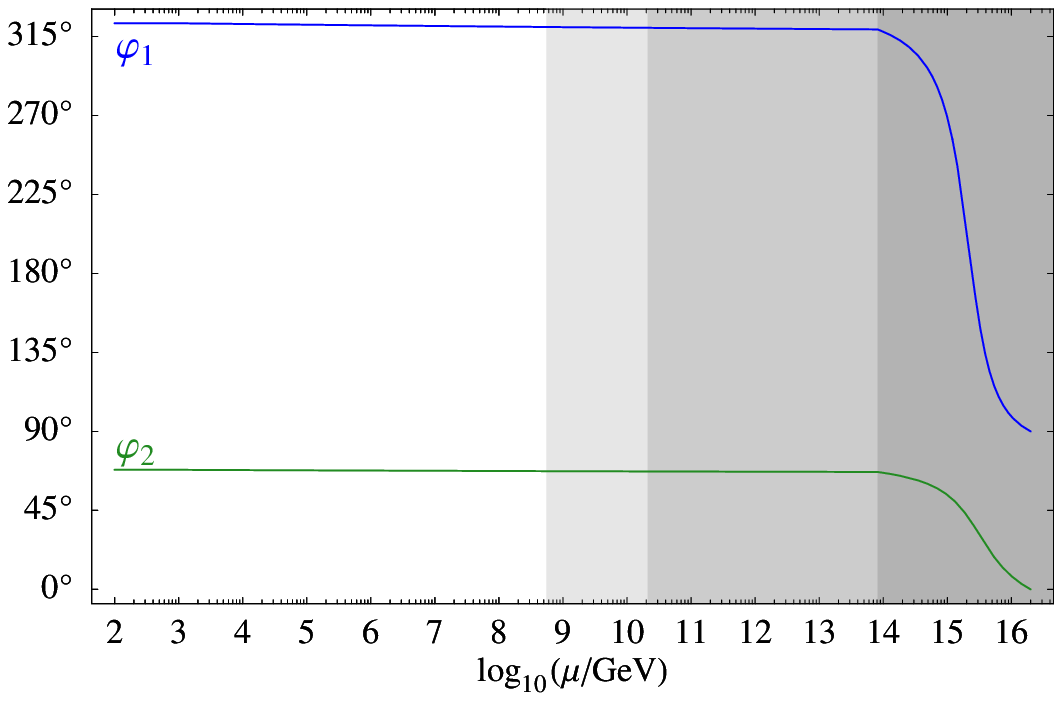}
 \caption{
  Highly non-linear running of $\theta_{12}$ and the Majorana phases in
  an example with large imaginary entries in the neutrino Yukawa matrix
  (see Eq.~\eqref{eq:YnuCrazyRunning}).  We used the MSSM with
  $\tan\beta=10$, $M_\mathrm{SUSY}=1\TeV$ and the following
  initial conditions at $M_\mathrm{GUT}=2\cdot10^{16}\GeV$:
  $\theta_{12}=\theta_{23}=\pi/4$, $\theta_{13}=0$,
  $\varphi_1=\pi/2$, $\varphi_2=0$,
  normal hierarchy, $m_1=0.08\eV$,
  $\Delta m^2_\mathrm{sol} = 1.1\cdot10^{-4}\eV^2$,
  $\Delta m^2_\mathrm{atm} = 4\cdot10^{-3}\eV^2$.
 }
 \label{fig:CrazyRunning}
\end{figure}

Adding another large imaginary entry in the 32-element,
\begin{equation} \label{eq:YnuCrazyRunning}
    Y_\nu(M_\mathrm{GUT}) =
    \begin{pmatrix} 0.001 & 0 & 0 \\ 0 & 0.01 & 0 \\
     -0.4\I & -0.5\I & 0.5
    \end{pmatrix} \Rightarrow
    Y_\nu^\dagger Y_\nu(M_\mathrm{GUT}) =
    \begin{pmatrix} 0.16 & 0.2 & 0.2\I \\ 0.2 & 0.25 & 0.25\I \\
     -0.2\I & -0.25\I & 0.25
    \end{pmatrix} ,
\end{equation}
yields a rather extreme behavior of $\theta_{12}$, as shown in
Fig.~\ref{fig:CrazyRunning}.
The highest see-saw scale lies at about $8\cdot10^{13}\GeV$ here, i.e.\ 
the turnaround in the running is not a threshold
effect.  Instead, it is due to the evolution of the Majorana phases,
c.f.\ the lower plot in Fig.~\ref{fig:CrazyRunning}.
Their difference initially equals $\pi/2$ but quickly starts to increase
as soon as $\theta_{12}$ has moved away from $\pi/4$.  The evolution is
dominated by the term proportional to $\im P_{31}$, which is largest for
$\varphi_1-\varphi_2=\pi$.  At this point, $\sin(\varphi_1-\varphi_2)$
changes its sign, causing a sign change in the contributions of the
imaginary parts of the off-diagonal Yukawa couplings to the RGE for
$\theta_{12}$.  This explains the minimum in the evolution of this
angle.  At lower energies, the difference of the Majorana phases reaches
a value of about $4.4$ and remains approximately constant afterwards.%
\footnote{This happens even if the heaviest singlet neutrino is not
 integrated out, i.e.\ even if the large Yukawa couplings are not
 removed from the theory.
}
From Tab.~\ref{tab:RGcorrections4MajPhases}, one would expect this value
to be closer to $2\pi$.  The difference is due to the subleading
contributions to the running (the terms proportional to
$\sin\theta_{13}$ and the charged lepton contribution), which become
relevant here because of the strong damping of the leading terms.

\subsection{Leptogenesis and RG Corrections}

Leptogenesis \cite{Fukugita:1986hr} is an attractive explanation of the 
observed baryon-to-photon ratio $n_\mathrm{B} /n_\gamma  = (6.5^{+0.4}_{-0.3})
\cdot 10^{-10}$ \cite{Spergel:2003cb}.  It typically operates at the mass scale
of the lightest right-handed neutrino.  In such a scenario, we have to deal with
three scales: the GUT scale where the predictions for the model parameters are
fixed, the scale of leptogenesis where the parameters have to be right for
successful baryogenesis, and the low scale at which the parameters can be
measured in experiments. In particular, one cannot use GUT scale parameters or
experimental results directly in order to test the viability of leptogenesis in
a given model, rather one has to take into account quantum corrections. In the
energy range between the leptogenesis scale $M_1$ and the electroweak scale
$M_\mathrm{EW}$, we can consider the running of the effective neutrino mass
operator.  For relating the see-saw parameters at the GUT scale with the ones at
$M_1$, the evolution above and between the see-saw scales has to be considered.

\subsubsection{Corrections to Decay Asymmetries and to the Neutrino Mass
Bound}
The decay asymmetry for leptogenesis in the SM \cite{Covi:1996wh}
can be written as 
\begin{eqnarray}\label{eq:DecayAs}
\varepsilon_1 \approx 
\frac{3}{8 \pi} \frac{M_1}{v^2} 
\frac{1}{(Y_\nu Y_\nu^\dagger)_{11}} 
\sum_{f,g}\mbox{Im}\, [(Y_\nu)_{1f} (Y_\nu)_{1g} 
 (m^*_{\nu})_{fg}]\;,
\end{eqnarray}
if $M_1 \ll M_2,M_3$. In the MSSM, it is a factor of 2 larger. In the case of
a type II see-saw and for $M_1 \ll M_\Delta$,  where $M_\Delta$ is the mass of
the SU(2)$_\mathrm{L}$-triplet Higgs, the decay asymmetry for type II
leptogenesis via the lightest right-handed neutrino coincides with the
result for the conventional see-saw \cite{Antusch:2004xy}. 
In the SM or for a moderate $\tan\beta$ in the MSSM, the RG running  from
$M_\mathrm{EW}$ to $M_1$ leads mainly to a scaling of the neutrino mass matrix
$m_\nu$.  Including the RG effects results in an  enhancement of the decay 
asymmetry for leptogenesis by roughly 20\% in the MSSM and 30\% -- 50\% in the
SM \cite{Barbieri:1999ma,Antusch:2003kp}. The decay asymmetry can be calculated
by the \package{REAP} package described in Sec.~\ref{sec:MathematicaPackages} as
a function of energy.  Thus, one can easily check if a particular high-energy
model for fermion masses is able to produce a large enough asymmetry. Let us
remark that also the running of the mixing angles can be very important for the
calculation of the baryon asymmetry, as has been shown recently for non-thermal
leptogenesis models \cite{Senoguz:2003hc}.

The requirement of successful baryogenesis via thermal leptogenesis imposes
constraints on fermion mass models and even places an upper bound on the mass of
the light neutrinos \cite{Buchmuller:2003gz}.  With respect to quantum
corrections to this mass bound, it turns out that there are two effects
operating in opposite directions,  which partially cancel each other
\cite{Antusch:2003kp,Giudice:2003jh,Buchmuller:2004nz}:  on the one hand, the
increase of the mass scale leads to a larger decay asymmetry compared to the one
at low energies.  On the other hand, it results in a stronger washout driven by
Yukawa couplings. Taking into account these effects and further corrections, one
finds that the upper bound on the neutrino mass scale becomes more restrictive.

\subsubsection{Models for Resonant Leptogenesis and RG Corrections}
As an example where the running above the lowest see-saw scale can have 
large effects,
we consider the RG corrections to the small mass splitting 
$\Delta M = |M_1 -M_2|$ for resonant leptogenesis 
\cite{Covi:1996wh,Flanz:1996fb,Pilaftsis:1997jf,Pilaftsis:2003gt}. Here,
the decay asymmetry is enhanced compared to Eq.~(\ref{eq:DecayAs}). 
For resonance effects in the decay asymmetries to be maximal, 
a mass splitting of $\tfrac{1}{2}$ times one of the decay 
widths (in the MSSM) 
\begin{eqnarray}
\Gamma_1 \approx \frac{M_0}{8 \pi} (Y_\nu Y^\dagger_\nu)_{11} \; , \quad 
\Gamma_2 \approx \frac{M_0}{8 \pi} (Y_\nu Y^\dagger_\nu)_{22} \;,
\end{eqnarray}
with $M_2 \approx M_1 := M_0$, is required.  Given a model for neutrino masses
with such a small mass splitting defined at $M_\mathrm{GUT}$, the decay rate can
be affected significantly by the RG evolution of the mass matrix of the heavy
right-handed neutrinos from $M_\mathrm{GUT}$ to $M_1 \approx M_2$. 
Resonant leptogenesis with exactly degenerate heavy singlets at 
$M_\mathrm{GUT}$ has been discussed, e.g., 
in \cite{GonzalezFelipe:2003fi,Turzynski:2004xy,Hambye:2004jf}.
The running of $M$ and $Y_\nu$ between $M_\mathrm{GUT}$ and $M_1$, taking
into account the effects between the see-saw thresholds, can be computed
conveniently using the software packages presented in 
Sec.~\ref{sec:MathematicaPackages}.

\section{Alternative Scenarios}\label{sec:AlternativeScenarios}
For the examples in Sec.~\ref{sec:Applications}, 
we have focused on the conventional see-saw
mechanism in the SM and in the MSSM. We now give a brief outlook on other
scenarios. Some of them are already implemented in the software packages
\package{REAP}/\package{MPT} introduced in Sec.~\ref{sec:MathematicaPackages}.

\subsection{Type II See-Saw}\label{sec:typeII}
A generalization of the conventional see-saw is the type II see-saw 
\cite{Magg:1980ut,Lazarides:1980nt,Mohapatra:1981yp}, 
where an additional contribution to the 
neutrino mass matrix, e.g.\ from an induced vev of a
$\mathrm{SU}(2)_\mathrm{L}$-triplet Higgs, is present. 
Below the additional see-saw scale given by the mass $M_\Delta$ of the triplet, 
it can be integrated out, only leaving an additional
contribution to the effective neutrino mass operator. 
The packages \package{REAP}/\package{MPT} and the analytic formulae 
for the running of the neutrino parameters 
can thus be applied for
analyzing type II see-saw scenarios below $M_\Delta$.  
Above $M_\Delta$, the RGEs are modified due to the additional interactions.

\subsection{Dirac Neutrinos} 

At present it is not known whether the nature of neutrino masses is Dirac or
Majorana. The RG evolution of Dirac neutrino masses is studied in \cite{LRS}.
The packages \package{REAP}/\package{MPT} can also be used in this case.

\subsection{Two Higgs Models} \label{sec:2HDMs}
We restrict our discussion to a class of 2HDMs where flavour changing neutral
currents (FCNCs) are naturally absent
\cite{Weinberg:1976hu,Glashow:1977nt,Paschos:1977ay}.
The Yukawa couplings of the theory are given by
\begin{eqnarray}
  \mathscr{L}_\mathrm{Yukawa}^\mathrm{2HDM}
  & = &
  -\sum\limits_{i=1}^2\left\{
   z_e^{(i)}\, \overline{e_\mathrm{R}} Y_e^{(i)}
        \ell_\mathrm{L} \phi^{(i)\dagger}
   +z_\nu^{(i)}\, \overline{N_\mathrm{R}} Y_\nu^{(i)}
   \ell_\mathrm{L} \widetilde\phi^{(i)\dagger}
  \right.\nonumber\\
  & &\left.
  {}\hphantom{-\sum\limits_{i=1}^2\left\{\right.}
  +z_d^{(i)}\, \overline{d_\mathrm{R}} Y_d^{(i)}
   Q_\mathrm{L} \phi^{(i)\dagger}
  +z_u^{(i)}\, \overline{u_\mathrm{R}} Y_u^{(i)}
        Q_\mathrm{L} \widetilde\phi^{(i)\dagger}
  \right\}
  +\text{h.c.} \;,
        \label{eq:2HDMYukawaCouplings}
\end{eqnarray}
where either $z_f^{(1)}$ or $z_f^{(2)}$ has to be zero for each
$f\in\{e,\nu,d,u\}$ in order to ensure the absence of FCNCs. In order to
generate masses via Yukawa couplings, $z_f^{(1)}=1$ for $z_f^{(2)}=0$ and vice
versa.  By convention, the right-handed charged leptons always couple to
the first Higgs, i.e.\ $z_e^{(1)}=1, z_e^{(2)}=0$.

It is known that in these kind of models there are (at least)
two effective neutrino mass operators. Furthermore, RG
effects are comparatively large, since one has both the $\tan\beta$ enhancement as
well as the absence of cancellations due to the SUSY non-renormalization
theorem.
An analytic understanding of the RG effects is more difficult to
obtain, since the two components of the effective neutrino mass matrix
\begin{equation}
 m_\nu
 \,=\,
 -\frac{v_1^2}{4}\kappa^{(11)}-\frac{v_2^2}{4}\kappa^{(22)}
\end{equation}
run differently. Here, more investigations are needed, which are beyond the
scope of this study. With the \package{REAP} package, an
extensive numerical analysis is possible.
Recently, the RGEs in multi-Higgs models have been derived
\cite{Grimus:2004yh}. The structure of the $\beta$-functions is very similar.

\subsection{Split SUSY}
Let us note that the RGEs for the effective neutrino mass operator in the SM
describe the running in the framework of `split supersymmetry'
\cite{Arkani-Hamed:2004fb,Giudice:2004tc} as well, except for a contribution to the
flavour-trivial part of the RGE (cf.\ App.~\ref{app:RGEsSplitSUSY}). This
implies in particular that running effects for the mixing angles are suppressed
compared to the MSSM (with not too small $\tan\beta$).  The negative $g_2^2$
contribution to the flavour-trivial part of the RGE gets replaced by 
a positive $g_1^1$ contribution. This effect increases
the running of the mass eigenvalues.

\subsection{Other Alternative Sources of Neutrino Masses}

If the dimension 5 neutrino mass operator does not give the leading 
contribution, possible alternative sources to the light neutrino  masses can
have interesting consequences. Neutrino masses can e.g.\   emerge from the
K{\"a}hler potential in supersymmetric theories.   It has been observed that in
this case, large mixing angles can be an infrared fixed point  of the
renormalization group \cite{Casas:2002sn,Casas:2003kh}.   
In the SM, effects of additional dimension 6 operators on the running of the
dimension 5 neutrino mass operator have been considered in
\cite{Broncano:2004tz}.

\section{Discussion and Conclusions}
\label{sec:Conclusions}

We have discussed the running of neutrino masses and leptonic mixing parameters
in see-saw models involving singlet neutrinos.  At energies above the masses of
these heavy particles, their Yukawa couplings to the left-handed leptons play an
important role.  As they may be of order 1, they can cause significant quantum
corrections.  We have derived approximate renormalization group equations (RGEs)
for the mixing angles, CP phases and mass eigenvalues.  Due to the large number
of parameters in the see-saw scenario, the details of the running strongly
depend on the specific model under consideration.  One is still able to obtain
an extensive analytic understanding of the RG effects. It is instructive to
compare the RGEs of the physical mixing parameters
$\{\psi_\ell\}=\{\theta_{12},\theta_{32},\theta_{23},\delta,\varphi_1,\varphi_2\}$
above the see-saw scales, 
\begin{equation}\label{eq:RGESeeSaw}
 \mu\frac{\D}{\D\mu}\psi_\ell~=~
 \frac{f_\ell(m_k,\text{phases})}{m_i^2-m_j^2}
 \times F_\ell^{(\nu)}(Y_\nu,Y_e,\{\psi_\ell\})
 +F_\ell^{(e)}(Y_\nu,Y_e,\{\psi_\ell\})
\end{equation}
to those describing the evolution below the see-saw scales. The latter are
obtained by replacing $F_\ell^{(\nu)}(Y_\nu,Y_e,\{\psi_\ell\})$ by
$F_\ell(Y_e,\{\psi_\ell\})$ and $F_\ell^{(e)}$ by zero in
Eq.~\eqref{eq:RGESeeSaw}. Most importantly, the structure of the RGEs of the
mixing parameters is the same above and below the see-saw scales. Hence, there
are features common to the evolution above and below. For a degenerate spectrum,
the first mass quotient in \eqref{eq:RGESeeSaw} becomes large, yielding strong
RG effects. There are, however, important differences as well. First, the dimensionless
function $F_\ell(Y_e,\{\psi_\ell\})$ vanishes for zero mixing, which is not the
case for $F_\ell^{(\nu)}(Y_\nu,Y_e,\{\psi_\ell\})$. Zero mixing angles are hence
not stable under the RG in the full see-saw framework. Second, in the SM or the
MSSM with small $\tan\beta$, RG effects are small below the see-saw scales. In
contrast, above the entries of $Y_\nu$ can be of order one and cause important
running effects. Third, the RGE contains the
$F_\ell^{(e)}(Y_\nu,Y_e,\{\psi_\ell\})$ term, which describes the radiative
rotation of $Y_e^\dagger Y_e$ in the presence of neutrino Yukawa couplings $Y_\nu$.
Finally, between the thresholds, there are important effects in
non-supersymmetric theories which stem from the different scaling of different
parts of the effective neutrino mass matrix.

We listed the leading order RG coefficients for the mixing parameter RGEs in
extensive tables. Our results allow to obtain a qualitative understanding of
generic effects such as the influence of the CP phases and that of the absolute
neutrino mass scale.  For example, non-zero phases often damp the running, but
some terms in the RGEs are actually enhanced by them.  A rough quantitative
estimate of the size of the RG effects is possible as well. Although the change
of the mixing angles is quite small for strongly hierarchical masses (in the
case of a normal hierarchy), it turns out that often it is still comparable to
the sensitivities of planned oscillation experiments.  Therefore, quantum
corrections should not be neglected in any study of fermion mass models if one
aims at theoretical predictions whose precision matches that of the
experiments.  The neutrino mass eigenvalues always change significantly due to
the RG evolution.  This means that a model predicting precisely the measured
value of $\Delta m^2_\mathrm{atm}=2.1\cdot10^{-3}\,\mathrm{eV}^2$ at the GUT
scale would actually be excluded by several standard deviations. Another
consequence is a correction to the mass bound from thermal leptogenesis. 
Furthermore, the running of the masses of the singlet neutrinos is important for
models of resonant leptogenesis.

In order to obtain precise quantitative results, the complete system of coupled
RGEs has to be solved.  Therefore, one has to resort to numerical calculations. 
For this purpose, we have developed a set of Mathematica packages, which are
available at the web page \url{http://www.ph.tum.de/~rge/}. The package
\package{REAP} solves the RGEs and thus provides the neutrino mass matrix as
well as the other parameters such as Yukawa couplings at each energy.  In models
with heavy singlet neutrinos, they are integrated out automatically at the
corresponding mass thresholds. Thus, the effects of non-degenerate singlet
masses, which are generally sizable, are correctly taken into account.   From the results of \package{REAP},
\package{MixingParameterTools} allows to extract the values of the mixing
angles, phases and mass eigenvalues.

\subsubsection*{Acknowledgements}
We would like to thank W.~Buchm\"uller, K.~Hamaguchi, F.R.~Joaquim,
S.F.~King, M.~Pl\"u\-ma\-cher, P.~Ramond, K.~Turzy\'nski and S.~West for
useful discussions.
One of us (M.R.) would like to thank the Aspen Center for Physics for support.
This work was partially supported by the EU 6th Framework Program
MRTN-CT-2004-503369 ``Quest for Unification'', MRTN-CT-2004-005104
``ForcesUniverse'', by the PPARC grant PPA/G/O/2002/ 00468,
by the ``Impuls- und Vernetzungsfonds'' of the 
Helmholtz Association, contract number VH-NG-006, and by the
``Sonderforschungsbereich~375 f\"ur Astro-{}Teil\-chen\-phy\-sik der 
Deutschen Forschungsgemeinschaft''.

\subsubsection*{Note added}
We have been made aware that F.~R.~Joaquim is finalizing a work on similar
issues.

\newpage
\section*{Appendix}
\appendix

\renewcommand{\thesection}{\Alph{section}}
\renewcommand{\thesubsection}{\Alph{section}.\arabic{subsection}}
\def\theequation{\Alph{section}.\arabic{equation}}
\renewcommand{\thetable}{\arabic{table}}
\renewcommand{\thefigure}{\arabic{figure}}
\setcounter{section}{0}
\setcounter{equation}{0}

\section{Conventions for Mixing Parameters and Experimental Data}

\subsection{Conventions}
\label{subsec:Conventions4MixingParameters}

Here, we describe our conventions concerning mixing angles and  phases. For a
general unitary matrix we choose the so-called  standard-parametrization
\begin{eqnarray}\label{eq:StandardParametrizationU}
 U & = &\diag(e^{\I\delta_{e}},e^{\I\delta_{\mu}},e^{\I\delta_{\tau}}) \cdot V \cdot 
 \diag(e^{-\I\varphi_1/2},e^{-\I\varphi_2/2},1)
\end{eqnarray}
where 
\begin{equation}
 V=\left(
 \begin{array}{ccc}
 c_{12}c_{13} & s_{12}c_{13} & s_{13}e^{-\I\delta}\\
 -c_{23}s_{12}-s_{23}s_{13}c_{12}e^{\I\delta} &
 c_{23}c_{12}-s_{23}s_{13}s_{12}e^{\I\delta} & s_{23}c_{13}\\
 s_{23}s_{12}-c_{23}s_{13}c_{12}e^{\I\delta} &
 -s_{23}c_{12}-c_{23}s_{13}s_{12}e^{\I\delta} & c_{23}c_{13}
 \end{array}
 \right)
\end{equation}
with $c_{ij}$ and $s_{ij}$ defined as $\cos\theta_{ij}$ and
$\sin\theta_{ij}$, respectively. 

The MNS mixing matrix $U_\mathrm{MNS}$ is defined to diagonalize the effective
neutrino mass matrix $m_\nu$ in the basis where 
$Y_e^\dagger Y_e=\diag(y_e^2,y_\mu^2,y_\tau^2)$,
\begin{equation}
 U_\mathrm{MNS}^T\,m_\nu\,U_\mathrm{MNS}
 \,=\,
 \diag\big(m_1,m_2,m_3\big)\;.
\end{equation}
The mass eigenvalues $m_i$ are positive, and $m_1<m_2<m_3$ for a normal
hierarchy or $m_3<m_1<m_2$ for an inverted hierarchy, respectively. For our
conventions for extracting the mixing parameters from the MNS matrix, we would
like to refer the reader to Ref.~\cite{Antusch:2003kp} and the documentation of the
\texttt{MixingParameterTools} package associated with this study.

\subsection{Experimental Data}
An overview over the best-fit values and allowed ranges for the neutrino
oscillation parameters resulting from a global fit to the experimental
data \cite{Maltoni:2004ei} is given in Tab.~\ref{tab:ExpResults}.
\begin{table}[bhtp]
\centering
\begin{tabular}{|c|c|c|}
\hline
Parameter & Best-fit value & $3\sigma$ range \\
\hline
$\theta_{12}$ & $33.2^\circ$ & $28.7^\circ \;..\; 38.1^\circ$ \\
$\theta_{23}$ & $45.0^\circ$ & $35.7^\circ \;..\; 55.6^\circ$ \\
$\theta_{13}$ & $0^\circ$ & $0^\circ \;..\; 13.1^\circ$ \\
$\Delta m^2_\mathrm{sol}$ & $7.9 \cdot 10^{-5}\eV^2$ &
 $(7.1 \;..\; 8.9) \cdot 10^{-5}\eV^2$ \\
$|\Delta m^2_\mathrm{atm}|$ & $2.1 \cdot 10^{-3}\eV^2$ &
 $(1.3 \;..\; 3.2) \cdot 10^{-3}\eV^2$ \\
\hline
\end{tabular}
\caption{Overview of experimental results for neutrino oscillation
 parameters \cite{Maltoni:2004ei}.}
\label{tab:ExpResults}
\end{table}

\section{Derivation of the Analytic Formulae}
\label{app:AnalyticFormulae}

This appendix contains a couple of technical details relevant for the derivation
of the analytic formulae discussed in the main part. Our derivation is based on
earlier works
\cite{Babu:1987im,Grzadkowski:1987tf,Casas:1999tg},
but differs from them by a few steps allowing to express the running of the
mixing parameters by the mixing parameters themselves rather than mixing matrix
elements \cite{Antusch:2003kp} (see also \cite{Chankowski:2001mx} for real
couplings).

\subsection{General Strategy} \label{app:GeneralStrategy}

In an arbitrary basis, one can define unitary matrices $U_\nu$ and $U_e$ by
\begin{subequations}
\begin{eqnarray}
 U_\nu(t)^T\,m_\nu(t)\,U_\nu(t) &=&
 \diag\bigl(m_1(t),m_2(t),m_3(t)\bigr)\;,
\\
 U_e(t)^\dagger\,Y_e^\dagger Y_e(t)\,U_e(t) &=&
 \diag\bigl(y_e^2(t),y_\mu^2(t),y_\tau^2(t)\bigr) \;,
\label{eq:DefUe}
\end{eqnarray}
\end{subequations}
where $m_\nu$ is the effective light neutrino mass matrix of
Eq.~\eqref{eq:mnuFullTheory}. The MNS matrix is then given by
\begin{equation} \label{eq:DefUMNS}
        U_\mathrm{MNS}(t) \,=\, U_e^\dagger(t) \, U_\nu(t) \;.
\end{equation}
For convenience, we choose to work in a basis,  called reference basis in the
following, where 
\begin{equation}
 Y_e^\dagger Y_e(t_0)
 \,=\,
 \diag\big(y_e^2(t_0),y_\mu^2(t_0),y_\tau^2(t_0)\big)\;.
\end{equation}
Obviously, $U_e(t_0)=\mathbbm{1}$ and $U_\mathrm{MNS}(t_0)=U_\nu(t_0)$.

Let us now consider the changes caused by changing the renormalization scale
according to $t_0\to t_0+\Delta t$ (with $\Delta t$ being small). The RGE
\eqref{eq:Betamnu} for $m_\nu$ induces a change 
\begin{eqnarray}
 m_\nu(t_0+\Delta t)
 & = &
 m_\nu(t_0)
 +\frac{\Delta t}{16\pi^2} \left[
 P(t_0)^T\, m_\nu(t_0) +
  m_\nu(t_0)\,P(t_0) + 
 \bar\alpha(t_0) \, m_\nu(t_0) \right]
 \nonumber\\
 & &{}
 +\mathscr{O}\big((\Delta t)^2\big)
\end{eqnarray}
with $P=(C_e \, Y_e^\dagger Y_e + C_\nu \, Y_\nu^\dagger Y_\nu)$
in the energy region above the highest see-saw scale.  We restrict our
derivation to this region.  As explained in
Sec.~\ref{sec:RunningBetween}, the results for the MSSM can also be
applied between the see-saw scales after replacing $Y_\nu$ by
$\accentset{(n)}{Y_\nu}$.  However, this is not possible in the SM.
Due to the change of $m_\nu$,
\begin{equation} \label{eq:ChangeUnu}
 U_\nu(t_0+\Delta t)
 \,=\,
 U_\nu(t_0)+\Delta t\,U_\nu(t_0)\,T +
 \mathscr{O}\big((\Delta t)^2\big)\;,
\end{equation}
where $T$ is to be calculated below.  This relation, however, does not give the
full RG change of $U_\mathrm{MNS}$, since also $Y_e^\dagger Y_e$ gets rotated,
\begin{eqnarray}
 Y_e^\dagger Y_e(t_0+\Delta t)
 & = & 
 Y_e^\dagger Y_e(t_0)
 +\frac{\Delta t}{16\pi^2}\left[ 
                F^\dagger(t_0)\, Y_e^\dagger Y_e(t_0)
        +
                Y_e^\dagger Y_e \,F(t_0)
        +f(t_0) \,Y_e^\dagger Y_e(t_0)\right]
 \nonumber\\
 & & {}
 +\mathscr{O}\big((\Delta t)^2\big) \;,
\end{eqnarray}
where $F=(D_e \, Y_e^\dagger Y_e + D_\nu \, Y_\nu^\dagger Y_\nu)$
and $f=2\re \alpha_e$.
Hence, $U_e(t_0+\Delta t)$ is different from $U_e(t_0)=\mathbbm{1}$ in general,
\begin{equation} \label{eq:ChangeUe}
 U_e(t_0+\Delta t)
 \,=\,
 U_e(t_0)+\Delta t\,U_e(t_0)\,X + \mathscr{O}\big((\Delta t)^2\big)\;,
\end{equation}
with $X$ to be calculated below. 

Using Eq.~\eqref{eq:DefUMNS} together with Eqs.~\eqref{eq:ChangeUnu} and
\eqref{eq:ChangeUe}, we thus get two contributions to the change of the
MNS matrix,
\begin{equation}
 U_\mathrm{MNS}(t_0+\Delta t)
 \,=\,
 U_\mathrm{MNS}(t_0)+
 \Delta t \left[ U_\mathrm{MNS}(t_0)\, T 
 + X^\dagger\,U_\mathrm{MNS}(t_0)\right]+\mathscr{O}\big((\Delta t)^2)\;.
\end{equation}
We call them the $U_\nu$ and the $U_e$ contribution.
Following the analysis of \cite{Antusch:2003kp}, this relation allows to 
derive RGEs for the mixing parameters.

Before going to the actual calculation, we would like to stress that to
derive the mixing parameter RGEs, it is useful to work in the reference basis.
The resulting equations, however, are basis-independent. Of course, if one
changes the basis, one needs to transform $P$ and $F$ accordingly, which
means that the tables in Sec.~\ref{sec:AnalyticalFormulae} and
App.~\ref{app:Tables} are changed as well.

\subsection{RG Corrections Induced by $\boldsymbol{P}$}

This part of the derivation coincides with the one performed in
\cite{Antusch:2003kp} except for the fact that we have to deal with a non-diagonal
$P$. Rather than repeating the analysis of \cite{Antusch:2003kp}, we just 
summarize the results: the evolution of $U_\nu$ is found to be described by
\begin{equation}\label{eq:EvolutionOfUnu}
 U_\nu^\dagger\, \Dot U_\nu = T\;,
\end{equation}
where the entries of $T$ are given by
\begin{subequations} \label{eq:TfromPprime}
\begin{eqnarray}
 16\pi^2\,\im T_{ij}
 & = &
 -\frac{m_i-m_j}{m_i+m_j}\,\im P'_{ij}
 \;,
 \\
 16\pi^2\,\re T_{ij}
 & = &
 -\frac{m_i+m_j}{m_i-m_j}\,\re P'_{ij}
 \;.
\end{eqnarray}
\end{subequations}
$m_i$ denote the eigenvalues of the effective neutrino mass matrix $m_\nu$
(cf.\ App.~\ref{subsec:Conventions4MixingParameters}), and
$P'=U_\nu^\dagger\,P\,U_\nu$.

\subsection{Contribution from the Running of $\boldsymbol{Y_e}$}

Let us now derive the $U_e$ contribution to the RGEs stemming from the fact that
$Y_e^\dagger Y_e$ changes its structure under the RG. To
calculate the corresponding change of the MNS matrix, we only need the running
of the unitary matrix $U_e$ which diagonalizes $Y_e^\dagger Y_e$. 
Using Eq.~\eqref{eq:dYedt}, it is easy to check that 
\begin{eqnarray}
 16\pi^2 \, \frac{\D}{\D t}
 Y_e^\dagger Y_e
 & = & 
 F^\dagger \, Y_e^\dagger Y_e
 + Y_e^\dagger Y_e \, F
 + 2\re\alpha_e \, Y_e^\dagger Y_e \;.
\end{eqnarray}
Plugging this into the inverse of Eq.~\eqref{eq:DefUe},
$Y_e^\dagger Y_e = U_e \, \diag(y_e^2,y_\mu^2,y_\tau^2) \, U_e^\dagger=:
 U_e D \, U_e^\dagger$,
we obtain
\begin{eqnarray}
 \frac{\D}{\D t}(U_e D \, U_e^\dagger)
& = &
 \Dot U_e D \, U_e^\dagger + U_e D \, \Dot U_e^\dagger 
 + U_e \Dot D \, U_e^\dagger
\nonumber\\ 
& = &
 \frac{1}{16\pi^2} \left( F^\dagger U_e D \, U_e^\dagger
 + U_e D \, U_e^\dagger F + 2\re\alpha_e \, U_e D \, U_e^\dagger\right) .          
\end{eqnarray}
Multiplying by $U_e^\dagger$ from the left and by $U_e$ from the right
yields
\begin{equation}
 U_e^\dagger \Dot U_e D + D \, \Dot U_e^\dagger U_e + \Dot D \,=\,
 \frac{1}{16\pi^2}
 \left( F^{\prime\,\dagger} \, D + D \, F' + 2\re\alpha_e \,D \right) ,
\end{equation}
where $F':=U_e^\dagger \, F \, U_e$.
The evolution of $U_e$ can be written as
\begin{equation}\label{eq:EvolutionOfUe}
 \frac{\D}{\D t} U_e \,=\, U_e \, X \;,
\end{equation}
where $X$ is anti-Hermitian.
Inserting this relation and using the anti-Hermiticity yields
\begin{equation}
 \Dot D \,=\,
 \frac{1}{16\pi^2}
 \left( F^{\prime\,\dagger} \, D + D \, F' + 2\re\alpha_e\,D \right)
 - X \, D + D \, X \;.
\end{equation}
By analyzing the off-diagonal parts, we find
\begin{equation}
 y_i^2\,X_{ij} - X_{ij}\,y_j^2
 \,=\,
 -\frac{1}{16\pi^2}\left[(F^{\prime\,\dagger})_{ij}\,y_j^2
 +y_i^2\,F'_{ij}\right] ,
\end{equation}
where $y_1 \equiv y_e$ etc.
For Hermitian $F$, this can be written as
\begin{equation} \label{eq:XfromFprime}
 16\pi^2 \, X_{ij} \,=\, \frac{y_j^2+y_i^2}{y_j^2-y_i^2} \, F_{ij}' \;.
\end{equation}
Due to the strong hierarchy of the charged lepton Yukawa couplings, the $y_i$
dependent factor is approximately $\pm 1$.  The corresponding equations for the
$U_\nu$ contribution, Eqs.~\eqref{eq:TfromPprime}, contain the light neutrino
mass eigenvalues, so that a significant enhancement of $T_{ij}$, the analogon of
$X_{ij}$, occurs for quasi-degenerate neutrino masses.  In this case, we expect
the $U_e$ contribution to give only a small correction, unless severe
cancellations occur in the $U_\nu$ contribution.  However, for a strong normal
neutrino mass hierarchy, both contributions are generically of the same order of
magnitude. The diagonal parts of $X$, which only influence the evolution of
the unphysical phases, remain undetermined.

\subsection{Combination of both Contributions} \label{app:Combination}

Inserting Eqs.~\eqref{eq:EvolutionOfUnu} and \eqref{eq:EvolutionOfUe} into
Eq.~\eqref{eq:DefUMNS}, we find at $t=t_0$ in the reference basis
\begin{equation}\label{eq:EvolutionOfUMNS}
 \frac{\D}{\D t}U_\mathrm{MNS} \,=\,
 U_\mathrm{MNS} \, T + X^\dagger \, U_\mathrm{MNS} 
\end{equation}
or
\begin{equation}
 U_\mathrm{MNS}^\dagger \, \Dot U_\mathrm{MNS}\, =\,
 T-U_\mathrm{MNS}^\dagger \, X \, U_\mathrm{MNS}  =: R_{TX} \;.
\end{equation}
Note that this is a relation for $U_\mathrm{MNS}$ where both $X$ and $T$ 
depend on how we split $U_\mathrm{MNS}$ into $U_e$ and $U_\nu$. Specifically, in
an arbitrary basis we have
\begin{equation}
 U_\mathrm{MNS}^\dagger \, \Dot U_\mathrm{MNS} \,=\,
 U_e\,T\,U_e^\dagger - U_\nu^\dagger\,X\,U_\nu\;.
\end{equation}
As both sides of the last equation are anti-Hermitian,
the derivatives of the mixing parameters are found from the system of
linear equations
\begin{equation}\label{eq:xiTX}
 \sum\limits_k A^{(k)}
 \,\Dot\xi_k + \I\,S^{(k)}
 \,\Dot\xi_k 
 \,=\,
 R_{TX} \;,
\end{equation}
where
$\{\xi_k\} := 
 \{\theta_{12},\theta_{13},\theta_{23},
   \delta,\delta_e,\delta_\mu,\delta_\tau,\varphi_1,\varphi_2\}
$.
The real matrices $A^{(k)}$ and $S^{(k)}$ are antisymmetric and
symmetric, respectively.  Hence, each $A^{(k)}$ has 3 characteristic
elements and each $S^{(k)}$ has 6, so that we can regard
Eq.~\eqref{eq:xiTX} as a system of 9 linear equations,
\begin{equation} \label{eq:LinEqSyst}
 \underbrace{
 \left(\begin{array}{ccc}
        A^{(1)}_{12} & \cdots & A^{(9)}_{12}\\  
        A^{(1)}_{13} & \cdots & A^{(9)}_{13}\\  
        A^{(1)}_{23} & \cdots & A^{(9)}_{23}\\  
        S^{(1)}_{11} & \cdots & S^{(9)}_{11}\\  
        S^{(1)}_{12} & \cdots & S^{(9)}_{12}\\  
        S^{(1)}_{13} & \cdots & S^{(9)}_{13}\\  
        S^{(1)}_{22} & \cdots & S^{(9)}_{22}\\  
        S^{(1)}_{23} & \cdots & S^{(9)}_{23}\\  
        S^{(1)}_{33} & \cdots & S^{(9)}_{33}
 \end{array}\right)
 }_{=:\,B}
 \,
 \underbrace{
 \left(\begin{array}{c}
        \Dot\theta_{12}\\[0.4mm] \Dot\theta_{13}\\[0.4mm] \Dot\theta_{23}\\[0.4mm] 
        \Dot\delta\\[0.4mm] \Dot\delta_e\\[0.4mm] \Dot\delta_\mu\\[0.4mm] 
                \Dot\delta_\tau\\[0.4mm] 
        \Dot\varphi_1\\[0.4mm] \Dot\varphi_2
 \end{array}\right)
 }_{=\,\Dot\xi}
 =
 \underbrace{
 \left(\begin{array}{c}
        \re (R_{TX})_{12}\\[0.6mm]
        \re (R_{TX})_{13}\\[0.6mm]
        \re (R_{TX})_{23}\\[0.6mm]
        \im (R_{TX})_{11}\\[0.6mm]
        \im (R_{TX})_{12}\\[0.6mm]
        \im (R_{TX})_{13}\\[0.6mm]
        \im (R_{TX})_{22}\\[0.6mm]
        \im (R_{TX})_{23}\\[0.6mm]
        \im (R_{TX})_{33}
 \end{array}\right)
 }_{=:\,v}
 \;.
\end{equation}
$v$ can be split into two parts, 
\begin{equation}
 v\,=\,v_{T}+v_{X}\;,
\end{equation}
where $v_T$ is built from $T$ and $v_X$ is built from
$-U_\mathrm{MNS}^\dagger \, X \, U_\mathrm{MNS}$. In particular, each
$\Dot{\xi}_k$, for instance $\Dot\theta_{12}$, is the sum of two
contributions, one from $T$ (i.e.\ from the running of $m_\nu$) and one
from $X$ (i.e.\ from the running of $Y_e$).

\subsection{Comment: `Unphysical' Phases}
\label{app:UnphysPhases}

The RGEs in the full theory contain the entries of $P$.  However, the phases
appearing in the off-diagonal elements of $P$ are not basis-independent, rather
they can be changed by a transformation using the `unphysical' phases
$\delta_e$, $\delta_\mu$ and $\delta_\tau$ only. To see this, let us perform (in
the basis where $Y_e^\dagger Y_e$ is diagonal) a transformation $K$,
\begin{equation}
 \ell_\mathrm{L}\,\xrightarrow{K}\, K\,\ell_\mathrm{L}\;,\qquad
 e_\mathrm{R}\,\xrightarrow{K}\,K\,e_\mathrm{R}\;,
\end{equation}
where $K=\diag(e^{\I\phi_1},e^{\I\phi_2},e^{\I\phi_3})$ is a diagonal phase
matrix. $Y_e^\dagger Y_e$ is invariant under this transformation, yet it changes
the effective neutrino mass matrix according to
\begin{equation}
 m_\nu\,\xrightarrow{K}\,K^*\,m_\nu\,K^\dagger\;.
\end{equation}
Hence, also $U_\mathrm{MNS}$ gets changed under this transformation,
\begin{equation}
 U_\mathrm{MNS}\,\xrightarrow{K}\,K\,U_\mathrm{MNS}\;,
\end{equation}
i.e.\ $K$ affects the phases $\delta_e$, $\delta_\mu$ and $\delta_\tau$ in
the standard parametrization \eqref{eq:StandardParametrizationU}.
Furthermore, it rotates the phases of the off-diagonal entries of $Y_\nu^\dagger
Y_\nu$ as
\begin{equation}
Y_\nu^\dagger Y_\nu\,\xrightarrow{K}\,K\,Y_\nu^\dagger Y_\nu\,K^\dagger\;.
\end{equation}
This shows that one has to specify both the phases $\delta_e$, $\delta_\mu$,
$\delta_\tau$ and the arguments of the off-diagonal entries of $Y_\nu^\dagger
Y_\nu$, as one set of parameters can be traded for the other. In other words,
two theories with equal $P$ but different phases $\delta_f$ are not equivalent.
In the main text, we use the convention
\begin{equation}
 \delta_e\,=\,\delta_\mu\,=\,\delta_\tau\,=\,0\;.
\end{equation}

As a technical comment, we would like to mention that in order to diagonalize
a general neutrino mass matrix $m_\nu$, the parameters $\delta_e$, $\delta_\mu$ and
$\delta_\tau$ are needed. Only after the transformation with
$K=\diag(e^{-\I\delta_e},e^{-\I\delta_\mu},e^{-\I\delta_\tau})$, one can write
the MNS matrix without $\delta_e$, $\delta_\mu$ and $\delta_\tau$. The step of
going to the basis where $\delta_e=\delta_\mu=\delta_\tau=0$ has often not been
mentioned explicitly in the literature. 
In this context, we would like to comment that, of course, $\delta_e$,
$\delta_\mu$ and $\delta_\tau$ are subject to quantum corrections with their
RGEs depending on the physical parameters. $\Dot{\delta}_e$ has a term proportional to
$1/\Delta m^2_\mathrm{sol}$ whereas $\Dot{\delta}_\mu$ and $\Dot{\delta}_\tau$
are both proportional to $1/\Delta m^2_\mathrm{atm}$.\footnote{The
corresponding formulae below the see-saw scales can be obtained from the web page
\url{http://www.physik.tu-muenchen.de/~mratz/AnalyticFormulae/}.
There, the RG
evolution of the $\delta_f$ phases depends on the physical parameters,
but the RGEs of the physical parameters are independent of the $\delta_f$
phases.}

\section{RGE Coefficients} \label{app:Tables}
In the following, we show the RGEs for the lepton mixing parameters
obtained from the derivation discussed above.  We give the first order
of the expansion in the small CHOOZ angle $\theta_{13}$.  We furthermore
use the abbreviation $\zeta$ for the ratio of the mass squared
differences, cf.\ Eq.~\eqref{eq:DefZeta}.

The results are presented in the form of tables which list the
coefficients of
$P_{fg} = (C_e \, Y_e^\dagger Y_e + C_\nu \, Y_\nu^\dagger Y_\nu)_{fg}$ 
in the RGEs.  Thus, if only a single element of $P$ is dominant, the
derivatives of the mixing parameters are found from the corresponding
rows in the tables.  Of course, if several entries of $P_{fg}$ are
relevant, their contributions simply add up.  While the complete RGEs
are basis-independent, the table entries do depend on the choice of the
basis, since $P$ is basis-dependent.  We use the basis where $Y_e$ is
diagonal and where the unphysical phases in the MNS matrix are zero.

\begin{table}[htbp]\centering\setlength{\bigstrutjot}{6pt}
  \begin{tabular}{|l||l|}\hline
  $\mathcal{Q}^\pm_{13}\,=\,\frac{|m_3 \pm m_1 e^{i\varphi_1}|^2}{\Delta
  m^2_\mathrm{atm}\left(1+\zeta\right)}$ &
$\mathcal{S}_{13}\,=\,\frac{m_1 m_3 \sin\varphi_1}{\Delta m^2_\mathrm{atm}\left(1+\zeta\right)}$ $\vphantom{\frac{1}{2}}$\bigstrut\\
  $\mathcal{Q}^\pm_{23}\,=\,\frac{|m_3 \pm m_2 e^{i\varphi_2}|^2}{\Delta
  m^2_\mathrm{atm}}$ &
$\mathcal{S}_{23}\,=\,\frac{m_2 m_3 \sin\varphi_2}{\Delta m^2_\mathrm{atm}}$\bigstrut\\
  $\mathcal{Q}^\pm_{12}\,=\,\frac{|m_2 e^{i\varphi_2} \pm m_1
  e^{i\varphi_1}|^2}{\Delta m^2_\mathrm{sol}}$ &
    $\mathcal{S}_{12}\,=\,\frac{m_1 m_2 \sin(\varphi_1-\varphi_2)}{\Delta
    m^2_\mathrm{sol}}$\bigstrut\\\hline\hline

  $\mathcal{A}^\pm_{13}\,=\,
    \frac{\left(m_1^2+m_3^2\right)\cos\delta\pm 2m_1m_3\cos(\delta-\varphi_1)}{\Delta
    m_\mathrm{atm}^2\left(1+\zeta\right)}$ &
    $\mathcal{B}^\pm_{13}\,=\,
    \frac{\left(m_1^2+m_3^2\right)\sin\delta\pm 2m_1m_3\sin(\delta-\varphi_1)}{\Delta
    m_\mathrm{atm}^2\left(1+\zeta\right)}$ 
 \bigstrut\\
     $\mathcal{A}^\pm_{23}\,=\,\frac{\left(m_2^2+m_3^2\right)\cos\delta\pm 2m_2m_3\cos(\delta-\varphi_2)}{\Delta
    m_\mathrm{atm}^2}$ &
  
    $\mathcal{B}^\pm_{23}\,=\,\frac{\left(m_2^2+m_3^2\right)\sin\delta\pm 2m_2m_3\sin(\delta-\varphi_2)}{\Delta
    m_\mathrm{atm}^2}$ \bigstrut\\\hline\hline
  
\multicolumn{2}{|l|}{$\mathcal{C}_{13}^{12}\,=\,\frac{m_1}{\Delta m_\mathrm{sol}^2
    \left(1+\zeta\right)}\left[\left(1+\zeta\right)m_2\sin\left(\varphi_1-\varphi_2\right)-\zeta
    m_3\sin\left(2\delta-\varphi_1\right)\right]$}\bigstrut\\
\multicolumn{2}{|l|}{    $\mathcal{C}_{13}^{23}\,=\,\frac{m_3}{\Delta m_\mathrm{atm}^2
    \left(1+\zeta\right)}\left[m_1\sin\left(2\delta-\varphi_1\right)+\left(1+\zeta\right)m_2\sin\varphi_2\right]$ }\bigstrut\\
\multicolumn{2}{|l|}{    $\mathcal{C}_{23}^{12}\,=\,\frac{m_2}{\Delta m_\mathrm{sol}^2
    }\left[m_1\sin\left(\varphi_1-\varphi_2\right)-\zeta
    m_3\sin\left(2\delta-\varphi_2\right)\right]$}\bigstrut\\
\multicolumn{2}{|l|}{    $\mathcal{C}_{23}^{13}\,=\,\frac{m_3}{\Delta m_\mathrm{atm}^2
    \left(1+\zeta\right)}\left[m_1\sin\varphi_1+\left(1+\zeta\right)m_2\sin\left(2\delta-\varphi_2\right)\right]$} \bigstrut\\\hline\hline

\multicolumn{2}{|l|}{$\mathcal{D}_1\,=\,\frac{m_3}{\Delta m_\mathrm{atm}^2
    \left(1+\zeta\right)}\left[m_1\cos\left(\delta-\varphi_1\right)-\left(1+\zeta\right)m_2\cos\left(\delta-\varphi_2\right)\right]\sin\delta$}\bigstrut\\
\multicolumn{2}{|l|}{$\mathcal{D}_2\,=\,\frac{m_3}{\Delta m_\mathrm{atm}^2
    \left(1+\zeta\right)}\left[m_1\cos\left(2\delta-\varphi_1\right)-\left(1+\zeta\right)m_2\cos\left(2\delta-\varphi_2\right)+\zeta
    m_3\right]$}\bigstrut\\\hline
  \end{tabular}
\caption{Definition of the abbreviations used in 
Tabs.~\ref{tab:RGcorrections4MajPhaseDiff} and \ref{tab:RGcorrections4angles}--\ref{tab:RGcorrections4MajPhases}
}
\label{tab:Coefficients}
\end{table}

\begin{table}[htbp]\centering\setlength{\bigstrutjot}{4pt}
  \begin{tabular}{|l|c|c|c|}\hline
& $32\pi^2 \, \Dot\theta_{12}$ & $64\pi^2 \, \Dot\theta_{13}$ & $32\pi^2 \,
\Dot\theta_{23}$\bigstrut\\\hline

$ P_{11}$ &
$\mathcal{Q}^+_{12}\sin 2\theta_{12}$ &
$0$ &
$0$ \bigstrut\\

 $P_{22}$ &
 $-\mathcal{Q}^+_{12}\sin 2\theta_{12}c_{23}^2$ &
 $\left( \mathcal{A}^+_{23} - \mathcal{A}^+_{13} \right) \sin 2\theta_{12}\sin 2\theta_{23}$  &
 $\left( \mathcal{Q}^+_{23}c_{12}^2 + \mathcal{Q}^+_{13}s_{12}^2 \right) \sin 2\theta_{23}$ \bigstrut\\

 $P_{33}$ &
 $-\mathcal{Q}^+_{12} \sin 2\theta_{12}s_{23}^2$ &
 $-\!\left( \mathcal{A}^+_{23} - \mathcal{A}^+_{13} \right) \sin 2\theta_{12} \sin
 2\theta_{23}$ &
 $-\!\left( \mathcal{Q}^+_{23}c_{12}^2 + \mathcal{Q}^+_{13}s_{12}^2 \right) \sin 2\theta_{23}$ \bigstrut\\

 $\re P_{21}$ &
 $2\mathcal{Q}^+_{12}\cos 2\theta_{12} c_{23}$ &
 $4 \left( \mathcal{A}^+_{13}c_{12}^2 + \mathcal{A}^+_{23}s_{12}^2 \right)
 s_{23}$ &
 $\left( \mathcal{Q}^+_{23} - \mathcal{Q}^+_{13} \right) \sin 2\theta_{12}s_{23}$ \bigstrut\\

 $\re P_{31}$ &
 $-2\mathcal{Q}^+_{12} \cos 2\theta_{12} s_{23}$ &
 $4 \left( \mathcal{A}^+_{13}c_{12}^2 + \mathcal{A}^+_{23}s_{12}^2
 \right)c_{23}$  &
 $\left( \mathcal{Q}^+_{23} - \mathcal{Q}^+_{13} \right)\sin 2\theta_{12}
 c_{23}$ \bigstrut\\
 
 $\re P_{32}$ &
 $\mathcal{Q}^+_{12} \sin 2\theta_{12}\sin 2\theta_{23}$ &
 $2\! \left( \mathcal{A}^+_{23} - \mathcal{A}^+_{13} \right) \sin 2\theta_{12}\cos
 2\theta_{23}$ &
 $2\! \left( \mathcal{Q}^+_{23}c_{12}^2 + \mathcal{Q}^+_{13}s_{12}^2 \right) \cos
 2\theta_{23}$ \bigstrut\\
 
 $\im P_{21}$ &
 $4 \mathcal{S}_{12}c_{23}$ &
 $4\left( \mathcal{B}^-_{13}c_{12}^2 + \mathcal{B}^-_{23}s_{12}^2 \right)
 s_{23}$ &
 $2 \left( \mathcal{S}_{23} - \mathcal{S}_{13} \right) \sin 2\theta_{12}s_{23} $
 \bigstrut\\
 
 $\im P_{31}$ &
 $-4\mathcal{S}_{12} s_{23}$ &
 $4 \left( \mathcal{B}^-_{13}c_{12}^2 + \mathcal{B}^-_{23}s_{12}^2
 \right)c_{23}$  &
 $2 \left( \mathcal{S}_{23} - \mathcal{S}_{13} \right) \sin 2\theta_{12}c_{23} $
 \bigstrut\\
 
 $\im P_{32}$ &
 $0$ &
 $2 \left( \mathcal{B}^-_{23} - \mathcal{B}^-_{13} \right) \sin 2\theta_{12}$ &
 $4 \left( \mathcal{S}_{23}c_{12}^2 + \mathcal{S}_{13}s_{12}^2 \right)$\bigstrut\\\hline
\end{tabular}
  
\caption{Coefficients of $P_{fg}$ in the RGEs of the mixing angles
$\theta_{ij}$ in the limit $\theta_{13}\to0$.  The abbreviations $\mathcal{A}^\pm_{ij}$,
$\mathcal{B}^\pm_{ij}$, $\mathcal{S}_{ij}$ and $\mathcal{Q}_{ij}^\pm$ depend on the mass eigenvalues and phases only, and enhance the
running for a degenerate mass spectrum, since they are of the form
$f_{ij}(m_i,m_j,\text{phases})/(m_j^2-m_i^2)$.  They are listed in
Tab.~\ref{tab:Coefficients}.}
\label{tab:RGcorrections4angles}
\end{table}

\begin{table}[htbp]\centering\setlength{\bigstrutjot}{4pt}
  \begin{tabular}{|l|c|}\hline
    &$64\pi^2 \Dot\delta^{(-1)}$   \bigstrut\\\hline 

$ P_{11}$ &
$0$ \bigstrut\\

 $P_{22}$ &
 $-\left( \mathcal{B}^+_{23} - \mathcal{B}^+_{13} \right) \sin 2\theta_{12} \sin
 2\theta_{23}$ \bigstrut\\
 
 $P_{33}$ &
 $\left( \mathcal{B}^+_{23} - \mathcal{B}^+_{13} \right) \sin 2\theta_{12} \sin
 2\theta_{23}$ \bigstrut\\
 
 $\re P_{21}$ &
 $-4 \left( \mathcal{B}^+_{13}c_{12}^2 + \mathcal{B}^+_{23}s_{12}^2 \right)
 s_{23}$ \bigstrut\\
 
 $\re P_{31}$ &
 $-4 \left( \mathcal{B}^+_{13}c_{12}^2 + \mathcal{B}^+_{23}s_{12}^2
 \right)c_{23}$  \bigstrut\\
 
 $\re P_{32}$ &
 $-2\left( \mathcal{B}^+_{23} - \mathcal{B}^+_{13} \right) \sin 2\theta_{12}\cos
 2\theta_{23}$ \bigstrut\\
 
 $\im P_{21}$ &
 $4\left( \mathcal{A}^-_{13}c_{12}^2 + \mathcal{A}^-_{23}s_{12}^2 \right)
 s_{23}$ \bigstrut\\

  $\im P_{31}$ &
 $4\left( \mathcal{A}^-_{13}c_{12}^2 + \mathcal{A}^-_{23}s_{12}^2 \right)c_{23}$
 \bigstrut\\
 
 $\im P_{32}$ &
 $2\left( \mathcal{A}^-_{23} - \mathcal{A}^-_{13} \right) \sin 2\theta_{12} $\bigstrut\\\hline
\end{tabular}

  \vspace{1ex}

\begin{tabular}{|l|c|}\hline
    &$64\pi^2 \Dot\delta^{(0)}$   \bigstrut\\\hline

    $ P_{11}$ &
    $-8 \left( \left( \mathcal{C}_{13}^{23} + \mathcal{S}_{12} -
    \mathcal{S}_{23} \right)c_{12}^2 + \left( \mathcal{C}_{23}^{13} +
    \mathcal{S}_{12} - \mathcal{S}_{13} \right)s_{12}^2   \right)$  \bigstrut\\

 $P_{22}$ &
 $8 \left(
   \left( \left( \mathcal{S}_{12} - \mathcal{S}_{23} \right)c_{23}^2
   + \mathcal{C}_{13}^{23}s_{23}^2 \right) c_{12}^2
 +   \left(  \left( \mathcal{S}_{12} - \mathcal{S}_{13}
 \right)c_{23}^2  + \mathcal{C}_{23}^{13}s_{23}^2 \right) s_{12}^2
\right)$  \bigstrut\\

 $P_{33}$ &
 $8 \left(
  \left(\mathcal{C}_{13}^{23}c_{23}^2
    + \left( \mathcal{S}_{12} -\mathcal{S}_{23} \right)s_{23}^2  \right)c_{12}^2
  + \left( \mathcal{C}_{23}^{13}c_{23}^2
    + \left(\mathcal{S}_{12} - \mathcal{S}_{13} \right) s_{23}^2 \right)s_{12}^2
\right)$  \bigstrut\\

 $\re P_{21}$ &
 $-16\mathcal{S}_{12}c_{23}\cot 2\theta_{12}
 +4 \left( 2\mathcal{D}_{1}c_{23} +\left( \mathcal{S}_{23} - \mathcal{S}_{13} \right)s_{23} \tan \theta_{23}
 \right)\sin 2\theta_{12}$  \bigstrut\\
 
 $\re P_{31}$ &
 $16\mathcal{S}_{12}s_{23}\cot 2\theta_{12}
 - 4 \left( 2\mathcal{D}_{1}s_{23} +
 \left( \mathcal{S}_{23} - \mathcal{S}_{13} \right)c_{23} \cot \theta_{23}
 \right) \sin 2\theta_{12}$  \bigstrut\\
 
 $\re P_{32}$ &
 $-16\left(  \mathcal{S}_{23}c_{12}^2 +\mathcal{S}_{13}s_{12}^2\right) \cos
 2\theta_{23}\cot 2\theta_{23}
 -8 \left( \mathcal{C}_{13}^{12}c_{12}^2 + \mathcal{C}_{23}^{12}s_{12}^2 \right) \sin 2\theta_{23}$  \bigstrut\\

 $\im P_{21}$ &
 $-8\mathcal{Q}^-_{12}c_{23}\csc 2\theta_{12}
 -2 \left( 2\mathcal{D}_{2}c_{23} + \left( \mathcal{Q}^-_{23} - \mathcal{Q}^-_{13} \right) \cos 2\theta_{23}\sec \theta_{23} \right) \sin 2\theta_{12}$  \bigstrut\\

 $\im P_{31}$ &
 $8\mathcal{Q}^-_{12}s_{23}\csc 2\theta_{12}
 + 2\left( 2\mathcal{D}_{2}s_{23} - \left( \mathcal{Q}^-_{23} - \mathcal{Q}^-_{13} \right) \cos 2\theta_{23}\csc
 \theta_{23} \right) \sin 2\theta_{12}$ \bigstrut\\

 $\im P_{32}$ &
 $-8 \left( \mathcal{Q}^-_{23}c_{12}^2 + \mathcal{Q}^-_{13}s_{12}^2 \right) \cot 2\theta_{23} $\bigstrut\\\hline
\end{tabular}
\caption{Coefficients of $P_{fg}$ in the derivative of the Dirac CP phase.  The
 complete RGE is given by $\Dot\delta = \theta_{13}^{-1} \Dot\delta^{(-1)} +
 \Dot\delta^{(0)} + \mathscr{O}{(\theta_{13})}$. The abbreviations
 $\mathcal{A}^\pm_{ij}$, $\mathcal{B}^\pm_{ij}$,  $\mathcal{Q}_{ij}^\pm$,
 $\mathcal{C}_{ij}^{kl}$ and $\mathcal{D}_i$ depend on the mass eigenvalues and
 phases only, and  are listed in Tab.~\ref{tab:Coefficients}}
\label{tab:RGcorrections4Delta}
\end{table}

\begin{table}[htbp]\centering\setlength{\bigstrutjot}{4pt}
  \begin{tabular}{|l|c|}\hline
    &$16\pi^2 \Dot\varphi_1$   \bigstrut\\\hline

$ P_{11}$ &
$-4 \mathcal{S}_{12}c_{12}^2$ \bigstrut\\

 $P_{22}$ &
 $4\mathcal{S}_{12}c_{12}^2 c_{23}^2
 - 4\left( \mathcal{S}_{23}c_{12}^2 +\mathcal{S}_{13}s_{12}^2\right) \cos 2\theta_{23}$ \bigstrut\\

 $P_{33}$ &
 $4 \mathcal{S}_{12}c_{12}^2s_{23}^2
   +4 \left( \mathcal{S}_{23}c_{12}^2+ \mathcal{S}_{13}s_{12}^2 \right) \cos 2\theta_{23}$  \bigstrut\\

 $\re P_{21}$ &
 $-4 \mathcal{S}_{12}c_{23}\cos 2\theta_{12}\cot \theta_{12}
 - 2\left( \mathcal{S}_{23} - \mathcal{S}_{13} \right) \cos 2\theta_{23}\sec \theta_{23} \sin 2\theta_{12}$ \bigstrut\\
 
 $\re P_{31}$ &
 $4\mathcal{S}_{12}s_{23}\cos 2\theta_{12}\cot \theta_{12}
 - 2\left(\mathcal{S}_{23} - \mathcal{S}_{13} \right) \cos 2\theta_{23}\csc\theta_{23}\sin 2\theta_{12}$ \bigstrut\\
 
 $\re P_{32}$ &
 $-8\left( \mathcal{S}_{23}c_{12}^2 +\mathcal{S}_{13}s_{12}^2\right) \cos 2\theta_{23}\cot 2\theta_{23}
 -4 \mathcal{S}_{12}c_{12}^2\sin 2\theta_{23}$ \bigstrut\\
 
 $\im P_{21}$ &
 $-2\mathcal{Q}^-_{12}c_{23}\cot \theta_{12}
 - \left( \mathcal{Q}^-_{23} - \mathcal{Q}^-_{13} \right) \cos 2\theta_{23}\sec \theta_{23}\sin 2\theta_{12}$
 \bigstrut\\
 
 $\im P_{31}$ &
 $2\mathcal{Q}^-_{12} s_{23}\cot \theta_{12}
 - \left( \mathcal{Q}^-_{23} - \mathcal{Q}^-_{13} \right) \cos 2\theta_{23}\csc \theta_{23} \sin 2\theta_{12}$
 \bigstrut\\
 
 $\im P_{32}$ &
 $-4\left( \mathcal{Q}^-_{23}c_{12}^2 + \mathcal{Q}^-_{13}s_{12}^2 \right) \cot 2\theta_{23} $\bigstrut\\\hline
\end{tabular}
  
  \vspace{1ex}

  \begin{tabular}{|l|c|}\hline
    &$16\pi^2 \Dot\varphi_2$\bigstrut\\\hline 

$ P_{11}$ &
$-4\mathcal{S}_{12}s_{12}^2$ \bigstrut\\

 $P_{22}$ &
 $4\mathcal{S}_{12}c_{23}^2 s_{12}^2
 - 4\left( \mathcal{S}_{23}c_{12}^2 + \mathcal{S}_{13}s_{12}^2 \right) \cos 2\theta_{23}$ \bigstrut\\

 $P_{33}$ &
 $4 \mathcal{S}_{12}s_{23}^2s_{12}^2
 +4 \left( \mathcal{S}_{23}c_{12}^2 +\mathcal{S}_{13}s_{12}^2\right) \cos 2\theta_{23}$  \bigstrut\\

 $\re P_{21}$ &
 $- 4\mathcal{S}_{12}c_{23}\cos 2\theta_{12} \tan
 \theta_{12}
 -2\left( \mathcal{S}_{23} - \mathcal{S}_{13} \right) \cos 2\theta_{23}\sec
 \theta_{23} \sin 2\theta_{12}$ \bigstrut\\
 
 $\re P_{31}$ &
 $4\mathcal{S}_{12}s_{23}\cos 2\theta_{12} \tan \theta_{12}
 -2 \left( \mathcal{S}_{23} - \mathcal{S}_{13} \right) \cos 2\theta_{23}\csc
 \theta_{23}\sin 2\theta_{12} $ \bigstrut\\
 
 $\re P_{32}$ &
 $-8\left( \mathcal{S}_{23}c_{12}^2 +\mathcal{S}_{13}s_{12}^2 \right)\cos 2\theta_{23}\cot 2\theta_{23}
 -4 \mathcal{S}_{12}s_{12}^2\sin 2\theta_{23}$ \bigstrut\\
 
 $\im P_{21}$ &
 $ - 2\mathcal{Q}^-_{12}c_{23}\tan \theta_{12}
 -\left( \mathcal{Q}^-_{23} - \mathcal{Q}^-_{13} \right) \cos 2\theta_{23}\sec \theta_{23} \sin 2\theta_{12}$ \bigstrut\\
 
 $\im P_{31}$ &
 $ 2\mathcal{Q}^-_{12}s_{23}\tan \theta_{12}
 -\left( \mathcal{Q}^-_{23} - \mathcal{Q}^-_{13} \right) \cos 2\theta_{23}\csc
 \theta_{23}\sin 2\theta_{12} $ \bigstrut\\
 
 $\im P_{32}$ &
 $-4\left( \mathcal{Q}^-_{23}c_{12}^2 + \mathcal{Q}^-_{13}s_{12}^2 \right) \cot 2\theta_{23} $\bigstrut\\\hline
\end{tabular}
\caption{Coefficients of $P_{fg}$ in the RGEs of the Majorana phases
 for $\theta_{13}=0$.
}
\label{tab:RGcorrections4MajPhases}
\end{table}

\clearpage

\section{RGEs for See-Saw Models} \label{app:RGEs}

In order to calculate the RG evolution of the effective neutrino mass
matrix, the RGEs for all the parameters of the theory have to be solved
simultaneously.  We therefore summarize the RGEs for the minimal see-saw
extensions of the SM, of the class of 2HDMs described in
Sec.~\ref{sec:2HDMs}, and of the MSSM.  We list the MS 1-loop results in
the SM and 2HDM, as well as the 2-loop RGEs for the effective neutrino
mass operator, the singlet mass matrix and the Yukawa couplings in the
MSSM.  For further RGEs and references, see e.g.\
\cite{Machacek:1983tz,Machacek:1984fi,Machacek:1985zw,Das:2000uk}. 
We use the notation defined in Sec.~\ref{sec:RunningInSeeSaw}.  In
particular, a superscript $(n)$ denotes 
a quantity between the $n$th and the $(n+1)$th mass threshold.
The RGEs for the SM, 2HDM or MSSM without singlet neutrinos can 
be recovered by setting the neutrino Yukawa coupling to zero. 
In the full theories above the highest see-saw scale, the superscript
$(n)$ has to be omitted.

The RGEs for the gauge couplings are well-known and not affected
 by the additional singlets at 1-loop order. They are given by 
\begin{eqnarray}
  16\pi^2\,\beta_{g_A} := 16\pi^2\,\mu \frac{\D g_A}{\D \mu} =b_A \, g_A^3 \;,
\end{eqnarray} 
with $(b_{\SU (3)_\mathrm{C}},b_{\SU (2)_\mathrm{L}},b_{\U (1)_\mathrm{Y}})
=(-7,-\tfrac{19}{6},\tfrac{41}{10})$ in the SM, 
$(-7,-3,\tfrac{21}{5})$ in the 2HDMs and $(-3,1,\tfrac{33}{5})$ in the MSSM.
For $\U(1)_\mathrm{Y}$, we use GUT charge normalization.

\subsection{The RGEs in the Extended SM} \label{app:RGEsSM}
In the SM extended by singlet neutrinos, the RG evolution is governed by
the $\beta$-functions \cite{Grzadkowski:1987tf,Antusch:2002rr} 
\begin{subequations}
\begin{eqnarray}
16\pi^2\accentset{(n)}{\beta}_\kappa & = & 
 -\frac{3}{2} (Y_e^\dagger Y_e)^T \:\accentset{(n)}{\kappa}
 -\frac{3}{2}\,\accentset{(n)}{\kappa} \, (Y_e^\dagger Y_e)
 + \frac{1}{2} \RaiseBrace{\bigl(} \accentset{(n)}{Y}^\dagger_\nu   
   \accentset{(n)}{Y}_\nu \RaiseBrace{\bigr)}^T \,
  \accentset{(n)}{\kappa}
 +\frac{1}{2}\,\accentset{(n)}{\kappa} \: \RaiseBrace{\bigl(}
 \accentset{(n)}{Y}^\dagger_\nu\accentset{(n)}{Y}_\nu\RaiseBrace{\bigr)}
\nonumber \\*
&& {}\vphantom{\frac{1}{2}}
 +2\,\Tr(Y_e^\dagger Y_e)\,\accentset{(n)}{\kappa} 
 +2\, \Tr \RaiseBrace{\bigl(} \accentset{(n)}{Y}^{\dagger}_\nu 
 \accentset{(n)}{Y}_\nu\RaiseBrace{\bigr)}\,\accentset{(n)}{\kappa} 
 +6\,\Tr(Y_u^\dagger Y_u)\,\accentset{(n)}{\kappa} 
  \nonumber \\*
 && {} \vphantom{\frac{1}{2}}
 +6\,\Tr(Y_d^\dagger Y_d)\,\accentset{(n)}{\kappa}
- 3 g_2^2\: \accentset{(n)}{\kappa}
 +\lambda\accentset{(n)}{\kappa}
 \;,\\
16\pi^2 \accentset{(n)}{\beta}_{M} &=&\vphantom{\frac{1}{2}}
 \RaiseBrace{\bigl(}\accentset{(n)}{Y}_\nu   
   \accentset{(n)}{Y}^\dagger_\nu \RaiseBrace{\bigr)}\, \accentset{(n)}{M} 
   + \accentset{(n)}{M}\,\RaiseBrace{\bigl(}\accentset{(n)}{Y}_\nu   
   \accentset{(n)}{Y}^\dagger_\nu \RaiseBrace{\bigr)}^T \;,\\
   16\pi^2 \accentset{(n)}{\beta}_{Y_\nu}
 & = &
 \accentset{(n)}{Y}_\nu \left\{ 
        \frac{3}{2} \RaiseBrace{\bigl(}
        \accentset{(n)}{Y}^\dagger_\nu\accentset{(n)}{Y}_\nu\RaiseBrace{\bigr)}
        - \frac{3}{2}(Y_e^\dagger Y_e)
+ \Tr \RaiseBrace{\bigl(}\accentset{(n)}{Y}^{\dagger}_\nu  
\accentset{(n)}{Y}_\nu\RaiseBrace{\bigr)} +\Tr (Y_e^\dagger Y_e) \right.
\nonumber \\
&&\hphantom{\accentset{(n)}{Y}_\nu \left[ \right.} \left. 
{}+ 3\,\Tr(Y_u^\dagger Y_u)+3\,\Tr(Y_d^\dagger Y_d)
-\frac{9}{20} g_1^2 -\frac{9}{4} g_2^2 \right\} ,
\\
        16\pi^2 \, \accentset{(n)}{\beta}_{Y_e}
        & = &
        Y_e
        \left\{ 
        \vphantom{\Tr\left[Y_e^\dagger Y_e
                +z_\nu^{(1)}\,\accentset{(n)}{Y}_\nu^\dagger \accentset{(n)}{Y}_\nu
                +3z_d^{(1)}\,Y_d^\dagger Y_d\
                +3z_u^{(1)}\,Y_u^\dagger Y_u\right]}
     \frac{3}{2} Y_e^\dagger Y_e 
         -\frac{3}{2}\, \accentset{(n)}{Y}_\nu^\dagger \accentset{(n)}{Y}_\nu 
         -       \frac{9}{4} g_1^2 - \frac{9}{4} g_2^2
         \right.\nonumber\\*
        & &\hphantom{Y_e\left\{ \right.}
         \left.{}+
         \Tr\left[Y_e^\dagger Y_e
                +\accentset{(n)}{Y}_\nu^\dagger \accentset{(n)}{Y}_\nu
                +3\,Y_d^\dagger Y_d\
                +3\,Y_u^\dagger Y_u\right]
        \right\} ,\\
        16\pi^2 \, \accentset{(n)}{\beta}_{Y_d}
        & = &
        Y_d
        \left\{ 
     \frac{3}{2} Y_d^\dagger Y_d 
         -\frac{3}{2}\, Y_u^\dagger Y_u          
         - \frac{1}{4} g_1^2 - \frac{9}{4} g_2^2 - 8\,g_3^2
         \right.\nonumber\\*
        & &\hphantom{Y_e\left\{ \right.}
         \left.{}+ \Tr\left[Y_e^\dagger Y_e
                +\accentset{(n)}{Y}_\nu^\dagger \accentset{(n)}{Y}_\nu
                +3\,Y_d^\dagger Y_d
                +3\,Y_u^\dagger Y_u\right]
        \right\} , 
        \\
        16\pi^2 \, \accentset{(n)}{\beta}_{Y_u}
        & = &
        Y_u
        \left\{ 
     \frac{3}{2} Y_u^\dagger Y_u 
         - \frac{3}{2}\, Y_d^\dagger Y_d  
         - \frac{17}{20} g_1^2 - \frac{9}{4} g_2^2 - 8\,g_3^2
         \right.\nonumber\\*
        & &\hphantom{Y_e\left\{ \right.}
         \left.{}+ \Tr\left[Y_e^\dagger Y_e
                +\accentset{(n)}{Y}_\nu^\dagger \accentset{(n)}{Y}_\nu
                +3\,Y_d^\dagger Y_d
                +3\,Y_u^\dagger Y_u\right]
        \right\} ,
\\
 16\pi^2\,\accentset{(n)}{\beta}_{\lambda}
 &=& 
 6\,\lambda^2 
 -3\,\lambda\,\left(3g_2^2+\frac{3}{5} g_1^2\right)
 +3\,g_2^4
 +\frac{3}{2}\,\left(\frac{3}{5} g_1^2+g_2^2\right)^2
 \nonumber
 \\
 & &{}
 +4\,\lambda\,
 \Tr\left[
       Y_e^\dagger Y_e
       +\accentset{(n)}{Y}_\nu^\dagger\accentset{(n)}{Y}_\nu
       +3\,Y_d^\dagger Y_d
       +3\,Y_u^\dagger Y_u
 \right]
 \\*
 & &{}
 -8\,\Tr\left[
  Y_e^\dagger Y_e\,Y_e^\dagger Y_e
  +    \accentset{(n)}{Y}_\nu^\dagger\accentset{(n)}{Y}_\nu
               \,\accentset{(n)}{Y}_\nu^\dagger\accentset{(n)}{Y}_\nu
       +3\,Y_d^\dagger Y_d\,Y_d^\dagger Y_d
       +3\,Y_u^\dagger Y_u\,Y_u^\dagger Y_u
 \right].\nonumber
\end{eqnarray}
\end{subequations}
We use the convention that the Higgs self-interaction term in the
Lagrangian is $-\frac{\lambda}{4} (\phi^\dagger \phi)^2$.

\subsection{The RGEs in Extended 2HDMs}\label{app:RGEs2HDMs}

Here, we list the $\beta$-functions for
the class of 2HDMs described in Sec.~\ref{sec:2HDMs}
\cite{Hill:1985tg,Grzadkowski:1987wr,Antusch:2001vn}.
The coefficients $z_f^{(i)}$ determine which fermion couples to
which Higgs, cf.\ Eq.~\eqref{eq:2HDMYukawaCouplings}.
\begin{subequations}
\begin{eqnarray}
16\pi^2 \accentset{(n)}{\beta}_{\kappa^{(ii)}} & = &
 \left(\tfrac{1}{2}-2\delta_{i1}\right)\,
  \left[\kappa^{(ii)}(Y_e^\dagger Y_e)
        +(Y_e^\dagger Y_e)^T\kappa^{(ii)}\right]
  +\left[z^{(i)}_\nu\,2\,
  \Tr(\accentset{(n)}{Y}_\nu^\dagger \accentset{(n)}{Y}_\nu)\right.
 \nonumber\\*
  & & {}
  +\left.\delta_{i1}\,2 \,\Tr(Y_e^\dagger Y_e) 
        +z^{(i)}_u\,6\,\Tr(Y_u^\dagger Y_u)
        +z^{(i)}_d\,6\,\Tr(Y_d^\dagger Y_d)
  \vphantom{\accentset{(n)}{Y}_\nu^\dagger \accentset{(n)}{Y}_\nu}
  \right]\kappa^{(ii)}
 \nonumber\\*
  & & {}\vphantom{\frac{1}{2}}
        +\lambda_i\kappa^{(ii)}+\delta_{i1}\lambda_5^*\kappa^{(22)}
        +\delta_{i2}\lambda_5\kappa^{(11)}
        -3g_2^2\kappa^{(ii)}
\; ,\\
 16\pi^2 \accentset{(n)}{\beta}_{M} &=& \vphantom{\frac{1}{2}}
   \RaiseBrace{\bigl(}\accentset{(n)}{Y}_\nu
   \accentset{(n)}{Y}^\dagger_\nu \RaiseBrace{\bigr)}\, \accentset{(n)}{M} 
   + \accentset{(n)}{M}\,\RaiseBrace{\bigl(}\accentset{(n)}{Y}_\nu   
   \accentset{(n)}{Y}^\dagger_\nu \RaiseBrace{\bigr)}^T ,\\*
16\pi^2 \accentset{(n)}{\beta}_{Y_\nu}
 &=&\vphantom{\frac{1}{2}}
 \accentset{(n)}{Y}_\nu
        \left\{ \vphantom{\sum_{i=1}^2z_\nu^{(i)}}
         \frac{3}{2} \accentset{(n)}{Y}_\nu^\dagger \accentset{(n)}{Y}_\nu +
         \left(\tfrac{1}{2}-2\,z_\nu^{(1)}\right) 
         \frac{3}{2} Y_e^\dagger Y_e 
         -       \frac{9}{20} g_1^2- \frac{9}{4} g_2^2
         \right. 
\nonumber\\*
        & & {}
         \left.+
         \sum_{i=1}^2\!z_\nu^{(i)}\,
         \Tr\!\left[\delta_{i1}Y_e^\dagger Y_e
                +\accentset{(n)}{Y}_\nu^\dagger \accentset{(n)}{Y}_\nu
                +3z_d^{(i)}\,Y_d^\dagger Y_d\
                +3z_u^{(i)}\,Y_u^\dagger Y_u \right]
        \right\} ,
\\
        16\pi^2 \, \accentset{(n)}{\beta}_{Y_e}
        & = &
        Y_e
        \left\{ 
        \vphantom{\Tr\left[Y_e^\dagger Y_e
                +z_\nu^{(1)}\,\accentset{(n)}{Y}_\nu^\dagger \accentset{(n)}{Y}_\nu
                +3z_d^{(1)}\,Y_d^\dagger Y_d\
                +3z_u^{(1)}\,Y_u^\dagger Y_u\right]}
     \frac{3}{2} Y_e^\dagger Y_e +
         \left(\tfrac{1}{2}-2\,z_\nu^{(1)}\right) \accentset{(n)}{Y}_\nu^\dagger \accentset{(n)}{Y}_\nu 
         -       \frac{9}{4} g_1^2 - \frac{9}{4} g_2^2
         \right.\\*
         & &
         \left.
         +
         \Tr\left[Y_e^\dagger Y_e
                +z_\nu^{(1)}\,\accentset{(n)}{Y}_\nu^\dagger \accentset{(n)}{Y}_\nu
                +3z_d^{(1)}\,Y_d^\dagger Y_d\
                +3z_u^{(1)}\,Y_u^\dagger Y_u\right]
        \right\} \!,
        \nonumber\\*
        16\pi^2 \, \accentset{(n)}{\beta}_{Y_d}
        & = &
        Y_d
        \left\{ 
        \vphantom{\sum_{i=1}^2z_d^{(i)}\,
         \Tr\left[\delta_{i1}\,Y_e^\dagger Y_e
                +z_{\nu}^{(1)}\accentset{(n)}{Y}_\nu^\dagger \accentset{(n)}{Y}_\nu
                +3\,Y_d^\dagger Y_d\
                +3z_u^{(i)}\,Y_u^\dagger Y_u\right]}
     \frac{3}{2} Y_d^\dagger Y_d 
         + \left(\frac{1}{2}-2\sum_{i=1}^2\! z_u^{(i)}z_d^{(i)}\right) Y_u^\dagger Y_u   
         - \frac{1}{4}  g_1^2- \frac{9}{4} g_2^2 - 8g_3^2
         \right.\nonumber \\*
        & &
         \left.+
         \sum_{i=1}^2\!z_d^{(i)}
         \Tr\!\left[\delta_{i1} Y_e^\dagger Y_e
                +z_{\nu}^{(i)}\accentset{(n)}{Y}_\nu^\dagger \accentset{(n)}{Y}_\nu
                +3Y_d^\dagger Y_d
                +3z_u^{(i)} Y_u^\dagger Y_u\right]
        \right\}, 
        \\
        16\pi^2 \, \accentset{(n)}{\beta}_{Y_u}
        & = &
        Y_u
        \left\{ 
        \vphantom{       \sum_{i=1}^2z_d^{(i)}\,
         \Tr\left[\delta_{i1}\,Y_e^\dagger Y_e
                +z_{\nu}^{(1)}\accentset{(n)}{Y}_\nu^\dagger \accentset{(n)}{Y}_\nu
                +3\,Y_d^\dagger Y_d\
                +3z_u^{(i)}\,Y_u^\dagger Y_u\right]}
     \frac{3}{2} Y_u^\dagger Y_u 
         + \left(\frac{1}{2}-2\sum_{i=1}^2 z_u^{(i)}z_d^{(i)}\right)
          Y_d^\dagger Y_d 
         \right.
        \left. 
         - \frac{17}{20}  g_1^2 - \frac{9}{4} g_2^2 - 8g_3^2
         \right.\nonumber\\*
        & &
         \left.+
         \sum_{i=1}^2\!z_u^{(i)}
         \Tr\!\left[\delta_{i1}Y_e^\dagger Y_e
                +z_{\nu}^{(i)}\accentset{(n)}{Y}_\nu^\dagger \accentset{(n)}{Y}_\nu
                +3z_d^{(i)}Y_d^\dagger Y_d
                +3Y_u^\dagger Y_u\right]
        \right\} .
 \end{eqnarray}
 \end{subequations}
 For the parameters of the 
 Higgs interaction Lagrangian, the $\beta$-functions are \cite{Hill:1985tg}
(Note that we use different conventions for the renormalizable Higgs
couplings, as specified in \cite{Antusch:2001vn}.)
\begin{subequations}
 \begin{eqnarray}
  16\pi^2\,\accentset{(n)}{\beta}_{\lambda_1}
  & = & 
  6 \lambda_1^2 + 8 \lambda_3^2+6 \lambda_3 \lambda_4
  +\lambda_5^2
  -3 \lambda_1\left(3g_2^2+\frac{3}{5} g_1^2\right)
  +3 g_2^4
  +\frac{3}{2} \left(\frac{3}{5} g_1^2+g_2^2\right)^2
  \nonumber\\*
  & &{}
  +4 \lambda_1 
  \Tr\left(
        Y_e^\dagger Y_e
        +z_\nu^{(1)}\,\accentset{(n)}{Y}_\nu^\dagger\accentset{(n)}{Y}_\nu
        +3 z_d^{(1)}\,Y_d^\dagger Y_d 
        +3 z_u^{(1)}\,Y_u^\dagger Y_u
  \right)
\nonumber\\*
  & &{} 
  -8 \Tr\left(\!
   Y_e^\dagger Y_e Y_e^\dagger Y_e
   +z_\nu^{(1)}
                \accentset{(n)}{Y}_\nu^\dagger\accentset{(n)}{Y}_\nu
                 \accentset{(n)}{Y}_\nu^\dagger\accentset{(n)}{Y}_\nu
        +3 z_d^{(1)} Y_d^\dagger Y_d Y_d^\dagger Y_d
        +3 z_u^{(1)} Y_u^\dagger Y_u Y_u^\dagger Y_u
  \!\right),
  \nonumber\\
  \\
  16\pi^2\,\accentset{(n)}{\beta}_{\lambda_2}
  & = & 
  6 \lambda_2^2 + 8 \lambda_3^2+6 \lambda_3 \lambda_4
  +\lambda_5^2
  -3 \lambda_2 \left(3g_2^2+\frac{3}{5} g_1^2\right)
  +3 g_2^4
  +\frac{3}{2} \left(\frac{3}{5} g_1^2+g_2^2\right)^2
  \nonumber\\*
  & &{}  
  +4 \lambda_2 
  \Tr\left(
        z_\nu^{(2)}\,\accentset{(n)}{Y}_\nu^\dagger\accentset{(n)}{Y}_\nu
        +3 z_d^{(2)}\,Y_d^\dagger Y_d
        +3 z_u^{(2)}\,Y_u^\dagger Y_u
  \right)
  \nonumber\\*
  & &{}
  -8\Tr\left(
    z_\nu^{(2)}\,
                \accentset{(n)}{Y}_\nu^\dagger\accentset{(n)}{Y}_\nu 
                \accentset{(n)}{Y}_\nu^\dagger\accentset{(n)}{Y}_\nu
        +3 z_d^{(2)}\,Y_d^\dagger Y_d Y_d^\dagger Y_d
        +3 z_u^{(2)}\,Y_u^\dagger Y_u Y_u^\dagger Y_u
  \right) ,
  \\
  16\pi^2\,\accentset{(n)}{\beta}_{\lambda_3}
  & = & 
  (\lambda_1+\lambda_2)\,(3\lambda_3+\lambda_4) 
  + 4\lambda_3^2+2\lambda_4^2
  +\frac{1}{2}\lambda_5^2
  -3\lambda_3 \left(3g_2^2+\frac{3}{5} g_1^2\right)
  +\frac{9}{4}g_2^4
  \nonumber\\*
  & & {}
  +\frac{27}{100}g_1^4
  -\frac{9}{10}g_1^2g_2^2
  +4\lambda_3
  \Tr\left(
        Y_e^\dagger Y_e
        +\accentset{(n)}{Y}_\nu^\dagger\accentset{(n)}{Y}_\nu
        +3 Y_d^\dagger Y_d
        +3 Y_u^\dagger Y_u
  \right)
  \nonumber\\*
  & &{}
  -4 \Tr\left(
    z_\nu^{(2)}\,Y_e^\dagger Y_e\,\accentset{(n)}{Y}_\nu^\dagger\accentset{(n)}{Y}_\nu
        +3 \left(z_d^{(1)}z_u^{(2)}+z_d^{(2)}z_u^{(1)}\right) Y_d^\dagger Y_d\,Y_u^\dagger Y_u
  \right) ,
  \\
  16\pi^2\,\accentset{(n)}{\beta}_{\lambda_4}
  & = & 
  2\,(\lambda_1+\lambda_2)\,\lambda_4 
  + 4\,(2 \lambda_3+\lambda_4)\,\lambda_4
  +8\lambda_5^2
  -3 \lambda_4 \left(3g_2^2+\frac{3}{5} g_1^2\right)
  +\frac{9}{5} g_1^2 g_2^2 
  \nonumber\\*
  & &{}
  +4 \lambda_4
  \Tr\left(
        Y_e^\dagger Y_e
        +\accentset{(n)}{Y}_\nu^\dagger\accentset{(n)}{Y}_\nu
        +3Y_d^\dagger Y_d
        +3Y_u^\dagger Y_u
  \right)
 \nonumber\\*
  & &{}
  +4 \Tr\left(
    z_\nu^{(2)}
    Y_e^\dagger Y_e\,\accentset{(n)}{Y}_\nu^\dagger\accentset{(n)}{Y}_\nu
        +3 \left(z_d^{(1)}z_u^{(2)}+z_d^{(2)}z_u^{(1)}\right) Y_d^\dagger Y_d\,Y_u^\dagger Y_u
  \right) ,
  \\
  16\pi^2\,\accentset{(n)}{\beta}_{\lambda_5}
  & = &
  \lambda_5 
  \left[\vphantom{\accentset{(n)}{Y}_\nu^\dagger\accentset{(n)}{Y}_\nu}
        \lambda_1+\lambda_2+8 \lambda_3+12 \lambda_4
        -6\left(\frac{3}{5} g_1^2+3 g_2^2\right)
        \right.
        \nonumber\\*
  && \hphantom{\lambda_5 \left[\right.} \left.{}
  +2 \Tr\left(
    Y_e^\dagger Y_e
        +\accentset{(n)}{Y}_\nu^\dagger\accentset{(n)}{Y}_\nu
        +3 Y_d^\dagger Y_d
        +3 Y_u^\dagger Y_u
  \right)
  \right] .
\end{eqnarray}
\end{subequations}

\subsection{Split Supersymmetry}
\label{app:RGEsSplitSUSY}

\begin{figure}
 \centerline{
        \subfigure{\CenterObject{\includegraphics{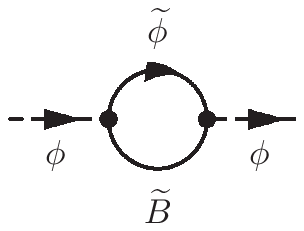}}}
        \quad
        \subfigure{\CenterObject{\includegraphics{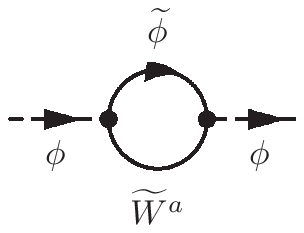}}}
 }
 \caption{Additional diagrams contributing to the wavefunction
 renormalization of the Higgs in split SUSY. The Higgsino is denoted by
 $\widetilde{\phi}$, and $\widetilde{B}$ and $\widetilde{W}^a$ represent the Bino
 and the Winos.}
 \label{fig:SplitSUSYadditionalWaveFunction}
\end{figure}

The $\beta$-functions for the renormalizable couplings in the framework
of split SUSY are listed in Ref.~\cite{Giudice:2004tc}. The diagrams
contributing to the RGE of the effective neutrino mass operator are
those relevant in the SM, amended by two diagrams involving Higgsinos
and gauginos (cf.\ Fig.~\ref{fig:SplitSUSYadditionalWaveFunction}).
These diagrams contribute to the flavour-trivial part of the RGE.
At 1 loop, we obtain for the divergent parts of the renormalization
constants in dimensional regularization and in the MS scheme
\begin{subequations}
\begin{eqnarray*}
 \delta Z_{\phi,1} 
 & = & 
 -\frac{1}{16\pi^2}\left[
        2\,\Tr\left(3Y_u^\dagger Y_u+3 Y_d^\dagger Y_d + Y_e^\dagger Y_e\right)
        +\frac{3}{10}(\xi_1-1) g_1^2 
        +\frac{3}{2}(\xi_2-1) g_2^2
 \right] ,
 \\
 \delta Z_{\ell_\mathrm{L},1} 
 & = & 
 -\frac{1}{16\pi^2}\left(
        Y_e^\dagger Y_e +\frac{3}{10}\xi_1\,g_1^2+\frac{3}{2}\xi_2\,g_2^2
 \right) ,
 \\
 \delta \kappa_{,1}
 & = &
 -\frac{1}{16\pi^2} \left\{
        2\kappa\,(Y_e^\dagger Y_e)+2(Y_e^\dagger Y_e)^T\kappa
        -\left[\lambda+\frac{3}{5}\left(\frac{3}{2}-\xi_1\right)g_1^2
        +       \left(\frac{3}{2}-3\xi_2\right)g_2^2\right] \kappa
 \right\} ,
\end{eqnarray*} 
\end{subequations}
where $\xi_1$ and $\xi_2$ are the gauge parameters in $R_\xi$ gauge.
$Z_i := \mathbbm{1} + \delta Z_{i,1} \, \frac{1}{\epsilon} + \dots$ 
($\epsilon:=4-d$) are wavefunction renormalization constants, and
$\delta\kappa := \delta\kappa_{,1} \, \frac{1}{\epsilon} + \dots$ is
defined via the counterterm for the dimension 5 operator,
\[
\mathscr{C}_\kappa \,=\,
    \frac{1}{4} \, \delta\kappa_{fg} \; 
    (\overline{\ell_\mathrm{L}^\ChargeC}{}^f\cdot \phi) \;
    (\ell_{\mathrm{L}}^g\cdot\phi)
    + \text{h.c.} \;.
\]
Using the method described in \cite{Antusch:2001ck}, we then find the 1-loop
$\beta$-function 
\begin{eqnarray}
 16\pi^2\,\beta_\kappa^\mathrm{Split\:SUSY}
 &=&
 -\frac{3}{2}(Y_e^\dagger Y_e)^T\kappa
 -\frac{3}{2} \kappa\,(Y_e^\dagger Y_e)
 \nonumber\\
 & &{}
 +\left[\lambda+\frac{3}{5}g_1^2
 +2\,\Tr\left(3Y_u^\dagger Y_u+3 Y_d^\dagger Y_d + Y_e^\dagger Y_e\right)\right]
 \kappa\;.
\end{eqnarray}
Clearly, the term involving the gauge couplings in the flavour-diagonal
part differs from the SM case.

\subsection{The RGEs in the MSSM Extended by Heavy Singlets}\label{app:RGEsMSSM}
We give the 2-loop RGEs for the quantities $Q\in\left\{
\accentset{(n)}{\kappa},\accentset{(n)}{M}
,\accentset{(n)}{Y}_\nu,{Y}_d,{Y}_u,{Y}_e
\right\}$ in the form 
\begin{equation}
 \mu\frac{\D  \accentset{(n)}{Q}}{\D \mu}
 =
 \accentset{(n)}{\beta}_{Q}^{\,(1)}+ \accentset{(n)}{\beta}_{Q}^{\,(2)}\;. 
\end{equation}
The 1-loop parts are given by \cite{Grzadkowski:1987wr,Antusch:2002rr}
\begin{subequations}
\begin{eqnarray}
(4\pi)^2 \accentset{(n)}{\beta}^{\,(1)}_\kappa & = & \vphantom{\frac{1}{2}}
 (Y_e^\dagger Y_e)^T \: \accentset{(n)}{\kappa}
 + \accentset{(n)}{\kappa} \, (Y_e^\dagger Y_e)
 + \RaiseBrace{\bigl(} \accentset{(n)}{Y}^\dagger_\nu   
   \accentset{(n)}{Y}_\nu \RaiseBrace{\bigr)}^T\,\accentset{(n)}{\kappa}
 + \accentset{(n)}{\kappa} \: \RaiseBrace{\bigl(}
 \accentset{(n)}{Y}^\dagger_\nu\accentset{(n)}{Y}_\nu\RaiseBrace{\bigr)}
\nonumber \\
&& {} + 2 \Tr \RaiseBrace{\bigl(} \accentset{(n)}{Y}^{\dagger}_\nu 
 \accentset{(n)}{Y}_\nu\RaiseBrace{\bigr)}\,\accentset{(n)}{\kappa}
 +6\Tr( Y_u^\dagger Y_u)\,\accentset{(n)}{\kappa} 
 -\frac{6}{5} g_1^2 \:\accentset{(n)}{\kappa}- 6 g_2^2 \:
 \accentset{(n)}{\kappa}
\;,
\\
\label{eq:BetaMintheMSSM} (4\pi)^2 \accentset{(n)}{\beta}^{\,(1)}_{M} &=& \vphantom{\frac{1}{2}}
   2\,\RaiseBrace{\bigl(}\accentset{(n)}{Y}_\nu
   \accentset{(n)}{Y}^\dagger_\nu \RaiseBrace{\bigr)}\, \accentset{(n)}{M} 
   + 2\,\accentset{(n)}{M}\,\RaiseBrace{\bigl(}\accentset{(n)}{Y}_\nu   
   \accentset{(n)}{Y}^\dagger_\nu \RaiseBrace{\bigr)}^T ,
\\
   (4\pi)^2 \accentset{(n)}{\beta}^{\,(1)}_{Y_\nu}
 &=&
 \accentset{(n)}{Y}_\nu
 \left\{ 3 \accentset{(n)}{Y}^\dagger_\nu
\accentset{(n)}{Y}_\nu + Y_e^\dagger Y_e
+ \Tr \RaiseBrace{\bigl(}\accentset{(n)}{Y}^{\dagger}_\nu  
\accentset{(n)}{Y}_\nu\RaiseBrace{\bigr)} +3\Tr (Y_u^\dagger Y_u)
- \frac{3}{5} g_1^2 - 3 g_2^2 \right\} ,
 \\
(4\pi)^2 \accentset{(n)}{\beta}_{Y_d}^{\,(1)} 
 & = & Y_d\left\{
 3Y_d^\dagger Y_d 
 + Y_u^\dagger Y_u 
 + 3\Tr(Y_d^\dagger Y_d) 
 + \Tr(Y_e^\dagger Y_e)
 - \frac{7}{15}g_1^2
 - 3g_2^2 
 - \frac{16}{3}g_3^2
 \right\},
 \nonumber\\
 \\
(4\pi)^2 \accentset{(n)}{\beta}_{Y_u}^{\,(1)} 
 & = & Y_u\left\{
 Y_d^\dagger Y_d 
 + 3 Y_u^\dagger Y_u 
 + \Tr( \accentset{(n)}{Y}_\nu ^\dagger  \accentset{(n)}{Y}_\nu ) 
 + 3\Tr(Y_u^\dagger Y_u)
 - \frac{13}{15}g_1^2 
 - 3g_2^2 
 - \frac{16}{3}g_3^2 
 \right\} ,
 \nonumber\\
 \\
(4\pi)^2\accentset{(n)}{\beta}_{Y_e}^{\,(1)} 
 & = & Y_e\left\{
 3Y_e^\dagger Y_e 
 +  \accentset{(n)}{Y}_\nu ^\dagger  \accentset{(n)}{Y}_\nu  
 + 3\Tr(Y_d^\dagger Y_d) 
 + \Tr(Y_e^\dagger Y_e)
 - \frac{9}{5}g_1^2 
 - 3g_2^2 
 \right\}.
\end{eqnarray}
\end{subequations}

The results for the 2-loop parts, which are an extension of the usual
2-loop $\beta$-functions for the MSSM \cite{Martin:1993yx}, are
\cite{Antusch:2002ek}
\begin{subequations}
\begin{eqnarray}\label{eq:NuMassOpRGE2LoopMSSM}
 \lefteqn{(4\pi)^4\, {\accentset{(n)}{\beta}^{\,(2)}_\kappa} =
 \left[ \vphantom{\frac{1}{2}} -6  \Tr (Y_u   Y_d^\dagger  Y_d  Y_u^\dagger) 
 - 18  \Tr (Y_u  Y_u^\dagger  Y_u  Y_u^\dagger)\right. 
 -  2  \Tr (\accentset{(n)}{Y}_\nu  Y_e^\dagger  Y_e \accentset{(n)}{Y}_\nu^\dagger )
 }\nonumber \\
 && \left. {} 
  - 6  \Tr (\accentset{(n)}{Y}_\nu^\dagger   
  \accentset{(n)}{Y}_\nu  \accentset{(n)}{Y}_\nu^\dagger  \accentset{(n)}{Y}_\nu) 
+ \frac{8}{5} g_1^2 \Tr (Y_u^\dagger  Y_u)
 +  32 g_3^2 \Tr (Y_u^\dagger  Y_u)+  \frac{207}{25}  g_1^4
 +  \frac{18}{5} g_1^2 g_2^2 
 + 15 g_2^4 \right]
  \accentset{(n)}{\kappa} 
 \nonumber \\
 && {}
 -
  \left[  2  (Y_e^\dagger  Y_e  Y_e^\dagger   Y_e )^T
 + 2 (\accentset{(n)}{Y}_\nu^\dagger   
  \accentset{(n)}{Y}_\nu  \accentset{(n)}{Y}_\nu^\dagger 
  \accentset{(n)}{Y}_\nu)^T
 +\left(
    \Tr (\accentset{(n)}{Y}_\nu \accentset{(n)}{Y}_\nu^\dagger) 
  + 3  \Tr (Y_u  Y_u^\dagger)
 \right)
  (\accentset{(n)}{Y}_\nu^\dagger \accentset{(n)}{Y}_\nu)^T 
  \right. \nonumber \\ 
 && \hphantom{-\left[\vphantom{\frac{1}{2}}\right.} \left. 
 + \left(
  -\frac{6}{5} g_1^2
  +  \Tr (Y_e  Y_e^\dagger) 
  + 3  \Tr (Y_d  Y_d^\dagger)
 \right)
  (Y_e^\dagger  Y_e)^T \right]
   \accentset{(n)}{\kappa} 
 \nonumber \\
 && {} - 
 \accentset{(n)}{\kappa}   \left[  
  2  Y_e^\dagger  Y_e  Y_e^\dagger  Y_e 
  + 2 \accentset{(n)}{Y}_\nu^\dagger 
  \accentset{(n)}{Y}_\nu  \accentset{(n)}{Y}_\nu^\dagger  \accentset{(n)}{Y}_\nu
 +\left(
    \Tr (\accentset{(n)}{Y}_\nu \accentset{(n)}{Y}_\nu^\dagger) 
   + 3  \Tr (Y_u  Y_u^\dagger)
 \right)
  \accentset{(n)}{Y}_\nu^\dagger \accentset{(n)}{Y}_\nu 
  \right. \nonumber \\
 && \hphantom{-\accentset{(n)}{\kappa} \left[\vphantom{\frac{1}{2}}\right.} \left. 
 +\left(
 - \frac{6}{5} g_1^2
  +  \Tr (Y_e  Y_e^\dagger) 
  + 3  \Tr (Y_d  Y_d^\dagger)
 \right)
 Y_e^\dagger  Y_e \right]  ,
\\
 \lefteqn{(4\pi)^4\accentset{(n)}{\beta}_M^{\,(2)} 
  = 
 \accentset{(n)}{M} \left[\vphantom{\frac{1}{2}}
 - 2 \accentset{(n)}{Y}_\nu ^*    Y_e^T    Y_e^*     \accentset{(n)}{Y}_\nu ^T 
 - 2 \accentset{(n)}{Y}_\nu ^*     \accentset{(n)}{Y}_\nu ^T     \accentset{(n)}{Y}_\nu ^*     \accentset{(n)}{Y}_\nu ^T 
 - 6 \accentset{(n)}{Y}_\nu ^*     \accentset{(n)}{Y}_\nu ^T\, \Tr(Y_u     Y_u^\dagger) \right.
 }\nonumber\\
 && {} \qquad\qquad\quad\:
 - 2 \accentset{(n)}{Y}_\nu ^*     \accentset{(n)}{Y}_\nu ^T\,   \Tr( \accentset{(n)}{Y}_\nu       \accentset{(n)}{Y}_\nu ^\dagger) 
 + \frac{6}{5} g_1^2\,  \accentset{(n)}{Y}_\nu ^*  \accentset{(n)}{Y}_\nu ^T
 + 6g_2^2\, \accentset{(n)}{Y}_\nu ^*     \accentset{(n)}{Y}_\nu ^T 
 \left.\vphantom{\frac{1}{2}} \right]
 \nonumber\\
 && {}
 +\left[\vphantom{\frac{1}{2}}
 - 2 \accentset{(n)}{Y}_\nu  Y_e^\dagger Y_e     \accentset{(n)}{Y}_\nu ^\dagger
 - 2 \accentset{(n)}{Y}_\nu      \accentset{(n)}{Y}_\nu ^\dagger     \accentset{(n)}{Y}_\nu   \accentset{(n)}{Y}_\nu ^\dagger
 - 6 \accentset{(n)}{Y}_\nu   \accentset{(n)}{Y}_\nu ^\dagger\, \Tr(Y_u     Y_u^\dagger)
 - 2 \accentset{(n)}{Y}_\nu   \accentset{(n)}{Y}_\nu ^\dagger\, \Tr( \accentset{(n)}{Y}_\nu       \accentset{(n)}{Y}_\nu ^\dagger) 
 \right.
 \nonumber\\
 && {} \hphantom{+ \left[\vphantom{\frac{1}{2}}\right.}
 + \frac{6}{5} g_1^2\, \accentset{(n)}{Y}_\nu   \accentset{(n)}{Y}_\nu ^\dagger
 + 6 g_2^2\, \accentset{(n)}{Y}_\nu   \accentset{(n)}{Y}_\nu ^\dagger 
 \left.\vphantom{\frac{1}{2}}\right] \accentset{(n)}{M}\;,
\\
 \lefteqn{(4\pi)^4 \accentset{(n)}{\beta}_{ Y_\nu }^{\,(2)}
  = 
  \accentset{(n)}{Y}_\nu \left\{\vphantom{\frac{1}{2}}
  \right.
 - 2 Y_e^\dagger Y_e Y_e^\dagger Y_e 
 - 2 Y_e^\dagger Y_e  \accentset{(n)}{Y}_\nu ^\dagger
          \accentset{(n)}{Y}_\nu  
 - 4  \accentset{(n)}{Y}_\nu ^\dagger  \accentset{(n)}{Y}_\nu 
          \accentset{(n)}{Y}_\nu ^\dagger  \accentset{(n)}{Y}_\nu  
 -  3 Y_e^\dagger Y_e\Tr(Y_d Y_d^\dagger) 
 }\nonumber\\*
 & & {}
 - Y_e^\dagger Y_e\Tr(Y_e Y_e^\dagger) 
 - 3 \accentset{(n)}{Y}_\nu ^\dagger  \accentset{(n)}{Y}_\nu 
   \Tr( \accentset{(n)}{Y}_\nu   \accentset{(n)}{Y}_\nu ^\dagger) 
 - 9 \accentset{(n)}{Y}_\nu ^\dagger  \accentset{(n)}{Y}_\nu 
        \Tr(Y_u Y_u^\dagger) \vphantom{\frac{1}{2}}
 - \Tr( \accentset{(n)}{Y}_\nu  Y_e^\dagger
       Y_e  \accentset{(n)}{Y}_\nu^\dagger )
 \nonumber\\*
 & & {}
 - 3\Tr( \accentset{(n)}{Y}_\nu ^\dagger  \accentset{(n)}{Y}_\nu  
         \accentset{(n)}{Y}_\nu ^\dagger   \accentset{(n)}{Y}_\nu ) 
 - 3\Tr(Y_u Y_d^\dagger Y_d Y_u^\dagger) 
 - 9\Tr(Y_u Y_u^\dagger Y_u Y_u^\dagger) \vphantom{\frac{1}{2}}
 + \frac{6}{5} g_1^2\,Y_e^\dagger Y_e
 + \frac{6}{5} g_1^2\, \accentset{(n)}{Y}_\nu^\dagger \accentset{(n)}{Y}_\nu 
 \nonumber\\*
 & & {}
 + 6 g_2^2\, \accentset{(n)}{Y}_\nu ^\dagger  \accentset{(n)}{Y}_\nu 
 + \frac{4}{5} g_1^2\, \Tr(Y_u^\dagger Y_u)
 +  16 g_3^2\,\Tr(Y_u^\dagger Y_u)+ \frac{207}{50} g_1^4
 + \frac{9}{5} g_1^2 g_2^2
 + \frac{15}{2} g_2^4
 \left.\vphantom{\frac{1}{2}}\right\},
\nonumber\\*
\\
\lefteqn{(4\pi)^4 \accentset{(n)}{\beta}_{Y_d}^{\,(2)}
  = 
 Y_d\left\{\vphantom{\frac{1}{2}}
 - 4Y_d^\dagger Y_d Y_d^\dagger Y_d 
 - 2Y_u^\dagger Y_u Y_d^\dagger Y_d 
 - 2Y_u^\dagger Y_u Y_u^\dagger Y_u
  - 9\Tr(Y_d Y_d^\dagger Y_d Y_d^\dagger)
 \right.
 }
 \nonumber\\ 
 & &\vphantom{\frac{1}{2}}  
 - 3 \Tr(Y_d Y_u^\dagger Y_u Y_d^\dagger) 
 - 3 \Tr(Y_e Y_e^\dagger Y_e Y_e^\dagger) 
 - \Tr(Y_e  \accentset{(n)}{Y}_\nu^\dagger   
               \accentset{(n)}{Y}_\nu  Y_e^\dagger) 
 - 9 Y_d^\dagger Y_d\,\Tr(Y_d Y_d^\dagger) 
 \nonumber\\
 & &\vphantom{\frac{1}{2}}  
 - 3 Y_d^\dagger Y_d\,\Tr(Y_e Y_e^\dagger) 
  - Y_u^\dagger Y_u\,\Tr( \accentset{(n)}{Y}_\nu  
         \accentset{(n)}{Y}_\nu ^\dagger) 
 - 3  Y_u^\dagger Y_u\,\Tr(Y_u Y_u^\dagger)
 + 6 g_2^2\,Y_d^\dagger Y_d 
 \nonumber\\
 & &\vphantom{\frac{1}{2}}  
 + \frac{4}{5} g_1^2\, Y_d^\dagger Y_d
 + \frac{4}{5} g_1^2\, Y_u^\dagger Y_u
 - \frac{2}{5} g_1^2\,\Tr(Y_d^\dagger Y_d)
 + \frac{6}{5} g_1^2\,\Tr(Y_e^\dagger Y_e) 
 + 16 g_3^2\, \Tr(Y_d^\dagger Y_d) 
 \nonumber\\
 & &\vphantom{\frac{1}{2}}  
 + \frac{287}{90} g_1^4 
 + g_1^2 g_2^2 
 + \frac{15}{2} g_2^4 
 + \frac{8}{9} g_1^2 g_3^2
 + 8 g_2^2 g_3^2 
 - \frac{16}{9} g_3^4
 \left.\vphantom{\frac{1}{2}}\right\},
\\
\lefteqn{(4\pi)^4 \accentset{(n)}{\beta}_{Y_u}^{\,(2)}
  = 
 Y_u\left\{
 - 2 Y_d^\dagger Y_d Y_d^\dagger Y_d
 - 2 Y_d^\dagger Y_d Y_u^\dagger Y_u 
 - 4 Y_u^\dagger Y_u Y_u^\dagger Y_u 
 - 3 Y_d^\dagger Y_d\Tr(Y_d Y_d^\dagger)
 \right.}
 \nonumber\\
 & & \vphantom{\frac{1}{2}}
 - Y_d^\dagger Y_d\Tr(Y_e Y_e^\dagger)
  - 9 Y_u^\dagger Y_u\Tr(Y_u Y_u^\dagger) 
 - 3Y_u^\dagger Y_u\Tr( \accentset{(n)}{Y}_\nu  
  \accentset{(n)}{Y}_\nu ^\dagger) 
 \nonumber\\
 & & \vphantom{\frac{1}{2}}
 - 3\Tr(Y_u Y_d^\dagger Y_d Y_u^\dagger) 
 - 9\Tr(Y_u Y_u^\dagger Y_u Y_u^\dagger) 
 - \Tr( \accentset{(n)}{Y}_\nu  Y_e^\dagger Y_e  \accentset{(n)}{Y}_\nu^\dagger ) 
 - 3\Tr( \accentset{(n)}{Y}_\nu   \accentset{(n)}{Y}_\nu^\dagger   \accentset{(n)}{Y}_\nu  \accentset{(n)}{Y}_\nu^\dagger ) 
 \nonumber\\
 & & \vphantom{\frac{1}{2}}
 + \frac{2}{5} g_1^2\, Y_d^\dagger Y_d 
 + \frac{2}{5} g_1^2\, Y_u^\dagger Y_u
 + 6 g_2^2 \,Y_u^\dagger Y_u
 + \frac{4}{5} g_1^2\,\Tr(Y_u^\dagger Y_u) 
 + 16 g_3^2\,\Tr(Y_u^\dagger Y_u)
 \nonumber\\
 & & \vphantom{\frac{1}{2}}
 + \frac{2743}{450} g_1^4
 + g_1^2 g_2^2 
 + \frac{15}{2} g_2^4
 + \frac{136}{45} g_1^2 g_3^2 
 + 8 g_2^2 g_3^2 
 - \frac{16}{9} g_3^4
 \left.\vphantom{\frac{1}{2}}\right\},
 \\
\lefteqn{(4\pi)^4  \accentset{(n)}{\beta}_{Y_e}^{\,(2)}
  = 
 Y_e\left\{
 \vphantom{\frac{1}{2}}
 - 4Y_e^\dagger Y_e Y_e^\dagger Y_e 
 - 2 \accentset{(n)}{Y}_\nu ^\dagger  \accentset{(n)}{Y}_\nu 
         Y_e^\dagger Y_e 
 - 2 \accentset{(n)}{Y}_\nu ^\dagger  \accentset{(n)}{Y}_\nu 
          \accentset{(n)}{Y}_\nu ^\dagger  \accentset{(n)}{Y}_\nu
- 9Y_e^\dagger Y_e\,\Tr(Y_d Y_d^\dagger)          
          \right.}  
 \nonumber\\*
 & &\vphantom{\frac{1}{2}} 
 - 3 Y_e^\dagger Y_e\,\Tr(Y_e Y_e^\dagger) 
 -  \accentset{(n)}{Y}_\nu ^\dagger  \accentset{(n)}{Y}_\nu \Tr( \accentset{(n)}{Y}_\nu   \accentset{(n)}{Y}_\nu ^\dagger) 
 - 3  \accentset{(n)}{Y}_\nu ^\dagger  \accentset{(n)}{Y}_\nu
 \Tr(Y_u Y_u^\dagger) 
 - 9\Tr(Y_d Y_d^\dagger Y_d Y_d^\dagger) 
 \nonumber\\*
 & & \vphantom{\frac{1}{2}}
 - 3\Tr(Y_d Y^\dagger_u Y_u Y_d^\dagger) 
 - 3\Tr(Y_e Y^\dagger_e Y_e Y_e^\dagger) 
 - \Tr(Y_e  \accentset{(n)}{Y}_\nu^\dagger   
                    \accentset{(n)}{Y}_\nu Y_e^\dagger) 
 + \frac{6}{5}g_1^2\Tr(Y_e^\dagger Y_e)
 \nonumber\\*
 & &\vphantom{\frac{1}{2}} 
 + 6 g_2^2\,Y_e^\dagger Y_e
 - \frac{2}{5} g_1^2\,\Tr(Y_d^\dagger Y_d) 
 + 16 g_3^2\, \Tr(Y_d^\dagger Y_d) 
 + \frac{27}{2} g_1^4
 + \frac{9}{5} g_1^2 g_2^2
 + \frac{15}{2} g_2^4
 \left.\vphantom{\frac{1}{2}}\right\}.
 \nonumber\\*
\end{eqnarray}
\end{subequations}
The 2-loop $\beta$-functions for the gauge couplings in the presence of
$Y_\nu$ can be found in \cite{Casas:2000pa}.

\clearpage

\bibliography{Running}

\providecommand{\bysame}{\leavevmode\hbox to3em{\hrulefill}\thinspace}
\begin{thebibliography}{10}

\bibitem{Minkowski:1977sc}
P.~Minkowski, Phys. Lett. \textbf{B67} (1977), 421.

\bibitem{Yanagida:1980}
T.~Yanagida, \emph{Horizontal gauge symmetry and masses of neutrinos}, in
  \emph{Proceedings of the Workshop on The Unified Theory and the Baryon Number
  in the Universe} (O.~Sawada and A.~Sugamoto, eds.), KEK, Tsukuba, Japan,
  1979, p.~95.

\bibitem{Glashow:1979vf}
S.~L. Glashow, \emph{The future of elementary particle physics}, in
  \emph{Proceedings of the 1979 Carg{\`e}se Summer Institute on Quarks and
  Leptons} (M.~L{\'e}vy, J.-L. Basdevant, D.~Speiser, J.~Weyers, R.~Gastmans,
  and M.~Jacob, eds.), Plenum Press, New York, 1980, pp.~687--713.

\bibitem{Gell-Mann:1980vs}
M.~Gell-Mann, P.~Ramond, and R.~Slansky, \emph{Complex spinors and unified
  theories}, in \emph{Supergravity} (P.~van Nieuwenhuizen and D.~Z. Freedman,
  eds.), North Holland, Amsterdam, 1979, p.~315.

\bibitem{Mohapatra:1980ia}
R.~N. Mohapatra and G.~Senjanovi{\'c}, Phys. Rev. Lett. \textbf{44} (1980),
  912.

\bibitem{Balaji:2000au}
K.~R.~S. Balaji, A.~S. Dighe, R.~N. Mohapatra, and M.~K. Parida, Phys. Lett.
  \textbf{B481} (2000), 33--38,  [hep-ph/0002177].

\bibitem{Miura:2000bj}
T.~Miura, E.~Takasugi, and M.~Yoshimura, Prog. Theor. Phys. \textbf{104}
  (2000), 1173--1187,  [hep-ph/0007066].

\bibitem{Antusch:2002fr}
S.~Antusch and M.~Ratz, JHEP \textbf{11} (2002), 010,  [hep-ph/0208136].

\bibitem{Mohapatra:2003tw}
R.~N. Mohapatra, M.~K. Parida, and G.~Rajasekaran, Phys. Rev. \textbf{D69}
  (2004), 053007,  [hep-ph/0301234].

\bibitem{Hagedorn:2004ba}
C.~Hagedorn, J.~Kersten, and M.~Lindner, Phys. Lett. \textbf{B597} (2004),
  63--72,  [hep-ph/0406103].

\bibitem{Antusch:2002hy}
S.~Antusch, J.~Kersten, M.~Lindner, and M.~Ratz, Phys. Lett. \textbf{B544}
  (2002), 1--10,  [hep-ph/0206078].

\bibitem{Miura:2003if}
T.~Miura, T.~Shindou, and E.~Takasugi, Phys. Rev. \textbf{D68} (2003), 093009,
  [hep-ph/0308109].

\bibitem{Shindou:2004tv}
T.~Shindou and E.~Takasugi, Phys. Rev. \textbf{D70} (2004), 013005,
  [hep-ph/0402106].

\bibitem{Chankowski:2000fp}
P.~H. Chankowski, A.~Ioannisian, S.~Pokorski, and J.~W.~F. Valle, Phys. Rev.
  Lett. \textbf{86} (2001), 3488--3491,  [hep-ph/0011150].

\bibitem{Chun:2001kh}
E.~J. Chun, Phys. Lett. \textbf{B505} (2001), 155--160,  [hep-ph/0101170].

\bibitem{Chen:2001gk}
M.-C. Chen and K.~T. Mahanthappa, Int. J. Mod. Phys. \textbf{A16} (2001),
  3923--3930,  [hep-ph/0102215].

\bibitem{Joshipura:2002xa}
A.~S. Joshipura, S.~D. Rindani, and N.~N. Singh, Nucl. Phys. \textbf{B660}
  (2003), 362--372,  [hep-ph/0211378].

\bibitem{Joshipura:2002gr}
A.~S. Joshipura and S.~D. Rindani, Phys. Rev. \textbf{D67} (2003), 073009,
  [hep-ph/0211404].

\bibitem{Singh:2004zu}
N.~N. Singh and M.~K. Das,  (2004),  hep-ph/0407206.

\bibitem{Antusch:2003kp}
S.~Antusch, J.~Kersten, M.~Lindner, and M.~Ratz, Nucl. Phys. \textbf{B674}
  (2003), 401--433,  [hep-ph/0305273].

\bibitem{Mei:2004rn}
J.-w. Mei and Z.-z. Xing, Phys. Rev. \textbf{D70} (2004), 053002,
  [hep-ph/0404081].

\bibitem{Antusch:2004yx}
S.~Antusch, P.~Huber, J.~Kersten, T.~Schwetz, and W.~Winter, Phys. Rev.
  \textbf{D70} (2004), 097302,  [hep-ph/0404268].

\bibitem{Tanimoto:1995bf}
M.~Tanimoto, Phys. Lett. \textbf{B360} (1995), 41--46,  [hep-ph/9508247].

\bibitem{Casas:1999tp}
J.~A. Casas, J.~R. Espinosa, A.~Ibarra, and I.~Navarro, Nucl. Phys.
  \textbf{B556} (1999), 3--22,  [hep-ph/9904395].

\bibitem{Casas:1999ac}
J.~A. Casas, J.~R. Espinosa, A.~Ibarra, and I.~Navarro, Nucl. Phys.
  \textbf{B569} (2000), 82--106,  [hep-ph/9905381].

\bibitem{King:2000hk}
S.~F. King and N.~N. Singh, Nucl. Phys. \textbf{B591} (2000), 3--25,
  [hep-ph/0006229].

\bibitem{Antusch:2002rr}
S.~Antusch, J.~Kersten, M.~Lindner, and M.~Ratz, Phys. Lett. \textbf{B538}
  (2002), 87--95,  [hep-ph/0203233].

\bibitem{Gogoladze:2003pp}
I.~Gogoladze, Y.~Mimura, S.~Nandi, and K.~Tobe, Phys. Lett. \textbf{B575}
  (2003), 66--74,  [hep-ph/0307397].

\bibitem{Casas:2000pa}
J.~A. Casas, J.~R. Espinosa, A.~Ibarra, and I.~Navarro, Phys. Rev. \textbf{D63}
  (2001), 097302,  [hep-ph/0004166].

\bibitem{Kielanowski:2003jg}
P.~Kielanowski and S.~R. Juarez~W.,  (2003),  hep-ph/0310122.

\bibitem{Grzadkowski:1987tf}
B.~Grzadkowski and M.~Lindner, Phys. Lett. \textbf{B193} (1987), 71.

\bibitem{Grzadkowski:1987wr}
B.~Grzadkowski, M.~Lindner, and S.~Theisen, Phys. Lett. \textbf{B198} (1987),
  64.

\bibitem{Pich:1998xt}
A.~Pich,  (1998),  hep-ph/9806303.

\bibitem{Chankowski:1993tx}
P.~H. Chankowski and Z.~Pluciennik, Phys. Lett. \textbf{B316} (1993), 312--317,
   [hep-ph/9306333].

\bibitem{Babu:1993qv}
K.~S. Babu, C.~N. Leung, and J.~Pantaleone, Phys. Lett. \textbf{B319} (1993),
  191--198,  [hep-ph/9309223].

\bibitem{Antusch:2001ck}
S.~Antusch, M.~Drees, J.~Kersten, M.~Lindner, and M.~Ratz, Phys. Lett.
  \textbf{B519} (2001), 238--242,  [hep-ph/0108005].

\bibitem{Antusch:2001vn}
S.~Antusch, M.~Drees, J.~Kersten, M.~Lindner, and M.~Ratz, Phys. Lett.
  \textbf{B525} (2002), 130--134,  [hep-ph/0110366].

\bibitem{Babu:1987im}
K.~S. Babu, Z. Phys. \textbf{C35} (1987), 69.

\bibitem{Casas:1999tg}
J.~A. Casas, J.~R. Espinosa, A.~Ibarra, and I.~Navarro, Nucl. Phys.
  \textbf{B573} (2000), 652,  [hep-ph/9910420].

\bibitem{Chankowski:1999xc}
P.~H. Chankowski, W.~Krolikowski, and S.~Pokorski, Phys. Lett. \textbf{B473}
  (2000), 109,  [hep-ph/9910231].

\bibitem{Maltoni:2004ei}
M.~Maltoni, T.~Schwetz, M.~A. Tortola, and J.~W.~F. Valle, New J. Phys.
  \textbf{6} (2004), 122,  [hep-ph/0405172].

\bibitem{Haba:2000tx}
N.~Haba, Y.~Matsui, and N.~Okamura, Eur. Phys. J. \textbf{C17} (2000),
  513--520,  [hep-ph/0005075].

\bibitem{Grimus:2004cj}
W.~Grimus and L.~Lavoura,  (2004),  hep-ph/0410279.

\bibitem{LRS}
M.~Lindner, M.~Ratz, and M.~A. Schmidt,  in preparation.

\bibitem{Giudice:2003jh}
G.~F. Giudice, A.~Notari, M.~Raidal, A.~Riotto, and A.~Strumia, Nucl. Phys.
  \textbf{B685} (2004), 89--149,  [hep-ph/0310123].

\bibitem{Mei:2003gn}
J.-w. Mei and Z.-z. Xing, Phys. Rev. \textbf{D69} (2004), 073003,
  [hep-ph/0312167].

\bibitem{Chun:1999vb}
E.~J. Chun and S.~Pokorski, Phys. Rev. \textbf{D62} (2000), 053001,
  [hep-ph/9912210].

\bibitem{Chankowski:2001hx}
P.~H. Chankowski and P.~Wasowicz, Eur. Phys. J. \textbf{C23} (2002), 249--258,
  [hep-ph/0110237].

\bibitem{Rodejohann:2000ne}
W.~Rodejohann, Nucl. Phys. \textbf{B597} (2001), 110--126,  [hep-ph/0008044].

\bibitem{Ferrandis:2004vp}
J.~Ferrandis and S.~Pakvasa,  (2004),  hep-ph/0412038.

\bibitem{Kang:2005as}
S.~K. Kang, C.~S. Kim, and J.~Lee,  (2005),  hep-ph/0501029.

\bibitem{Raidal:2004iw}
M.~Raidal, Phys. Rev. Lett. \textbf{93} (2004), 161801,  [hep-ph/0404046].

\bibitem{Minakata:2004xt}
H.~Minakata and A.~Y. Smirnov, Phys. Rev. \textbf{D70} (2004), 073009,
  [hep-ph/0405088].

\bibitem{Frampton:2004vw}
P.~H. Frampton and R.~N. Mohapatra,  (2004),  hep-ph/0407139.

\bibitem{Huber:2002mx}
P.~Huber, M.~Lindner, and W.~Winter, Nucl. Phys. \textbf{B645} (2002), 3--48,
  [hep-ph/0204352].

\bibitem{Huber:2003ak}
P.~Huber and W.~Winter, Phys. Rev. \textbf{D68} (2003), 037301,
  [hep-ph/0301257].

\bibitem{Huber:2004ug}
P.~Huber, M.~Lindner, M.~Rolinec, T.~Schwetz, and W.~Winter, Phys. Rev.
  \textbf{D70} (2004), 073014,  [hep-ph/0403068].

\bibitem{Ashie:2004mr}
Super-Kamiokande Collaboration, Y.~Ashie et~al., Phys. Rev. Lett. \textbf{93}
  (2004), 101801,  [hep-ex/0404034].

\bibitem{King:1998jw}
S.~F. King, Phys. Lett. \textbf{B439} (1998), 350--356,  [hep-ph/9806440].

\bibitem{King:1999mb}
S.~F. King, Nucl. Phys. \textbf{B576} (2000), 85--105,  [hep-ph/9912492].

\bibitem{Fukugita:1986hr}
M.~Fukugita and T.~Yanagida, Phys. Lett. \textbf{174B} (1986), 45.

\bibitem{Spergel:2003cb}
WMAP, D.~N. Spergel et~al., Astrophys. J. Suppl. \textbf{148} (2003), 175,
  [astro-ph/0302209].

\bibitem{Covi:1996wh}
L.~Covi, E.~Roulet, and F.~Vissani, Phys. Lett. \textbf{B384} (1996), 169--174,
   [hep-ph/9605319].

\bibitem{Antusch:2004xy}
S.~Antusch and S.~F. King, Phys. Lett. \textbf{B597} (2004), 199--207,
  [hep-ph/0405093].

\bibitem{Barbieri:1999ma}
R.~Barbieri, P.~Creminelli, A.~Strumia, and N.~Tetradis,  (2002),
  hep-ph/9911315 v3.

\bibitem{Senoguz:2003hc}
V.~N. Senoguz and Q.~Shafi, Phys. Lett. \textbf{B582} (2004), 6--14,
  [hep-ph/0309134].

\bibitem{Buchmuller:2003gz}
W.~Buchm{\"u}ller, P.~Di~Bari, and M.~Pl{\"u}macher, Nucl. Phys. \textbf{B665}
  (2003), 445--468,  [hep-ph/0302092].

\bibitem{Buchmuller:2004nz}
W.~Buchm{\"u}ller, P.~Di~Bari, and M.~Pl{\"u}macher,  (2004),  hep-ph/0401240.

\bibitem{Flanz:1996fb}
M.~Flanz, E.~A. Paschos, U.~Sarkar, and J.~Weiss, Phys. Lett. \textbf{B389}
  (1996), 693--699,  [hep-ph/9607310].

\bibitem{Pilaftsis:1997jf}
A.~Pilaftsis, Phys. Rev. \textbf{D56} (1997), 5431--5451,  [hep-ph/9707235].

\bibitem{Pilaftsis:2003gt}
A.~Pilaftsis and T.~E.~J. Underwood, Nucl. Phys. \textbf{B692} (2004),
  303--345,  [hep-ph/0309342].

\bibitem{GonzalezFelipe:2003fi}
R.~Gonzalez~Felipe, F.~R. Joaquim, and B.~M. Nobre, Phys. Rev. \textbf{D70}
  (2004), 085009,  [hep-ph/0311029].

\bibitem{Turzynski:2004xy}
K.~Turzynski, Phys. Lett. \textbf{B589} (2004), 135--140,  [hep-ph/0401219].

\bibitem{Hambye:2004jf}
T.~Hambye, J.~March-Russell, and S.~M. West, JHEP \textbf{07} (2004), 070,
  [hep-ph/0403183].

\bibitem{Magg:1980ut}
M.~Magg and C.~Wetterich, Phys. Lett. \textbf{B94} (1980), 61.

\bibitem{Lazarides:1980nt}
G.~Lazarides, Q.~Shafi, and C.~Wetterich, Nucl. Phys. \textbf{B181} (1981),
  287.

\bibitem{Mohapatra:1981yp}
R.~N. Mohapatra and G.~Senjanovi{\'c}, Phys. Rev. \textbf{D23} (1981), 165.

\bibitem{Weinberg:1976hu}
S.~Weinberg, Phys. Rev. Lett. \textbf{37} (1976), 657.

\bibitem{Glashow:1977nt}
S.~L. Glashow and S.~Weinberg, Phys. Rev. \textbf{D15} (1977), 1958.

\bibitem{Paschos:1977ay}
E.~A. Paschos, Phys. Rev. \textbf{D15} (1977), 1966.

\bibitem{Grimus:2004yh}
W.~Grimus and L.~Lavoura,  (2004),  hep-ph/0409231.

\bibitem{Arkani-Hamed:2004fb}
N.~Arkani-Hamed and S.~Dimopoulos,  (2004),  hep-th/0405159.

\bibitem{Giudice:2004tc}
G.~F. Giudice and A.~Romanino, Nucl. Phys. \textbf{B699} (2004), 65--89,
  [hep-ph/0406088].

\bibitem{Casas:2002sn}
J.~A. Casas, J.~R. Espinosa, and I.~Navarro, Phys. Rev. Lett. \textbf{89}
  (2002), 161801,  [hep-ph/0206276].

\bibitem{Casas:2003kh}
J.~A. Casas, J.~R. Espinosa, and I.~Navarro, JHEP \textbf{09} (2003), 048,
  [hep-ph/0306243].

\bibitem{Broncano:2004tz}
A.~Broncano, M.~B. Gavela, and E.~Jenkins, Nucl. Phys. \textbf{B705} (2005),
  269--295,  [hep-ph/0406019].

\bibitem{Chankowski:2001mx}
P.~H. Chankowski and S.~Pokorski, Int. J. Mod. Phys. \textbf{A17} (2002),
  575--614,  [hep-ph/0110249].

\bibitem{Machacek:1983tz}
M.~E. Machacek and M.~T. Vaughn, Nucl. Phys. \textbf{B222} (1983), 83.

\bibitem{Machacek:1984fi}
M.~E. Machacek and M.~T. Vaughn, Nucl. Phys. \textbf{B236} (1984), 221.

\bibitem{Machacek:1985zw}
M.~E. Machacek and M.~T. Vaughn, Nucl. Phys. \textbf{B249} (1985), 70.

\bibitem{Das:2000uk}
C.~R. Das and M.~K. Parida, Eur. Phys. J. \textbf{C20} (2001), 121--137,
  [hep-ph/0010004].

\bibitem{Hill:1985tg}
C.~T. Hill, C.~N. Leung, and S.~Rao, Nucl. Phys. \textbf{B262} (1985), 517.

\bibitem{Martin:1993yx}
S.~P. Martin and M.~T. Vaughn, Phys. Lett. \textbf{B318} (1993), 331--337,
  [hep-ph/9308222].

\bibitem{Antusch:2002ek}
S.~Antusch and M.~Ratz, JHEP \textbf{07} (2002), 059,  [hep-ph/0203027].

\end{thebibliography}
\bibliographystyle{ArXiv}

\end{document}